\documentclass[12pt]{article}
\usepackage{newinutile}
\usepackage{amsmath,amsfonts,amsthm,amssymb}
\usepackage{epsfig,graphics,color,calc,graphicx}
\usepackage{cite}
\usepackage{mathrsfs}

\newlength{\pecettawidth}
\begin{document}
\title{A lattice model approach to the morphology formation 
from ternary mixtures during the  evaporation of one component}

\author{Emilio N.M.\ Cirillo}
\email{emilio.cirillo@uniroma1.it}
\affiliation{Dipartimento di Scienze di Base e Applicate per 
             l'Ingegneria, Sapienza Universit\`a di Roma, 
             via A.\ Scarpa 16, I--00161, Roma, Italy.}

\author{Matteo Colangeli}
\email{matteo.colangeli1@univaq.it}
\affiliation{Dipartimento di Ingegneria e Scienze dell'Informazione e 
Matematica, Universit\`a degli Studi dell'Aquila, 67100 L'Aquila, Italy}

\author{Ellen Moons}
\email{ellen.moons@kau.se}
\affiliation{Department of Engineering and Physics, Karlstad University, 
Sweden}

\author{Adrian Muntean}
\email{adrian.muntean@kau.se}
\affiliation{Department of Mathematics and Computer Science, 
Karlstad University, Sweden}

\author{Stela--Andrea Muntean}
\email{andrea.muntean@kau.se}
\affiliation{Department of Engineering and Physics, Karlstad University, 
Sweden}

\author{Jan van Stam}
\email{jan.van.stam@kau.se}
\affiliation{Department of Engineering and Chemical Sciences, 
Karlstad University, Sweden}


\begin{abstract}
Stimulated by experimental evidence in the field of solution--born 
thin films, we study the morphology formation in a three state lattice 
system subjected to the evaporation of one component. The practical problem 
that we address is the understanding of the parameters that govern morphology 
formation from a ternary mixture upon evaporation, as is the case in the 
fabrication of thin films from solution for organic photovoltaics.
We use, as a tool, a generalized version of the Potts and Blume--Capel 
models in 2D, with the Monte Carlo Kawasaki--Metropolis algorithm, to 
simulate the phase behaviour of a ternary mixture upon evaporation of one 
of its components. The components with spin $+1$, $-1$ and $0$ 
in the Blume--Capel dynamics correspond to the electron--acceptor, 
electron--donor and solvent molecules, respectively, in a ternary mixture 
used in the preparation of the active layer films in an organic solar cell. 
Further, we introduce parameters that account for the relative composition of 
the mixture, temperature, and interaction between the species in the system. 
We identify the parameter regions that are prone to facilitate the phase 
separation.  Furthermore, we study qualitatively the types of formed 
configurations. We show that even a relatively simple model, as the present 
one, can generate key morphological features, similar to those observed in 
experiments, which proves the method valuable for the study of complex systems.
\end{abstract}





\maketitle


\section{Introduction}
\label{intro}
\par\noindent
Multi--state spin systems on a lattice have been widely studied in the Statistical Mechanics literature, we refer the reader for instance to \cite{P2008} where efforts have been invested in connecting microscopic dynamics to dynamics at mesoscopic and/or continuum scales. 
The celebrated Blume--Capel model \cite{B1966,BEG1971,C1966} is a three--state model which has been originally introduced to study the He$^3$--He$^4$ mixture low temperature properties. More recently, the model has received  a lot of interest also in the mathematics literature, since it is a prototype to study the effects of multiple metastable states \cite{CO1996,CN2013,CNS2017,LL2016}.
The Hamiltonian of the Blume--Capel setting is characterized by a nearest neighbor spin--spin interaction term, a chemical potential contribution, and, finally, a term describing the interaction with an external magnetic field. 
In the Blume--Capel model the spin can take the values $-1,0,+1$ and the three different interface pairs $-+$, $0-$, and $0+$ have different energetic costs. 
More precisely, the two pairs containing a zero have the same cost, whereas  the $-+$ one has a larger cost (four times). Pairs with equal spins $--$, $++$, and $00$ have no energetic cost. 

A different multi--state model is the Potts model \cite{P1952,Wu1982} --  a straightforward generalization of the Ising model in which the spin has $q>2$ states and in the Hamiltonian the nearest neighbor interaction and the external field contribution are taken into account. 
In the standard version of the model, all the pairs of spin interact similarly, but in its generalized version different interaction weights can be assigned to different spin pairs. 
In any case, in the Potts model, pairs with differing spins have the same energy cost, whereas pairs with equal spins can be differently favoured. 

In our study, we  use a generalized version of these models (Blume--Capel and Potts) with no external field. 
In the sequel we  consider three states, denoted respectively as $-1$, $0$, and $+1$. The ``$0$'' spin will be interpreted as a molecule of solvent on the site, whereas $\pm1$ will represent the other two components. No constrained symmetry will be assumed on the interaction between nearest neighbor spins, thus, to define the model we shall have to fix six interaction parameters that will be denoted by $J_{\alpha\gamma}$ with $\alpha,\gamma=-1,0,+1$. The number $J_{\alpha\gamma}$ will be the energy cost of two neighboring spins equal to $\alpha$ and $\gamma$. 
With this model at hand, we study the phase separation in a ternary mixture upon evaporation of one of the components. This is a crucial issue in the morphology formation in solution borne thin films, as for example in the preparation of the active layer in organic photovoltaics. In this specific case, the components of the ternary mixture are usually an electron-donor molecule, an electron-acceptor molecule, and a solvent. This subject is of large interest in the community and is discussed in many experimental  and computational \cite{Alison,Battaile,Rolf,Lowengrub,Michels,Du,Lyons2,Lyons1} studies; see also \cite{Olle} for related work done for stochastic models for competitive growth of phases. 

To tackle the problem briefly described above, we need to consider the stochastic version of our spin system and study its non--equilibrium properties. More precisely, we start the system with the initial configuration chosen at random with uniform probability, but with a fixed fraction of the three different spin species. 
Then we follow the evolution until all (or a vast majority of) the zeroes evaporate. At this point we stop the dynamics and analyze the morphology of the final configuration.

The dynamics will be of the Kawasaki--type \cite{K1972}, namely, nearest neighboring spins will be allowed to exchange their positions according to the Metropolis rule\cite{MRRTT1953}. Moreover, the zeroes in the first row of the lattice will evaporate from the system and will be replaced by a plus or a minus with probability chosen proportionally to the initial fractions. 
Finally, a parameter called \emph{volatility} will control the way in which we force upward vertical motion of zeroes.
In computing the energy differences associated to possible spin exchanges periodic boundary conditions will be used. 

Our study is, to some extent, connected to the spinodal decomposition problem \cite{B1994}. 
Indeed, the high temperature initial configuration (uniform random choice of the initial spins), after an abrupt quench, is let to evolve at a low temperature and the phase separation phenomenon is observed. The difference, in our dynamics, is that one species 
(the zeroes) is let to evaporate from the system. We mention, in passing, that, for the Potts model, the domain growth in the spinodal decomposition regime is not completely understood. The scenario is quite clear for the Glauber dynamics at $q=3,4$, where the standard $1/2$ growth exponent is recovered \cite{SM1995}, whereas the understanding is still partial for $q>4$. We refer to \cite{BFCLP2007} for more details. We were not able 
to find any reference for the Potts model with conservative dynamics and for the Blume--Capel model. 
The lattice gas model we consider in this work includes only short range, nearest neighbor-like, interactions and, in this respect, is hence a refined version of the paradigmatic 2D Ising model, which undergoes a phase transition when the temperature falls below a critical temperature computed by Onsager \cite{Onsager1944}.

In spite of its relatively simple microscopic dynamics, our model permits to capture some of the relevant physical features of the thin film evaporation phenomenon that are experimentally accessible. The main results obtained based on our numerical investigations are explained in section  \ref{sec:2}.

The paper is organized as follows: In section \ref{sec:1}, we introduce our Blume-Capel like model in the spirit of \cite{CO1996} to describe the evaporation of the solvent in a ternary reactive mixture. We show our main results on the effects of the evaporation on the morphology formation and discuss them in section \ref{sec:2}. We conclude the paper with a summary of our results in section \ref{sec:4}. 

\section{The model}
\label{sec:1}
\par\noindent
Consider the $2D$ rectangular torus  
$\Lambda=\{1,\dots,L_1\}\times\{1,\dots,L_2\}$ endowed with periodic 
boundary conditions. An element of $\Lambda$ is called \emph{site}.
Two sites are said to be \emph{nearest neighbors} if their 
Euclidean distance is one. 
A pair of nearest neighboring sites is called a \emph{bond}.

Associate the \emph{spin variable} $\sigma_x\in\{-1,0,+1\}$ with each 
site $x=(x_1,x_2)\in\Lambda$.
Fix the reals $J_{\alpha\gamma}$ for any 
$\alpha,\gamma=-1,0,+1$, such that 
$J_{\alpha\gamma}=J_{\gamma\alpha}$. 
The \emph{energy} associated with 
any configuration $\sigma\in\{-1,0,+1\}^\Lambda$ 
is 
\begin{equation}
\label{mod000}
H(\sigma)
=
\frac{1}{2}
\sum_{x,y\in\Lambda}J_{\sigma(x)\sigma(y)}.
\;\;
\end{equation}
The function $H$ is called the \emph{Hamiltonian}.

Consider the integer time variable $t\ge0$. 
Fix the parameter $\beta>0$ and refer to $1/\beta$ as the \emph{temperature}.
Fix the parameter $\phi\in[0,1]$ and call it \emph{volatility}.
Fix $n_0\ge0$.
Fix $p_{-1},p_0, p_{+1}\in[0,1)$ such that $p_{-1}+p_0+p_{+1}=1$ and 
choose the inital $t=0$ configuration by setting any spin $\sigma_x$, for 
any $x\in\Lambda$, equal to $-1$, $0$, or $+1$, respectively, with 
probability $p_{-1}$, $p_0$, and $p_{+1}$. 
At each time the following is repeated $2L_1L_2$ times: 
\begin{itemize}
\item[i)]
choose a bond at random with uniform probability;
\item[ii)]
if the bond is of the type $((x_1,L_2),(x_1,1))$ and $\sigma(x_1,L_2)=0$,
then replace the spin zero at the site $(x_1,L_2)$ with 
$+1$ with probability $p_{+1}/(1-p_0)$ and 
$-1$ with probability $1-p_{+1}/(1-p_0)=p_{-1}/(1-p_0)$ (say that 
the zero \emph{evaporated});
\item[iii)]
if the bond is of the type $((x_1,x_2),(x_1,x_2+1))$ 
with $x_2<L_2$,  
$\sigma(x_1,x_2)=0$,
and $\sigma(x_1,x_2+1)\neq0$,
then
exchange the two spins at the sites of the bond with 
probability $\phi$;
\item[iv)]
otherwise let $\Delta$ be the difference 
of energy between the configuration obtained by exchanging the spins 
at the two sites of the bond and the actual configuration
and
exchange the two spins at the sites of the bond with 
probability $1$ if $\Delta<0$ and $\exp\{-\beta\Delta\}$ 
otherwise.
\end{itemize}
Stop the dynamics when the total number of zeroes in the system becomes smaller than $n_0$ (note that the dynamics never stops if 
$n_0=0$).

The dynamics is, in spirit, the Kawasaki dynamics complemented with a Metropolis updating rule, with the addition of the evaporation rule (step ii) of the algorithm) and the forced zeroes vertical motion (step iii) of the algorithm).

\section{Simulation results and discussion}
\label{sec:2}
\par\noindent
In this section we report our main remarks on the effect of a number of model parameters (including the temperature and the strength of the interactions) on the onset of the phase transitions leading to the strong spatial separation of the two phases. We also investigate the parameter effects on the shape of the obtained morphologies.  

We simulate the process when the characteristic time scale of evaporation is close to the characteristic time scale of diffusion. We show a lateral view with the evaporation taking place along the upper boundary, while the bottom and the lateral sides are reflecting. This is motivated by the experimental condition that in a coated layer on a substrate the solvent evaporates through the surface of the thin film. 

Unless otherwise specified, we shall fix the initial proportion of the three different species of spins, namely $0,+1,-1$, to $40:30:30$. That is, the initial configuration considered in our simulations is the one in which each spin on the lattice is drawn at random from the set $\{0,+1,-1\}$ with probability, respectively, equal to $0.4$, $0.3$ and $0.3$. Our model contains a number of tunable parameters, that allow one to capture a rich variety of morphologies as a result of the annihilation of the ``$0$'' spins (namely, the evaporation). 

It seems that the morphology formation is essentially controlled by the temperature and by the strength of the interaction parameter $J_{+1,-1}$.
If the temperature is high, no phase separation appears. It is worth noting that if $\beta=0.6$ or larger and if a phase separation appears, then this will happen within the interval $[0.4,0.3]$ of the ratio of residual solvent. It seems that this phenomenon happens for all the investigated interaction parameters. Microstructures and morphologies appear at low temperatures, which, then, signals the likely presence of a phase transition of the kind of those typically met with the 2D Ising model.

\begin{figure}[h!]
\centering
\begin{tabular}{ccc}
\includegraphics[width = 0.33\textwidth]{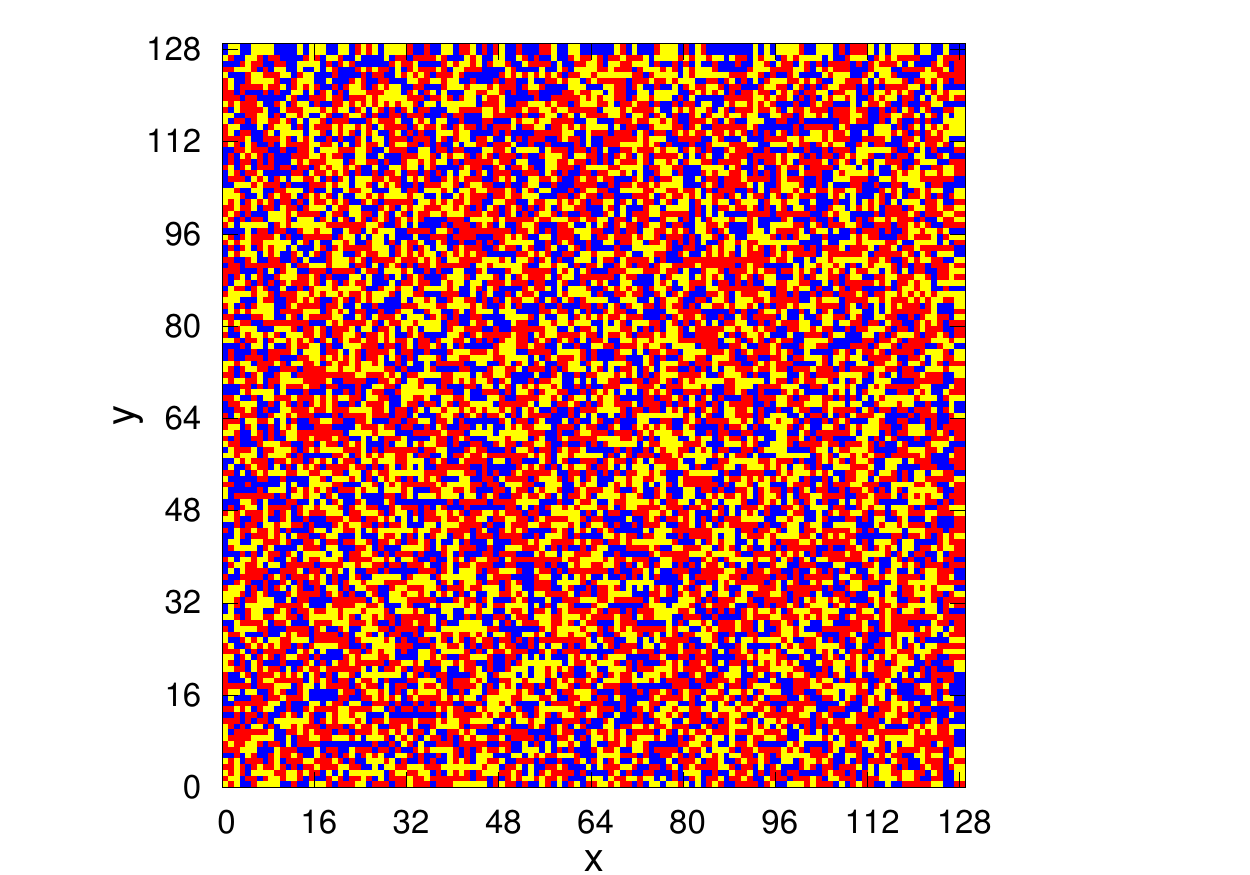} &
\includegraphics[width = 0.33\textwidth]{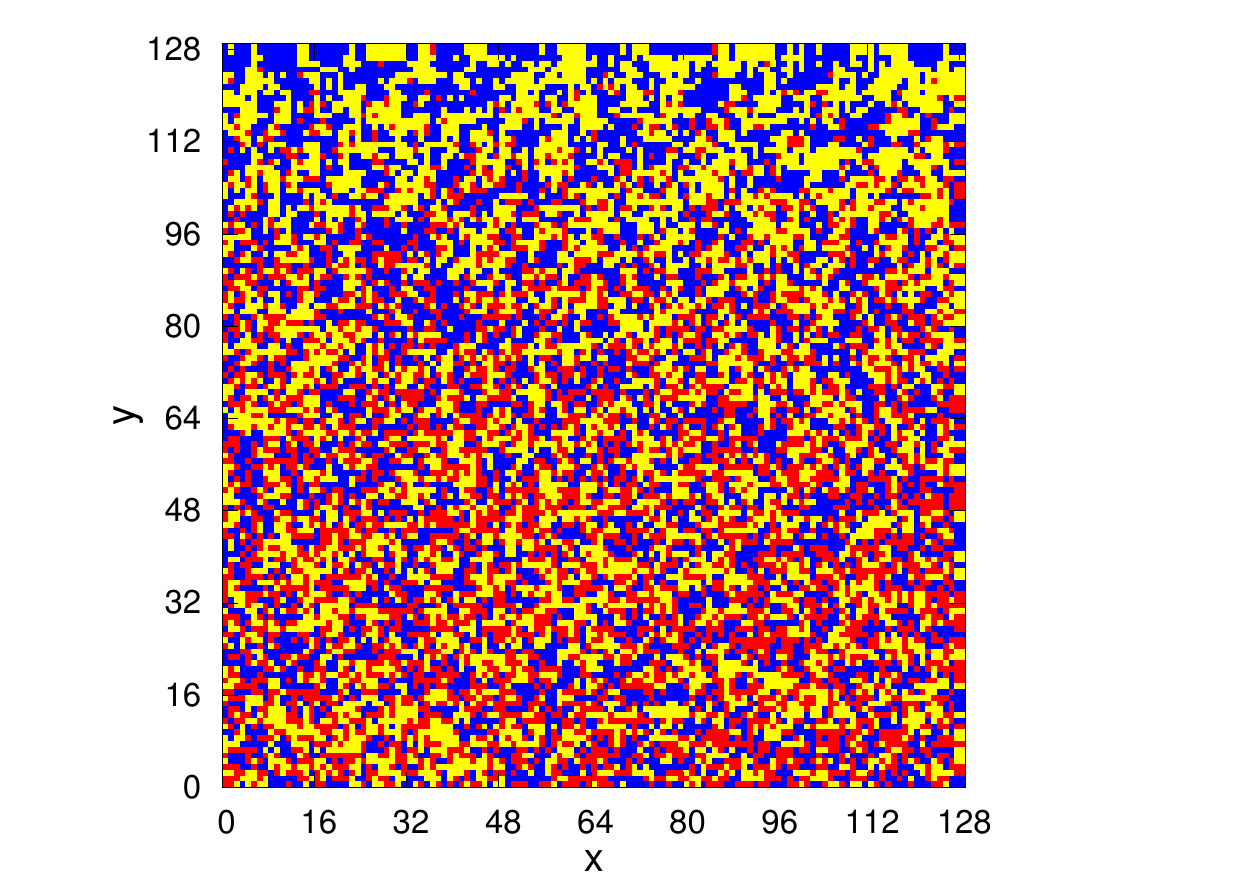}  & 
\includegraphics[width = 0.33\textwidth]{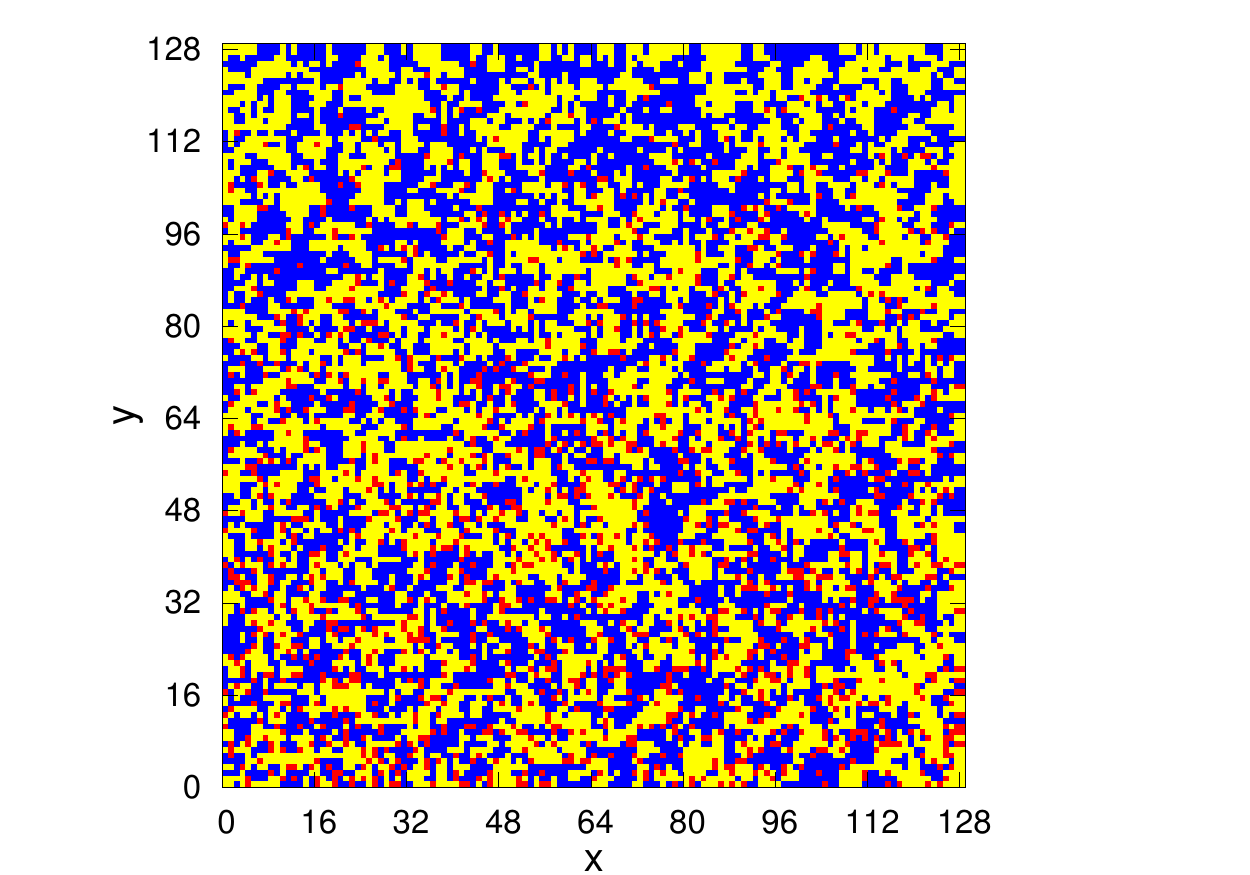}
\\[0.5cm]
\includegraphics[width = 0.33\textwidth]{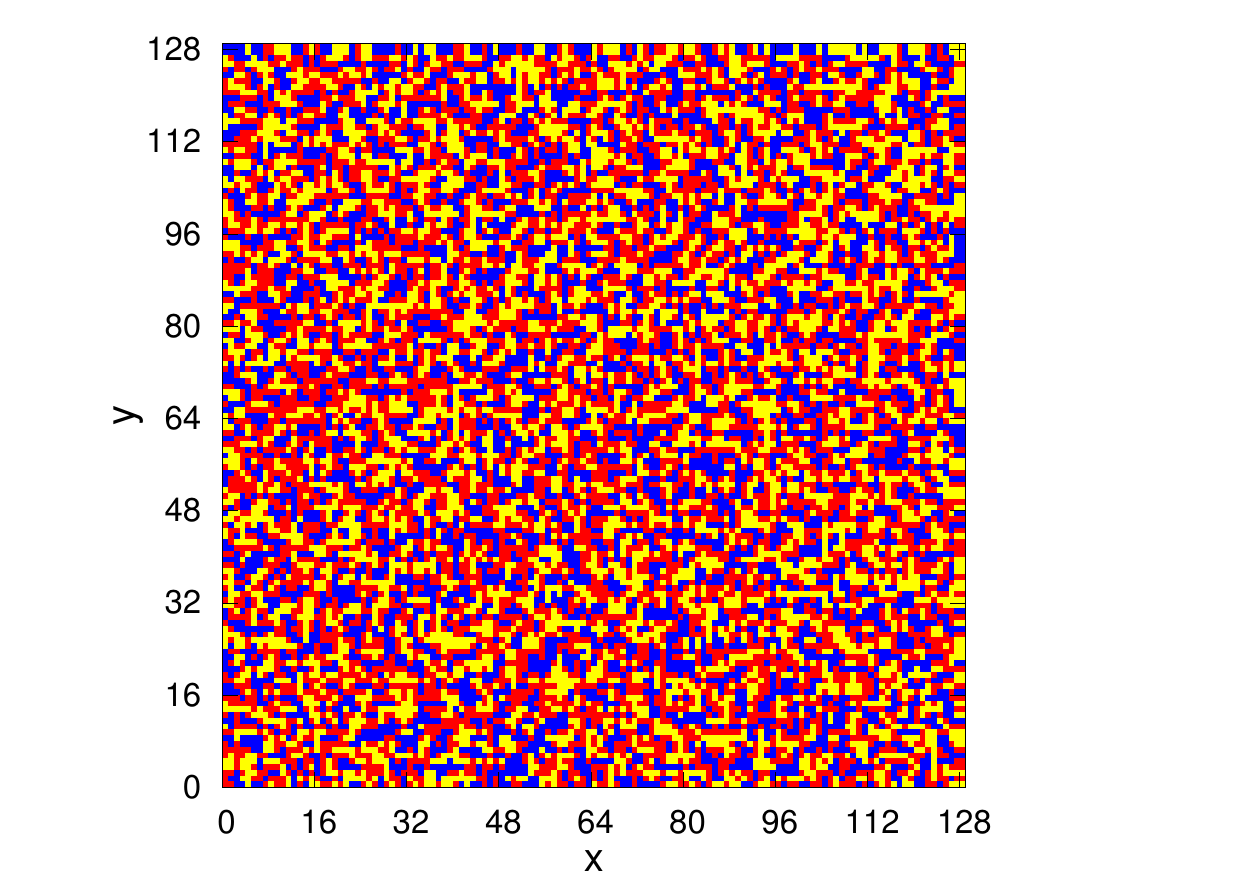} &
\includegraphics[width = 0.33\textwidth]{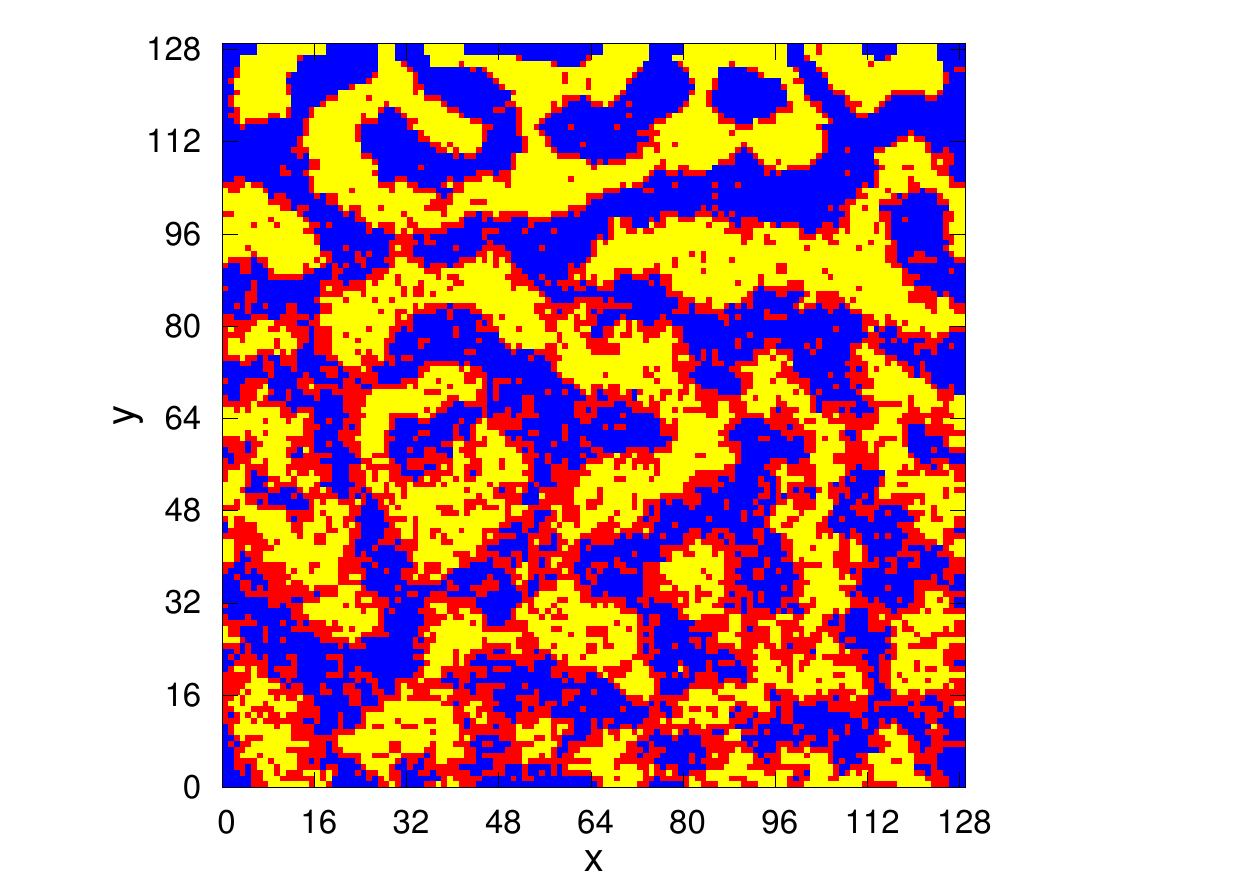}  & 
\includegraphics[width = 0.33\textwidth]{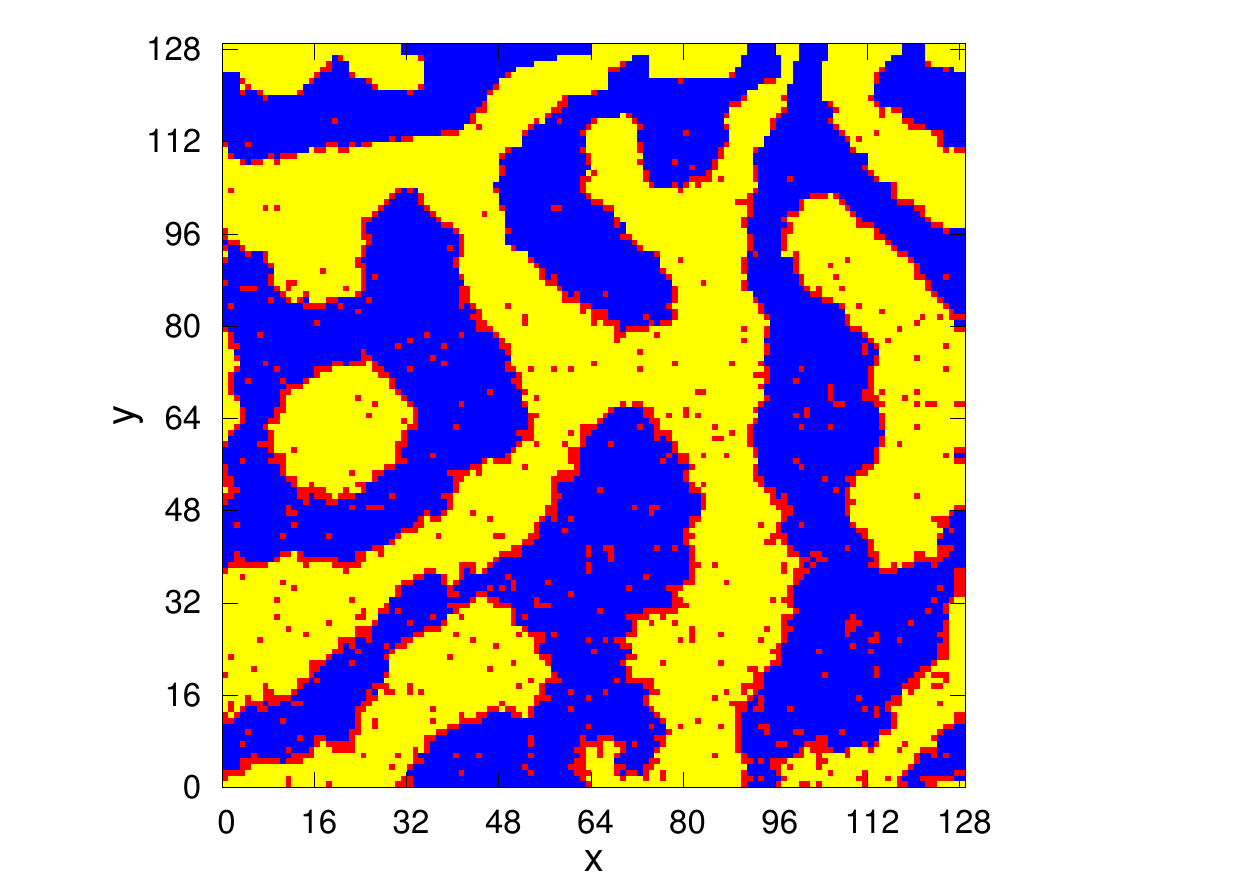}
\\[0.5cm]
\includegraphics[width = 0.33\textwidth]{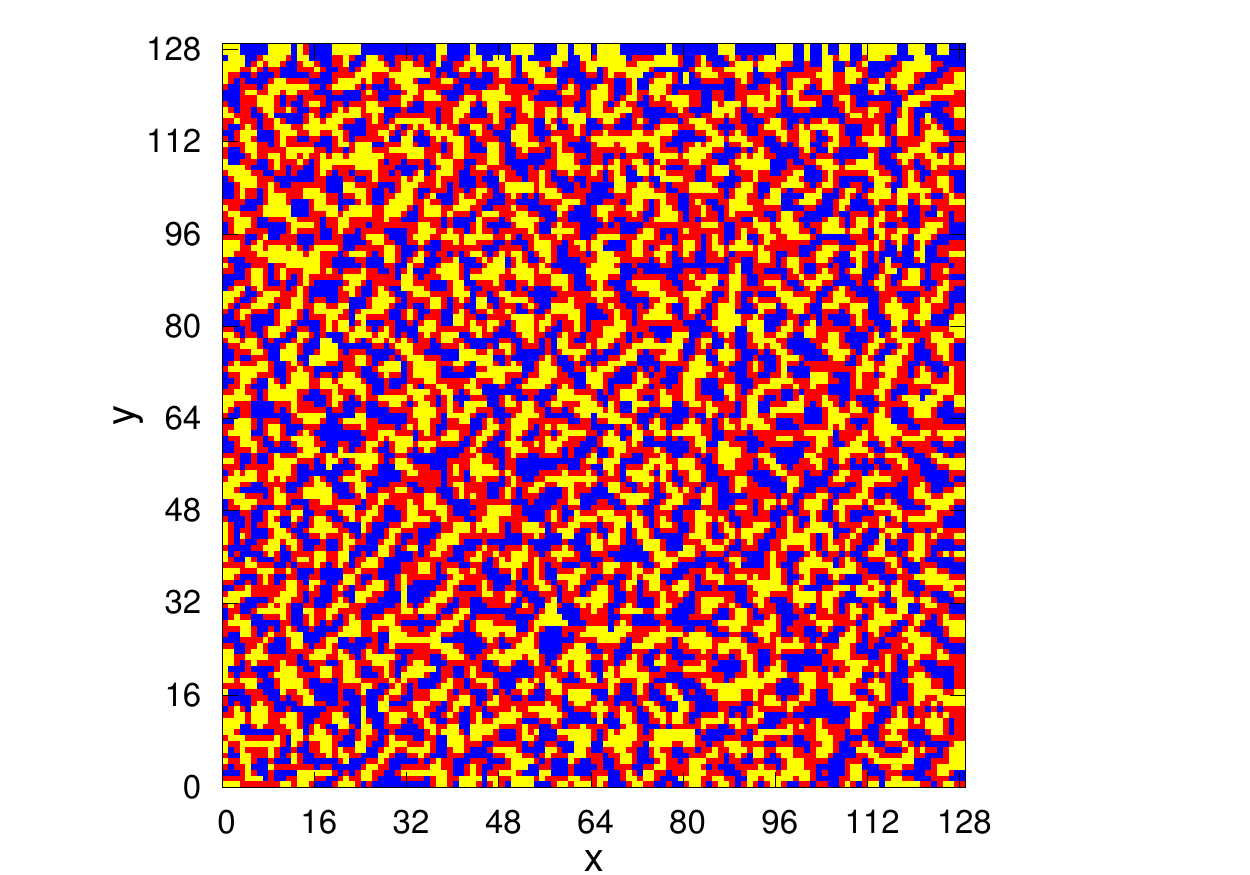} &
\includegraphics[width = 0.33\textwidth]{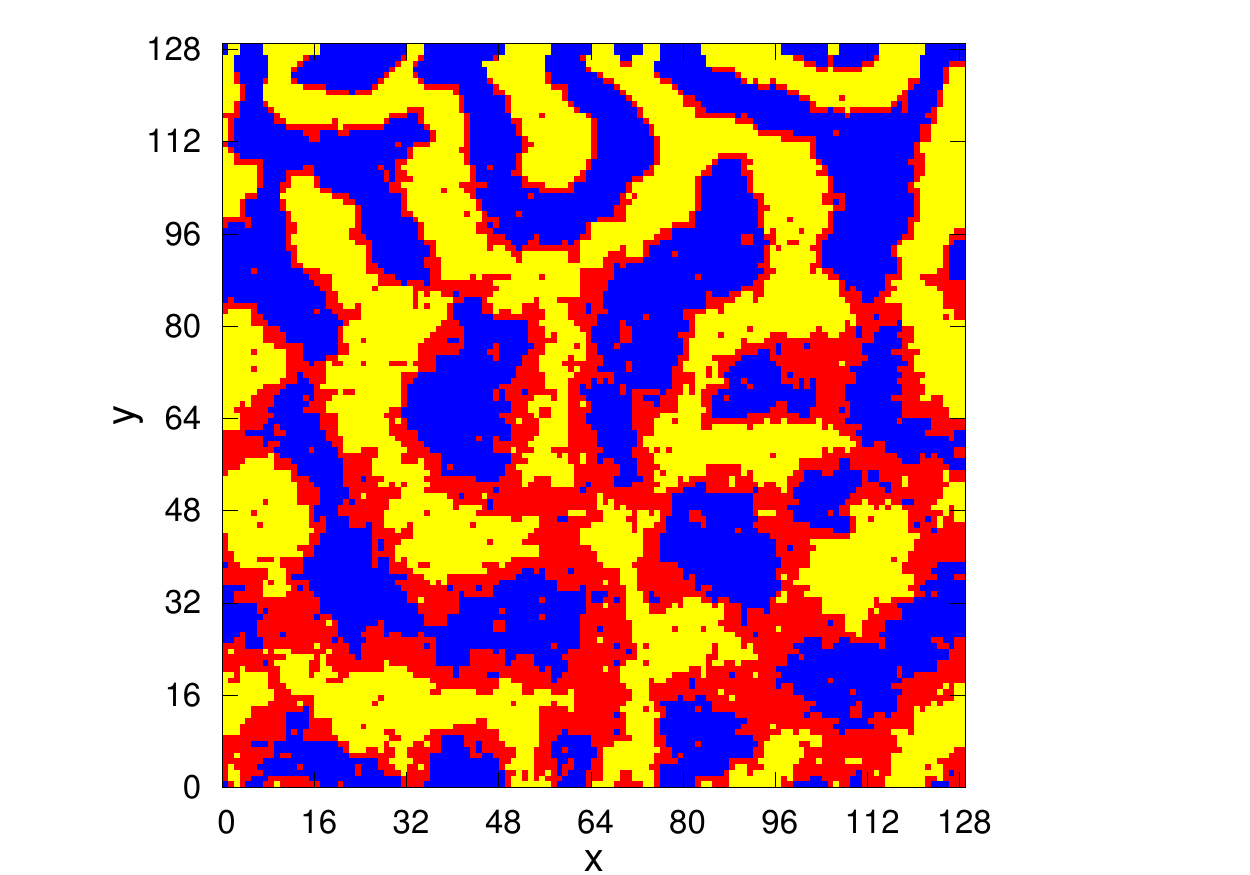}  & 
\includegraphics[width = 0.33\textwidth]{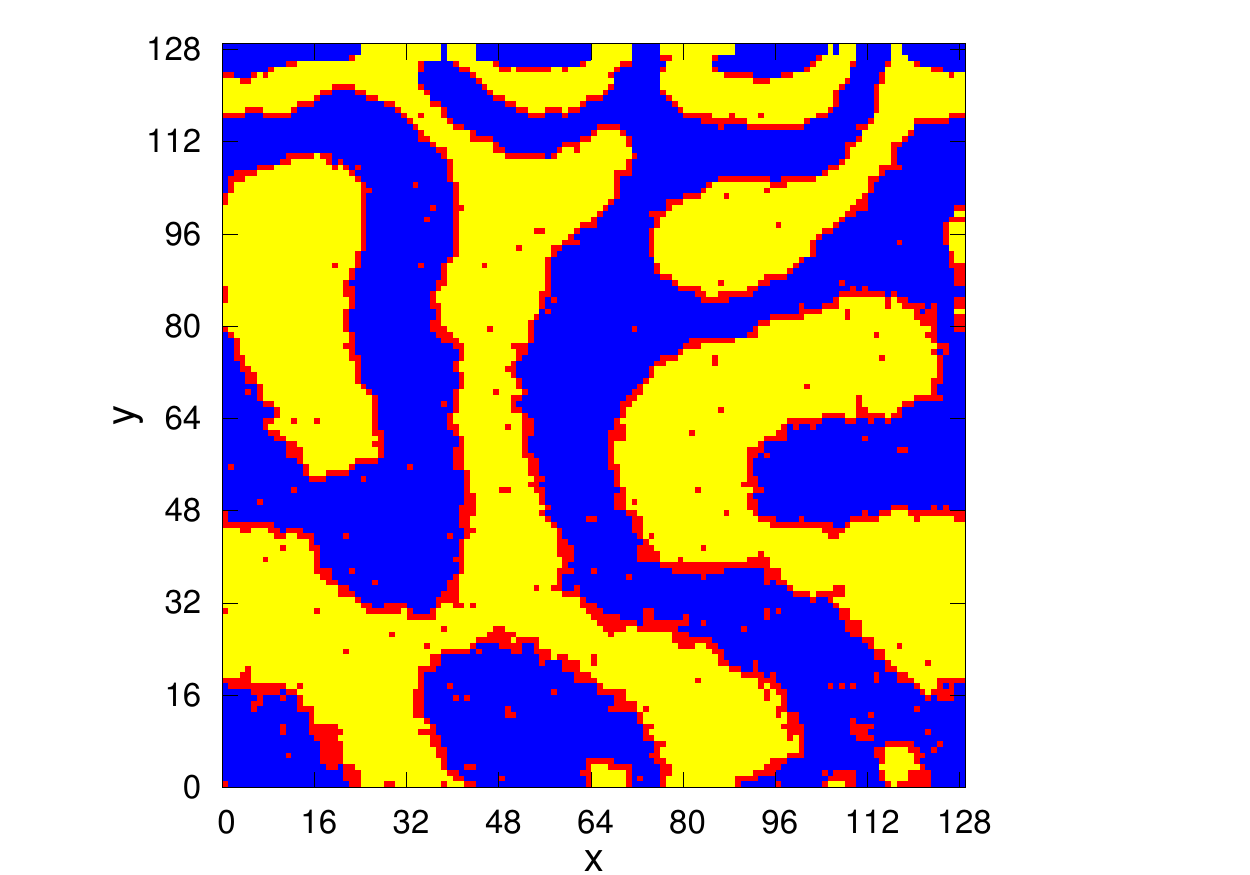}
\\[0.5cm]
\end{tabular}
\caption{Effect of temperature on microscopic configurations: (i) First row with $\beta=0.1$ (high temperature); (ii) Second row with $\beta=0.6$;  (iii) Third row with $\beta=1.0$ (low temperature). In all rows we have $J_{0,0}=J_{+1,+1}=J_{-1,-1}=0$, $J_{+1,0}=J_{-1,0}=1$, $J_{+1,-1}=6$ and the fraction of residual solvent is equal to $0.4$, $0.3$ and $0.1$, respectively (from left to right). The blue, yellow and red pixels represent the sites occupied by a ``$+1$'', ``$-1$'' or ``$0$'' spin, respectively.}
\label{fig:fig1}
\end{figure}

The effect of the temperature is well visible in Fig. \ref{fig:fig1}, which shows, for three different values of $\beta$, the microscopic configurations corresponding to different values of the fraction of solvent (namely, the ratio of the actual number of spins ``$0$'' to the total number of spins in $\Lambda$). For small values of $\beta$ (first row of Fig. \ref{fig:fig1}) no significant microstructures appear, the resulting configurations resembling the paramagnetic phase observed in the standard 2D Ising model for values of temperatures above the critical one \cite{Onsager1944}. When the temperature is lowered, a rich microstructure is observed to appear, characterized by two different phases, mostly constituted by either spins ``$+1$'' or ``$-1$'' (represented, respectively, by the blue and the yellow pixels in the figure), and containing  a small amount of solvent, displayed in red, dispersed inside the sea of the ``$+1$'' or ``$-1$'' spins. These features can be seen also in the experimental results in \cite{Sprenger}, e.g. Remarkably, the chosen set of parameters makes the presence of an interface between the two phases energetically unfavorable: hence, the residual solvent is mainly concentrated along the whirling interface of the blue and yellow regions, acting as a sort of ``shield'' between the two.

\begin{figure}[h!]
\centering
\begin{tabular}{ccc}
\includegraphics[width = 0.33\textwidth]{con-128-128-0600-zz0-mz1-pz1-mm0-pp0-mp6-04-eps-converted-to.pdf} &
\includegraphics[width = 0.33\textwidth]{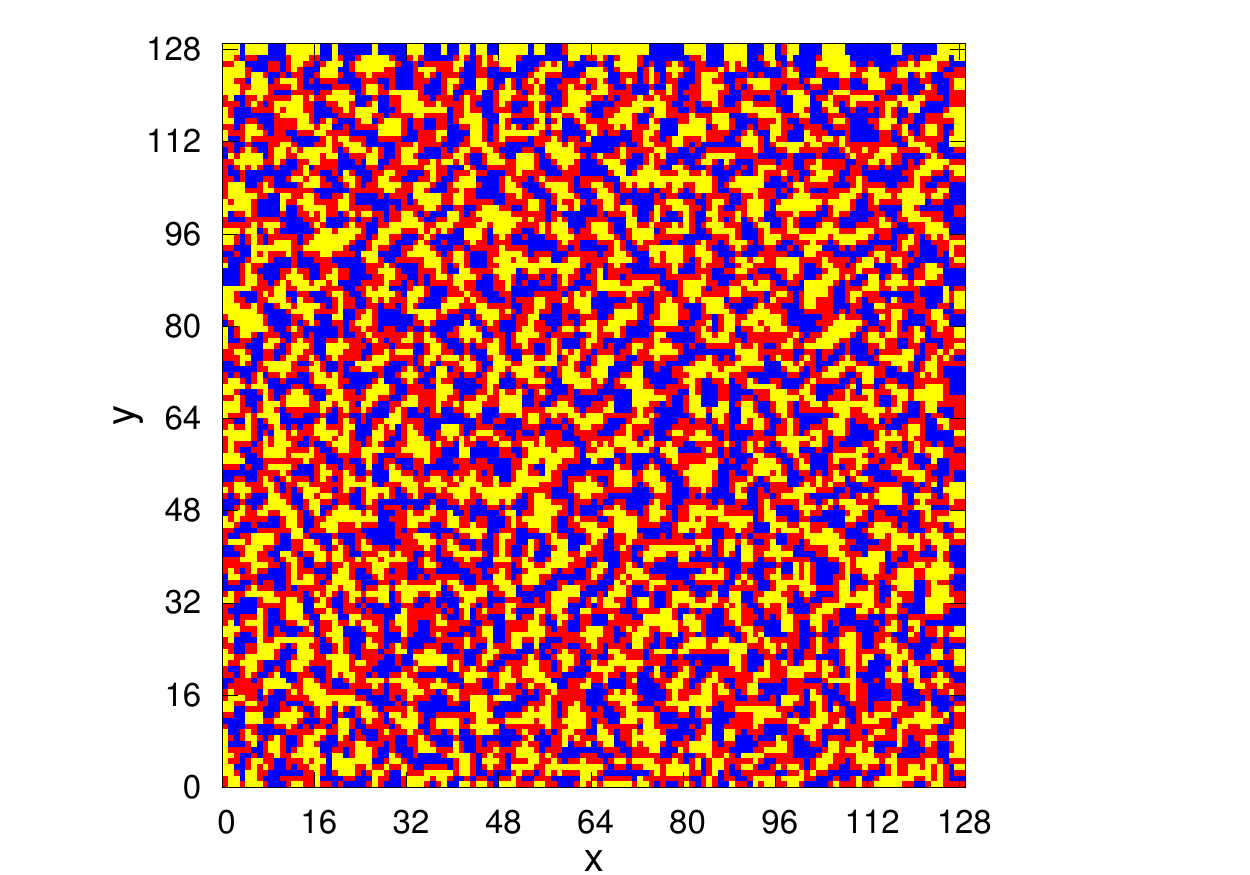}  & 
\includegraphics[width = 0.33\textwidth]{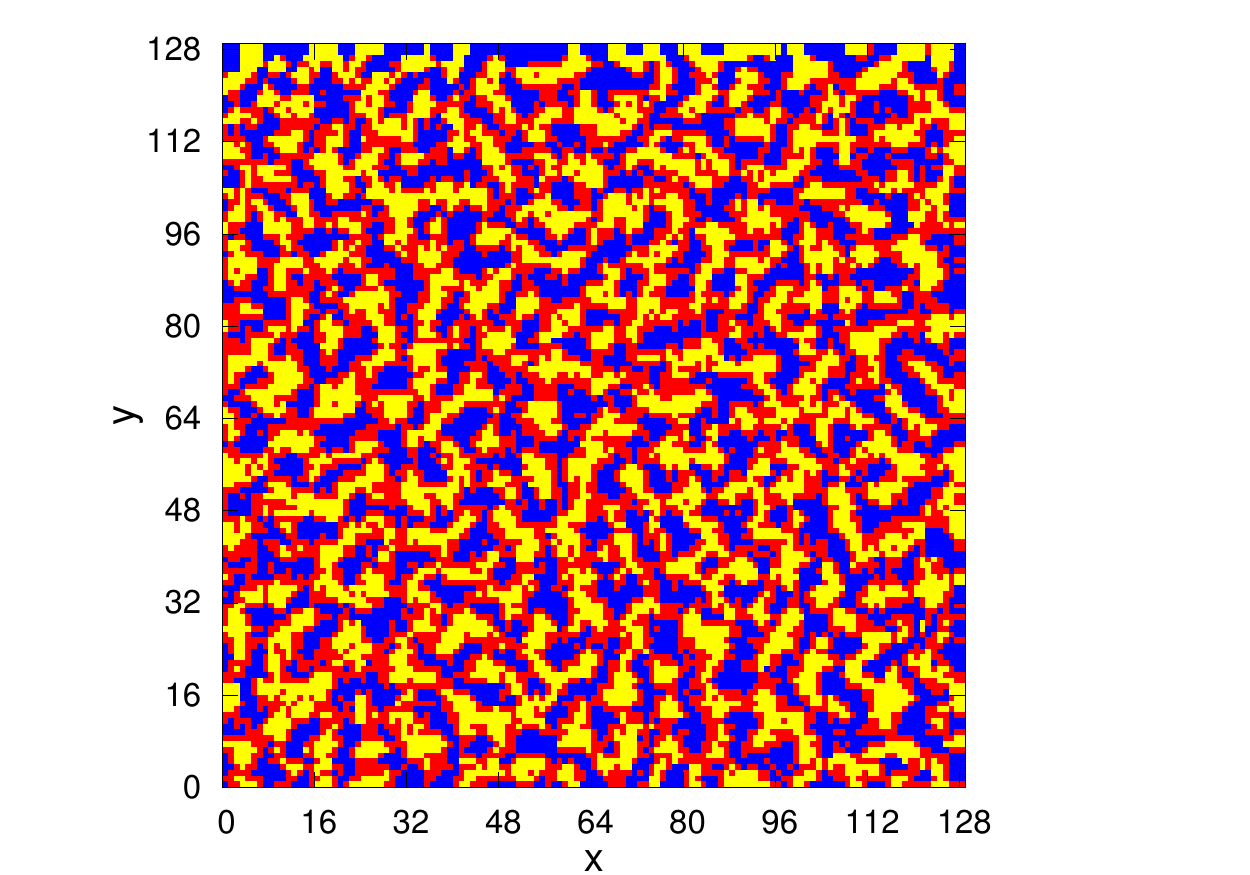}
\\[0.5cm]
\includegraphics[width = 0.33\textwidth]{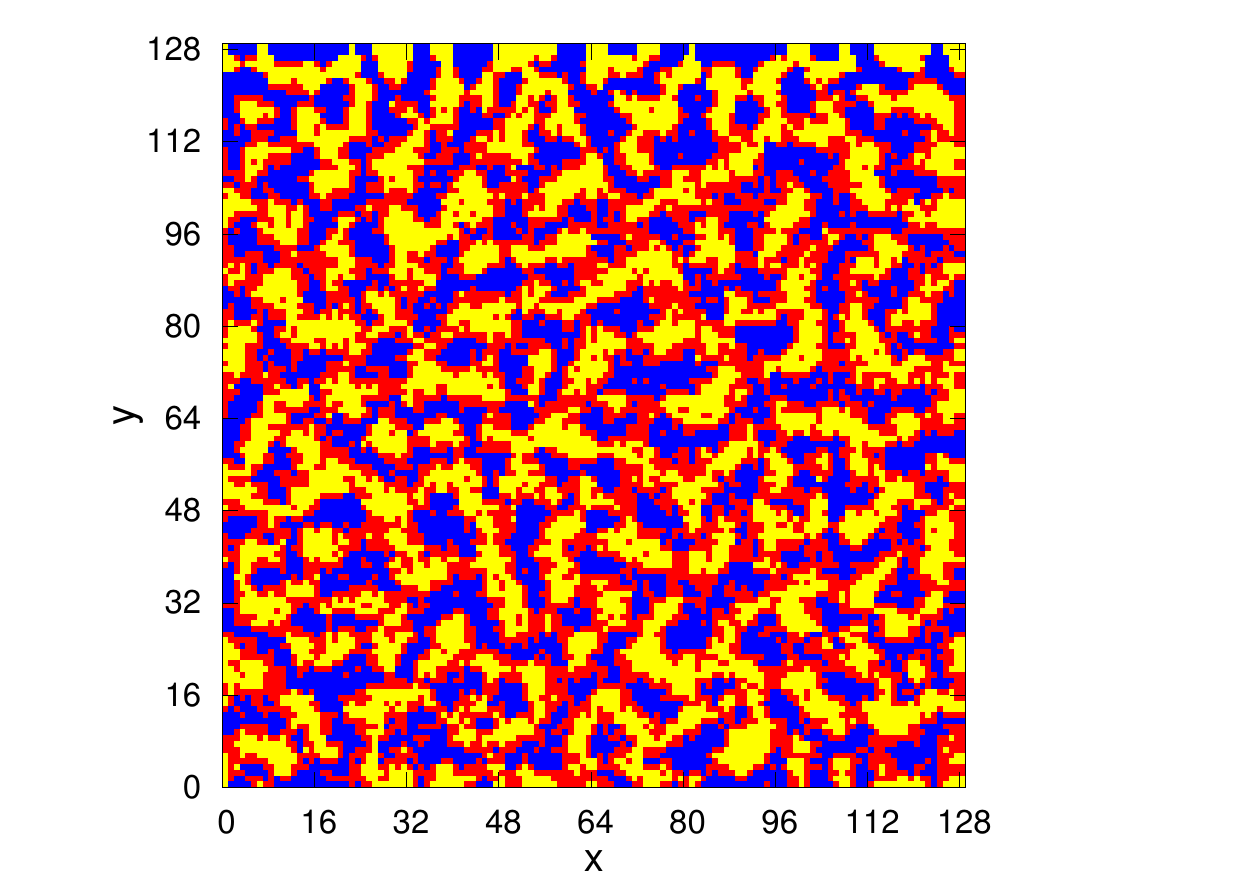} &
\includegraphics[width = 0.33\textwidth]{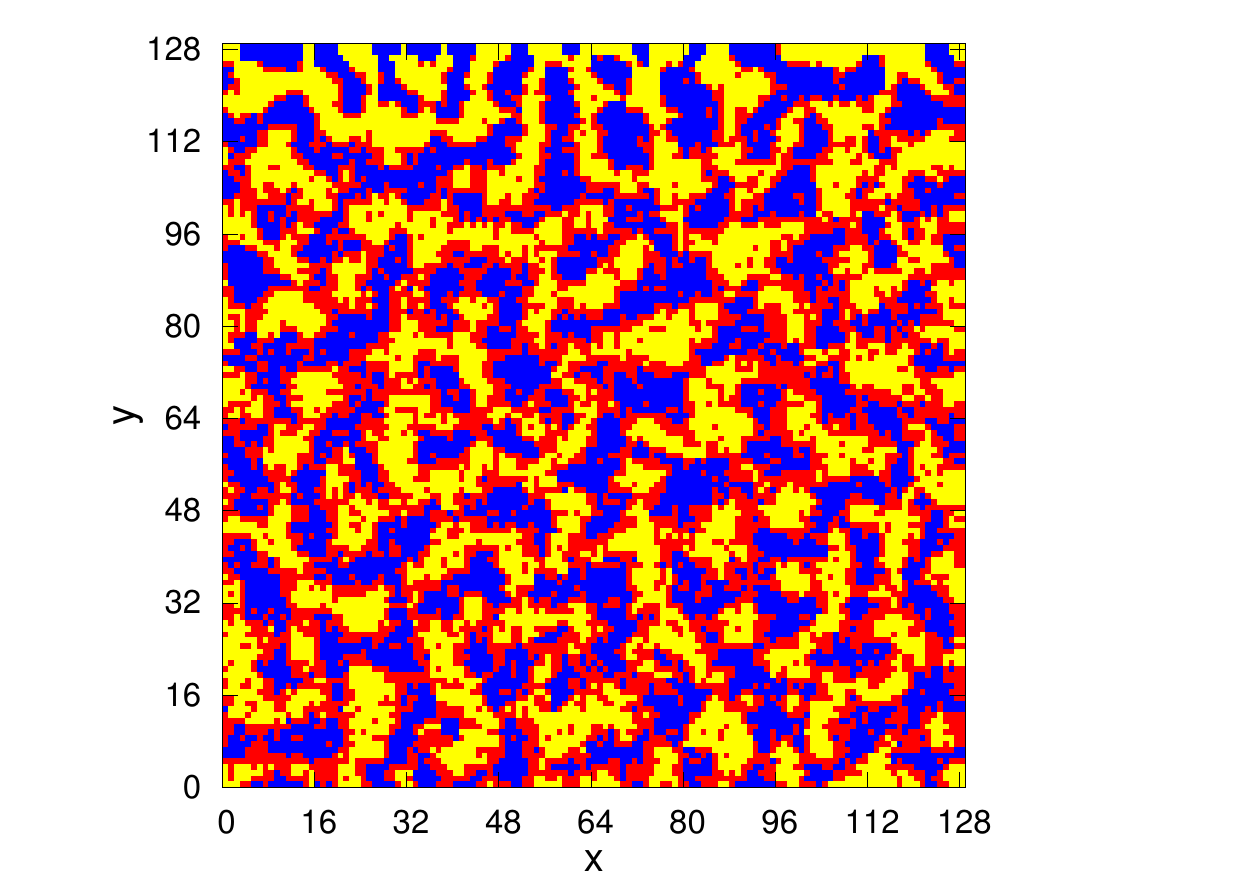} &
\includegraphics[width = 0.33\textwidth]{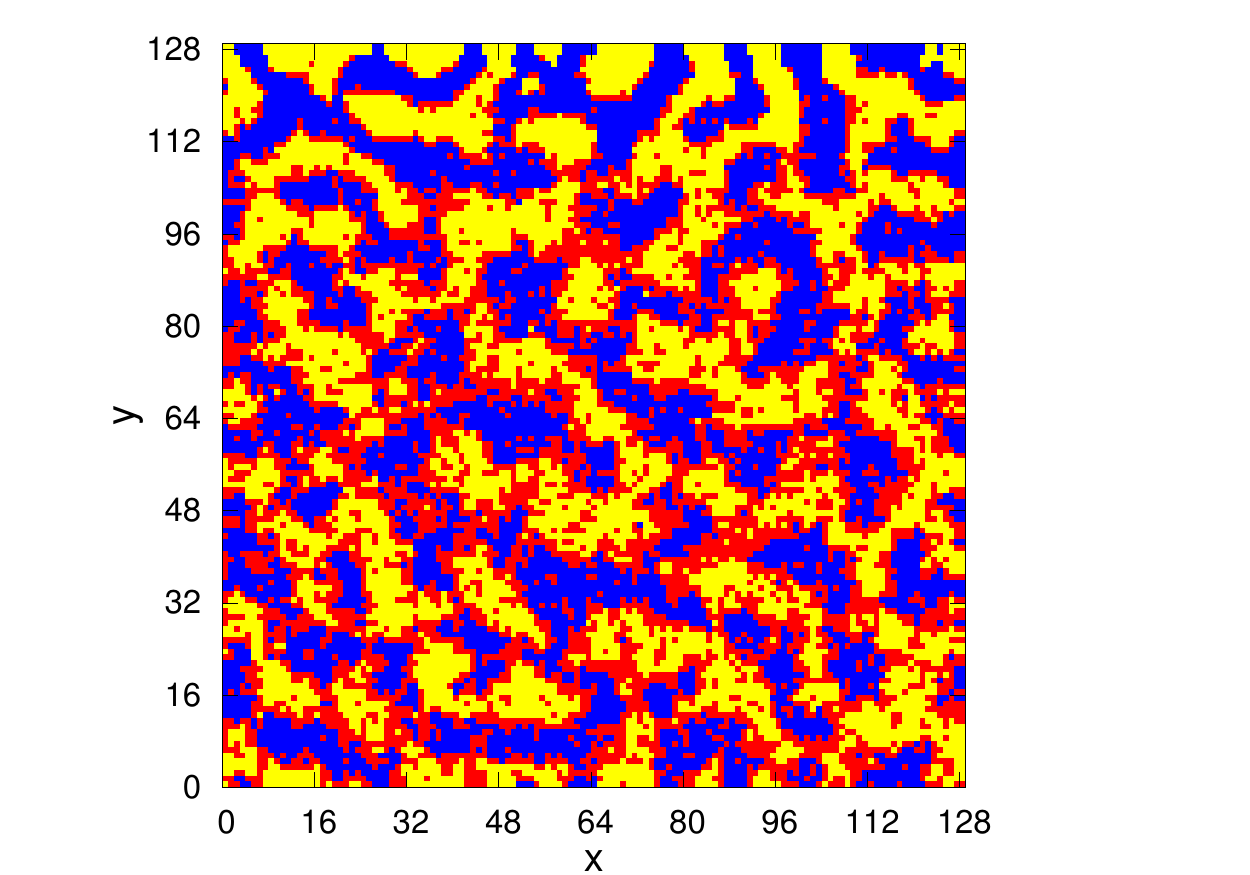} 
\\[0.5cm]
\end{tabular}
\caption{Capturing the phase separation at $\beta=0.6$ (below the critical temperature): (i) First row with $0.4$, $0.395$ and $0.39$ residual solvent and (ii) Second row with $0.38$, $0.37$ and $0.35$ residual solvent. This pictures are intermediate steps between the disordered and ordered states captured in Fig.~\ref{fig:fig1}, second row. The blue, yellow and red pixels represent the sites occupied by a ``$+1$'', ``$-1$'' or ``$0$'' spin, respectively.}
\label{fig:fig2}
\end{figure}

In Fig. \ref{fig:fig2} we explore in more detail the dynamics with $\beta=0.6$ and report the microscopic configurations obtained for values of the fraction of residual solvent between $0.4$ and $0.35$, in an attempt to capture the moment of phase separation. We observe that, in a relatively short interval of residual solvent, the ``$+1$'' and ``$-1$'' spins have the tendency to minimize the interface between them by creating larger phases of ``$+1$'' or ``$-1$'' spins and by ``pushing'' the ``$0$'' spins to the interface between these two phases.

\begin{figure}[h!]
\centering
\begin{tabular}{ccc}
\includegraphics[width = 0.33\textwidth]{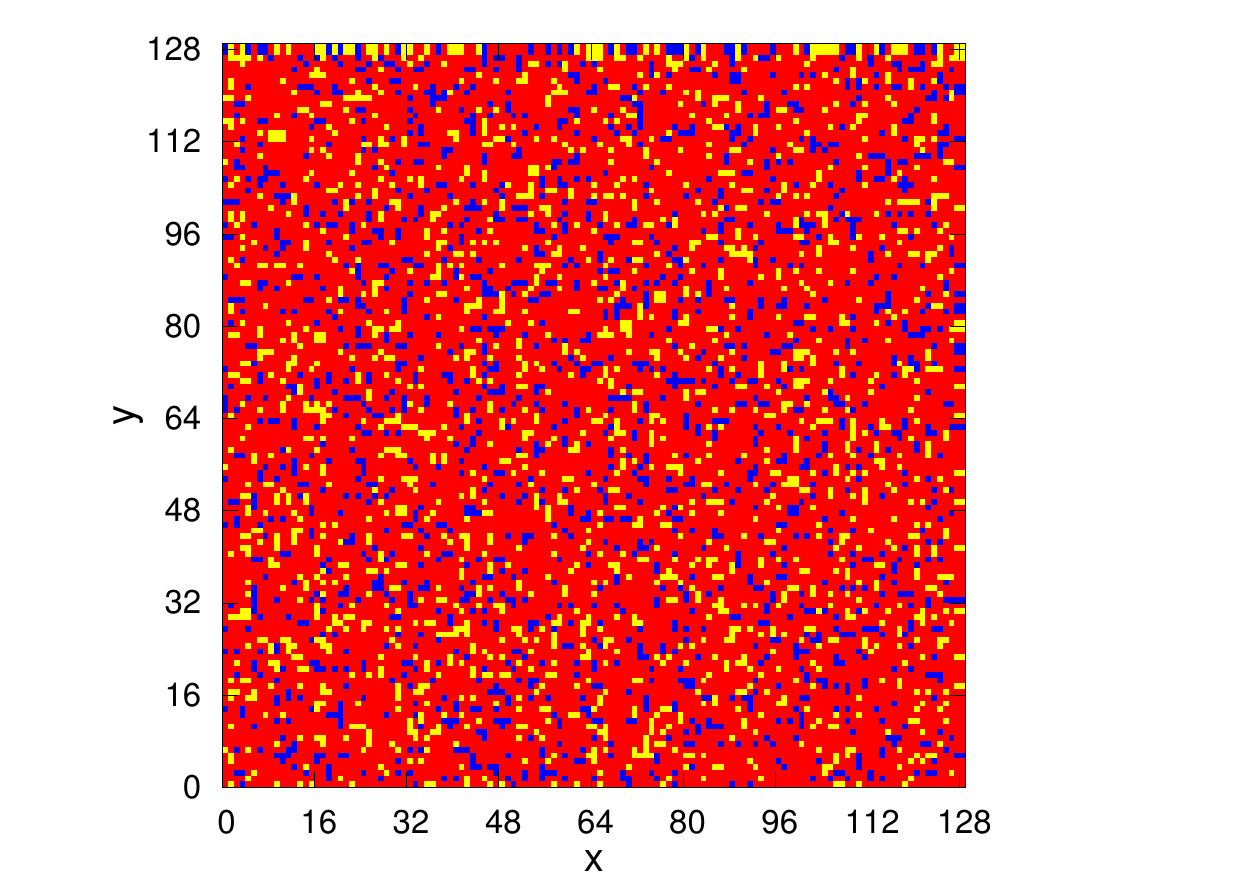} &
\includegraphics[width = 0.33\textwidth]{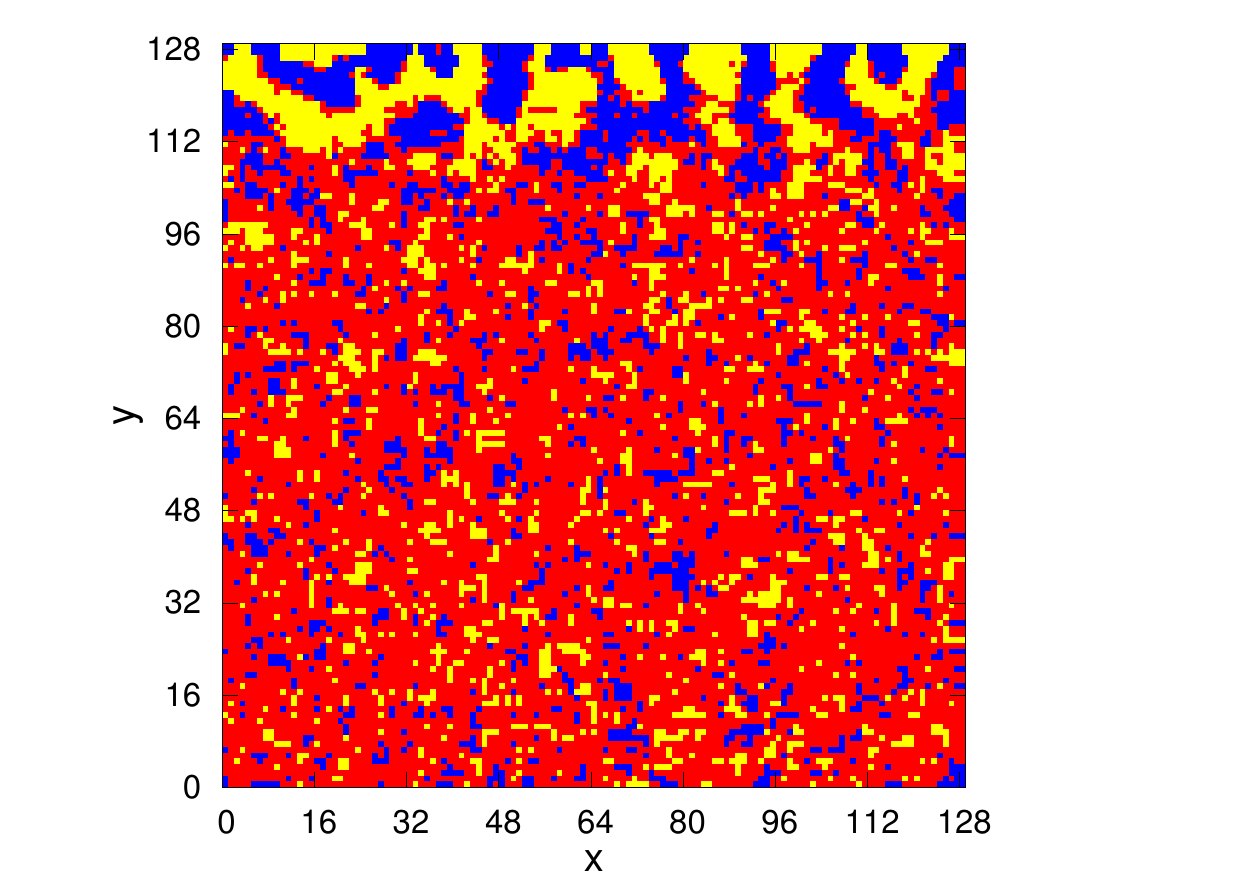}  & 
\includegraphics[width = 0.33\textwidth]{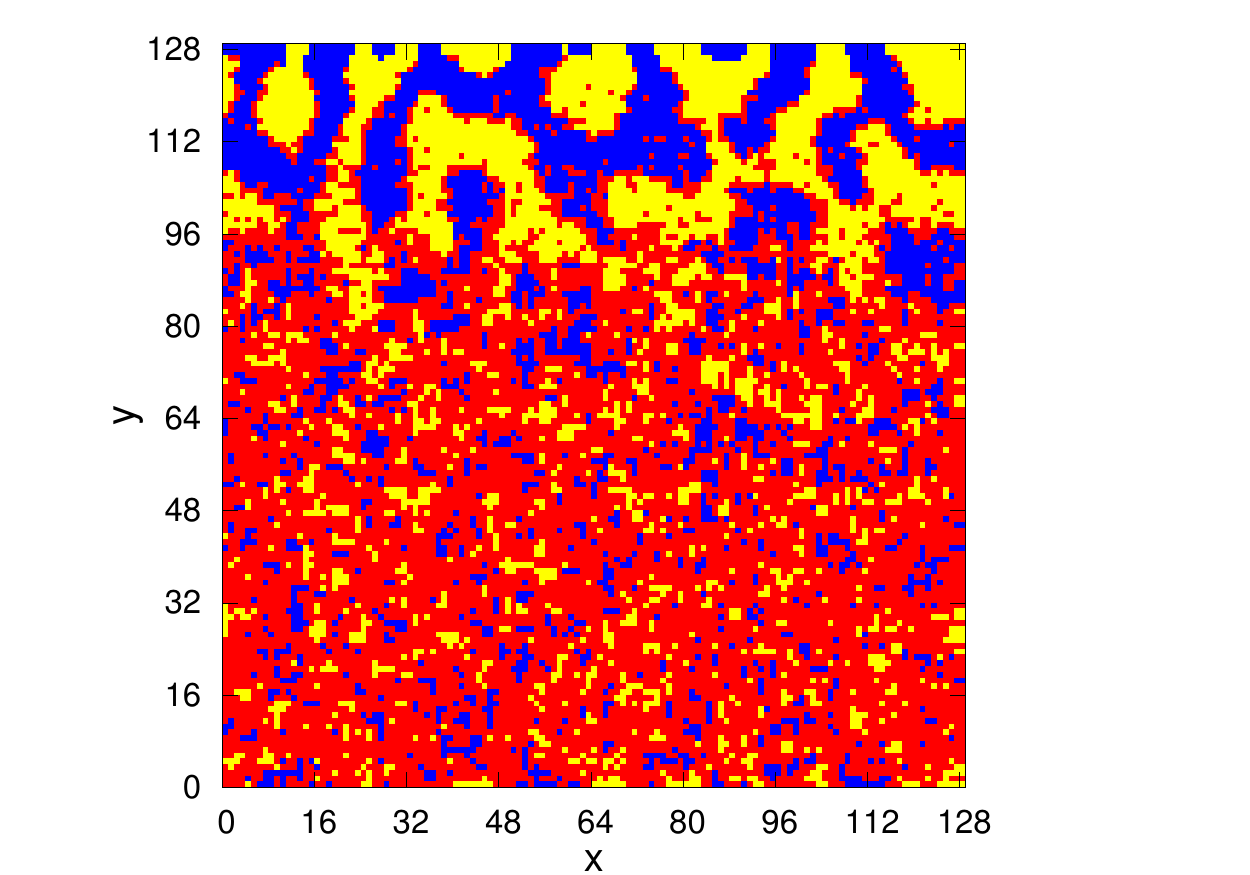}
\\[0.5cm]
\includegraphics[width = 0.33\textwidth]{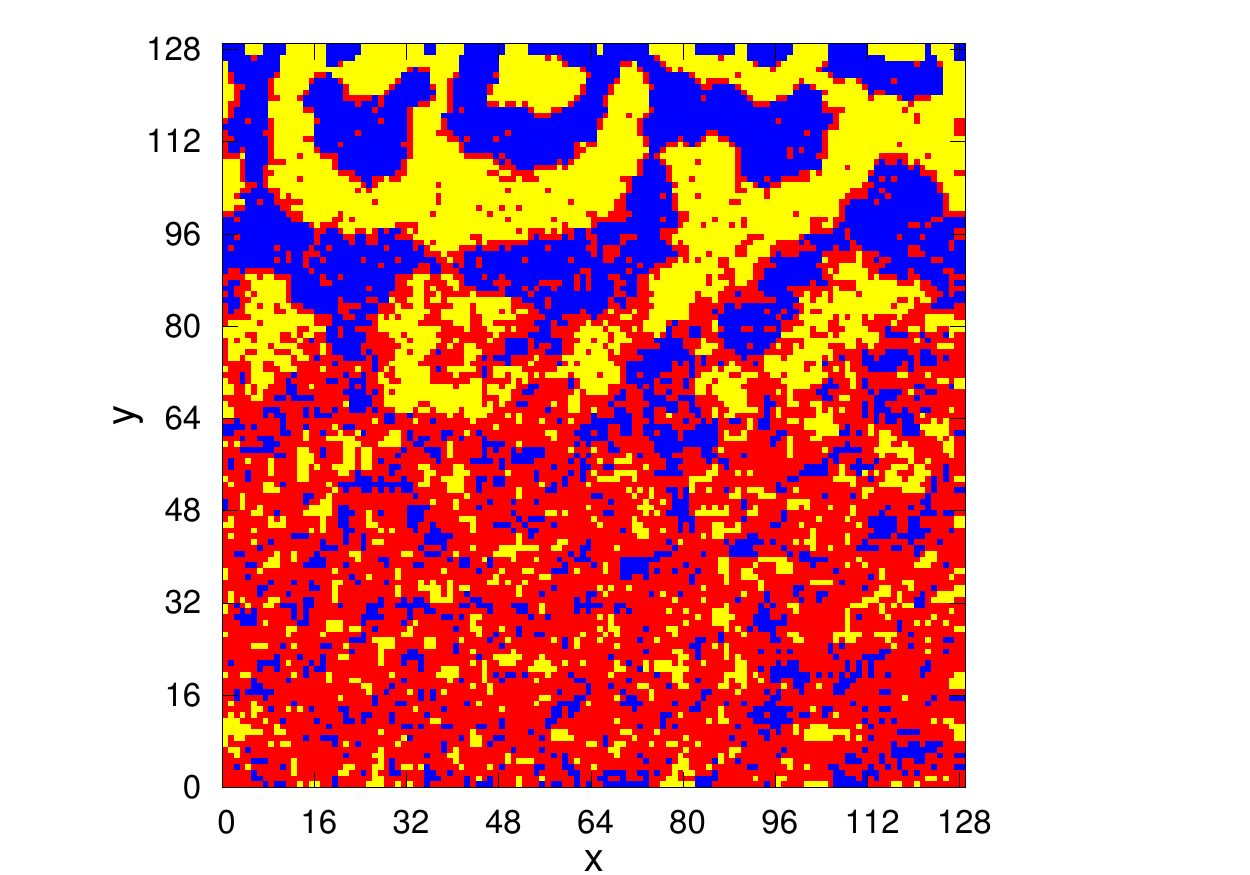} &
\includegraphics[width = 0.33\textwidth]{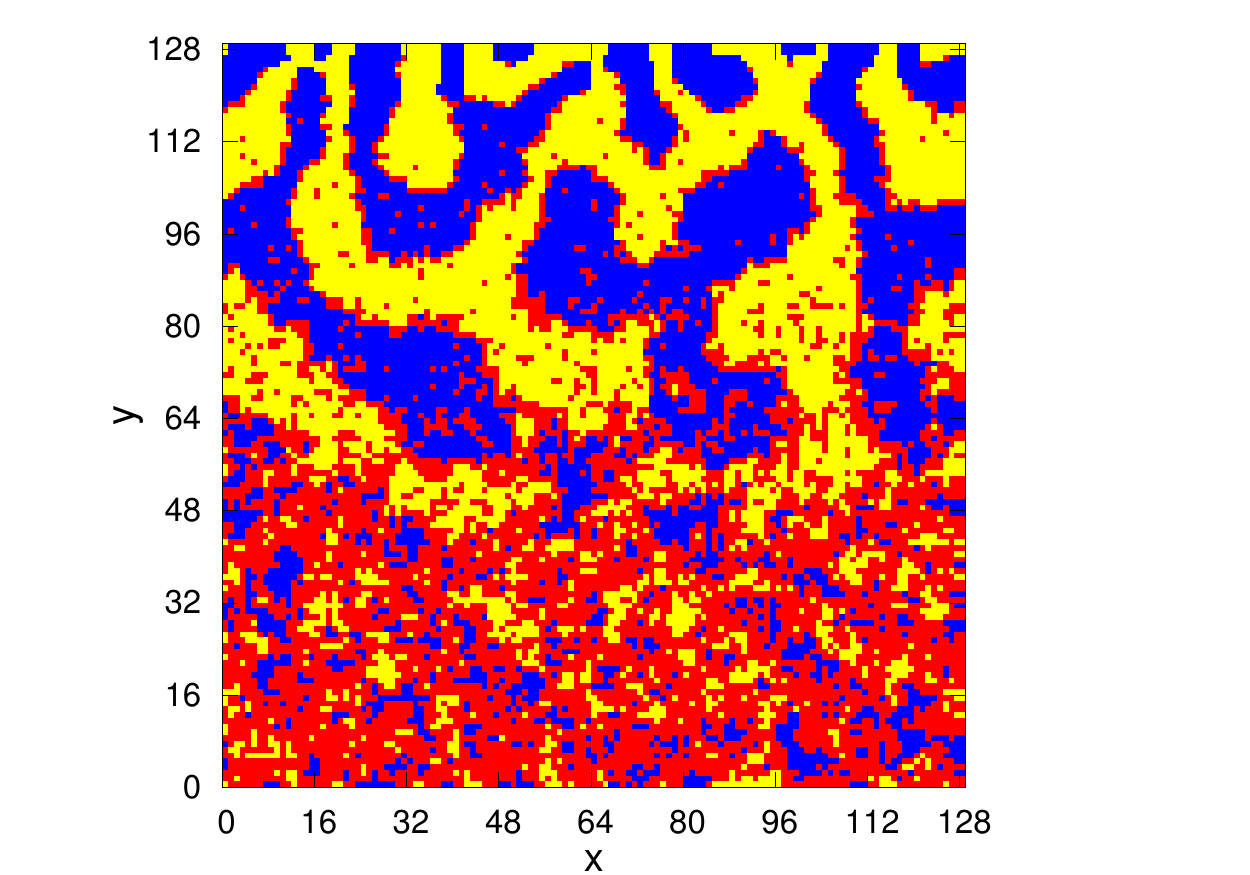} &
\includegraphics[width = 0.33\textwidth]{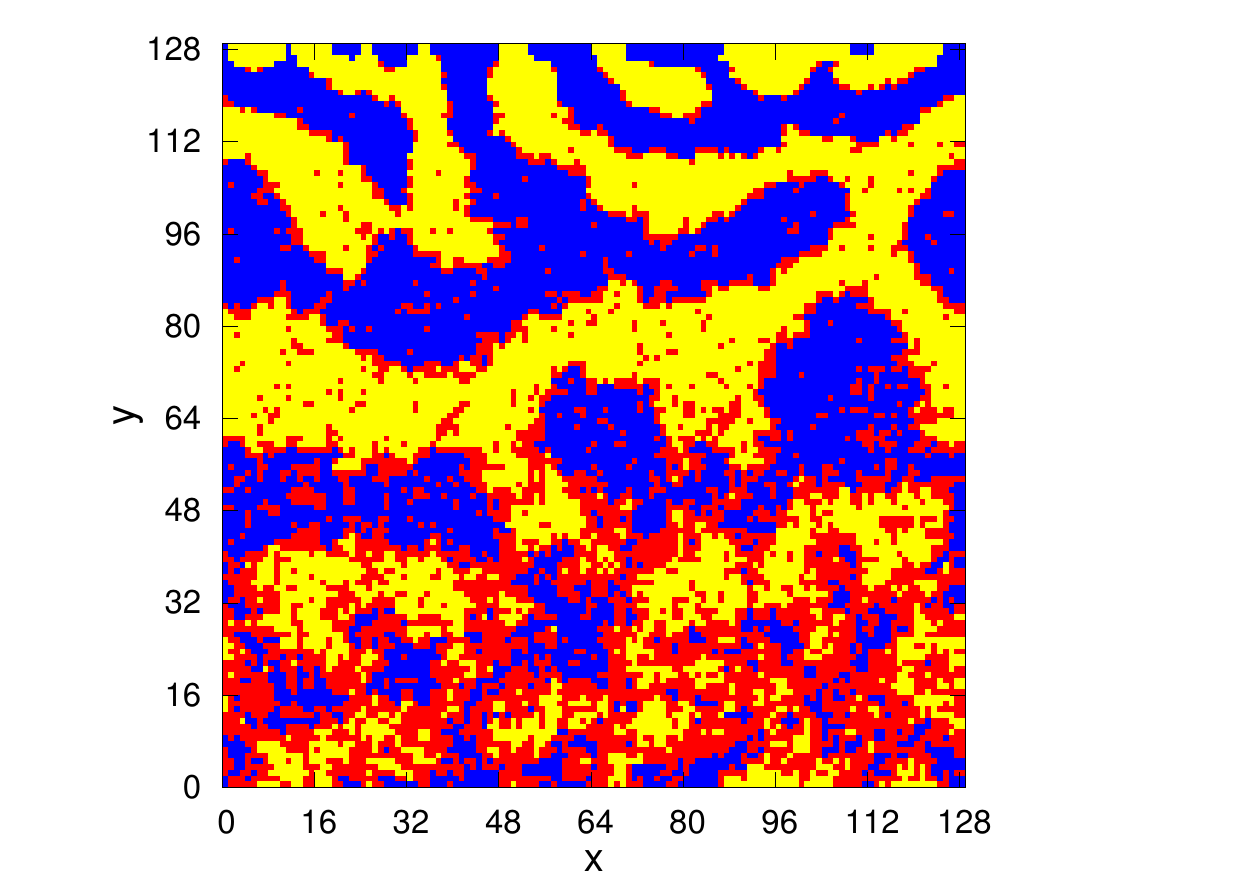} 
\\[0.5cm]
\includegraphics[width = 0.33\textwidth]{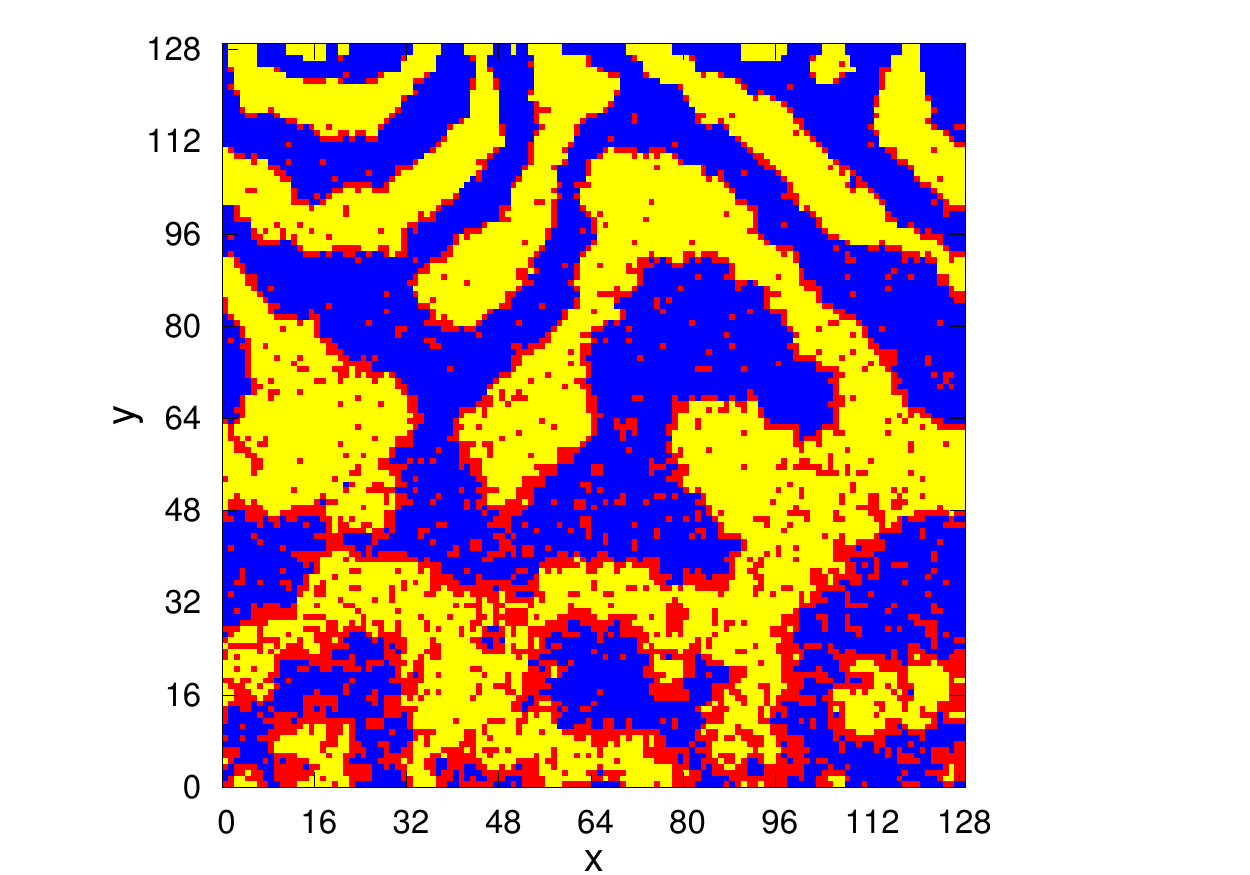} &
\includegraphics[width = 0.33\textwidth]{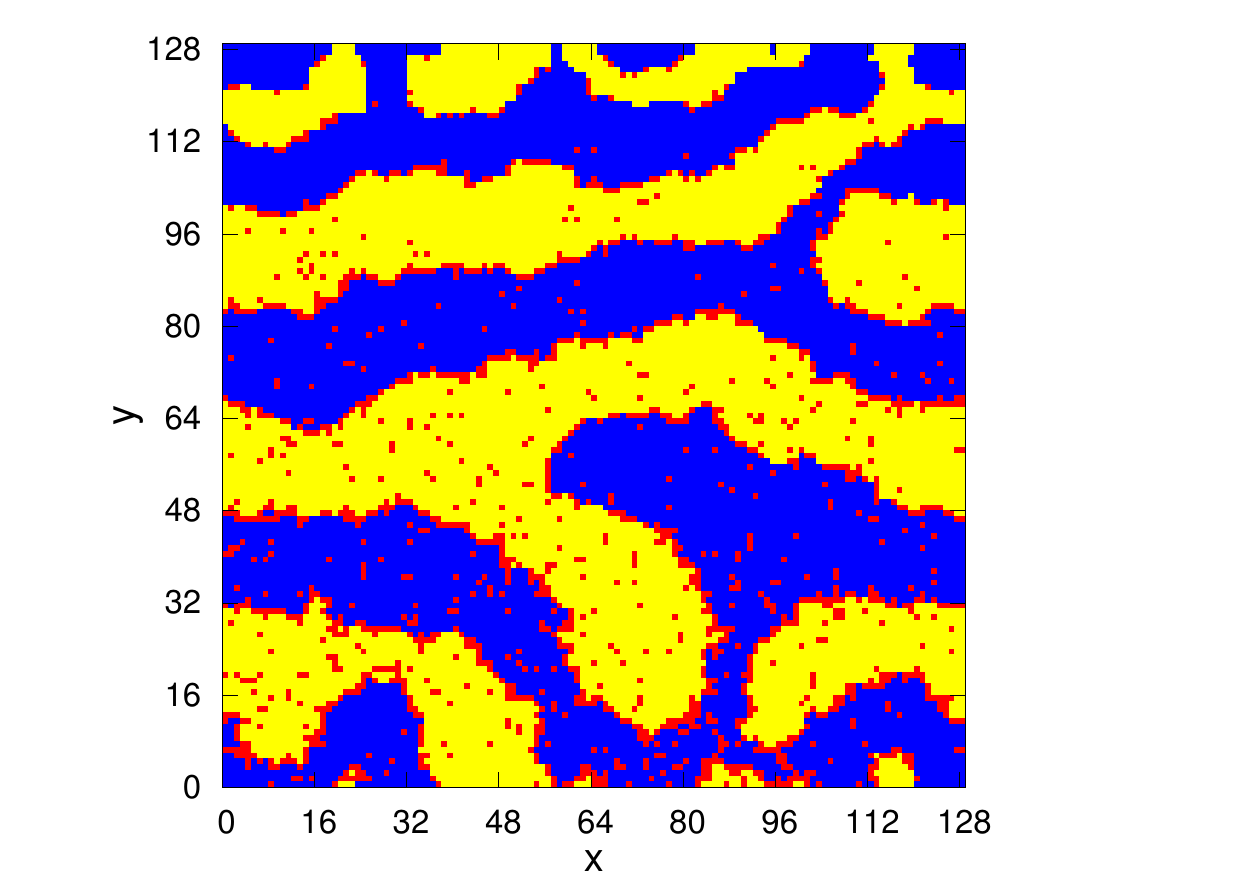}
\\[0.5cm]
\end{tabular}
\caption{The evaporation process in the case of dilute solution, with starting composition 0.8:0.1:0.1 (i) First row with the fraction of spins 0 being $0.8$, $0.7$, $0.6$; (ii) Second row with the fraction of spins 0 being $0.5$, $0.4$, $0.3$; (ii) Third row with the fraction of spins 0 being $0.2$ and  $0.1$. The other parameters are the same as in Fig. \ref{fig:fig2}. The blue, yellow and red pixels represent the sites occupied by a ``$+1$'', ``$-1$'' or ``$0$'' spin, respectively.}
\label{fig:fig3}
\end{figure}

In Fig. \ref{fig:fig3} we attempt to capture the phase separation in a similar way as in Fig. \ref{fig:fig2}, but with the initial proportion of the three species being $80:10:10$. Since the initial quantity of solvent is much larger, as compared to the previous case, the diffusion of the solvent require more time. We observe the formation of a depletion zone for the solvent and the formation of phases rich in ``$+1$'' and ``$-1$'' spins as soon as a critical concentration of ``0'' spins is reached. 

\begin{figure}[h!]
\centering
\begin{tabular}{ccc}
\includegraphics[width = 0.33\textwidth]{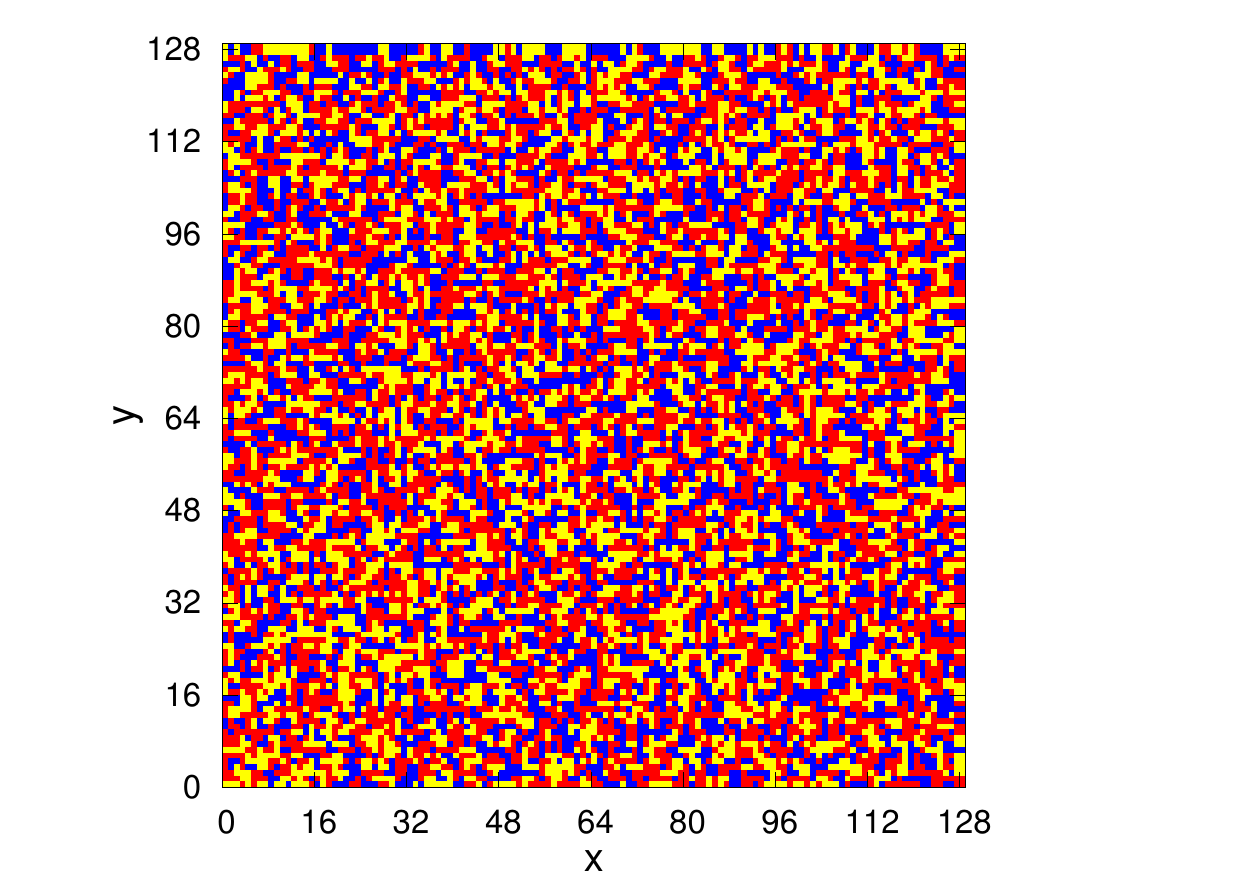} &
\includegraphics[width = 0.33\textwidth]{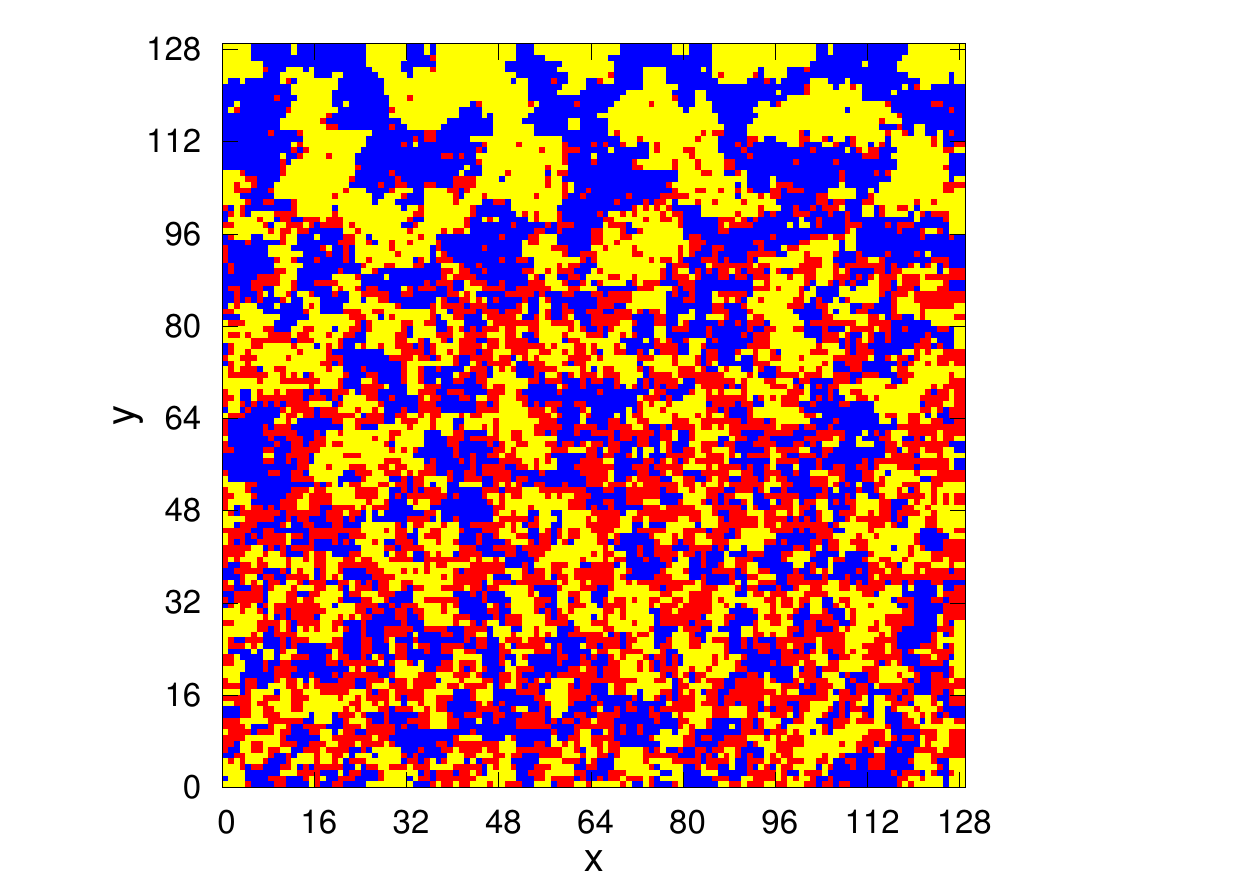}  & 
\includegraphics[width = 0.33\textwidth]{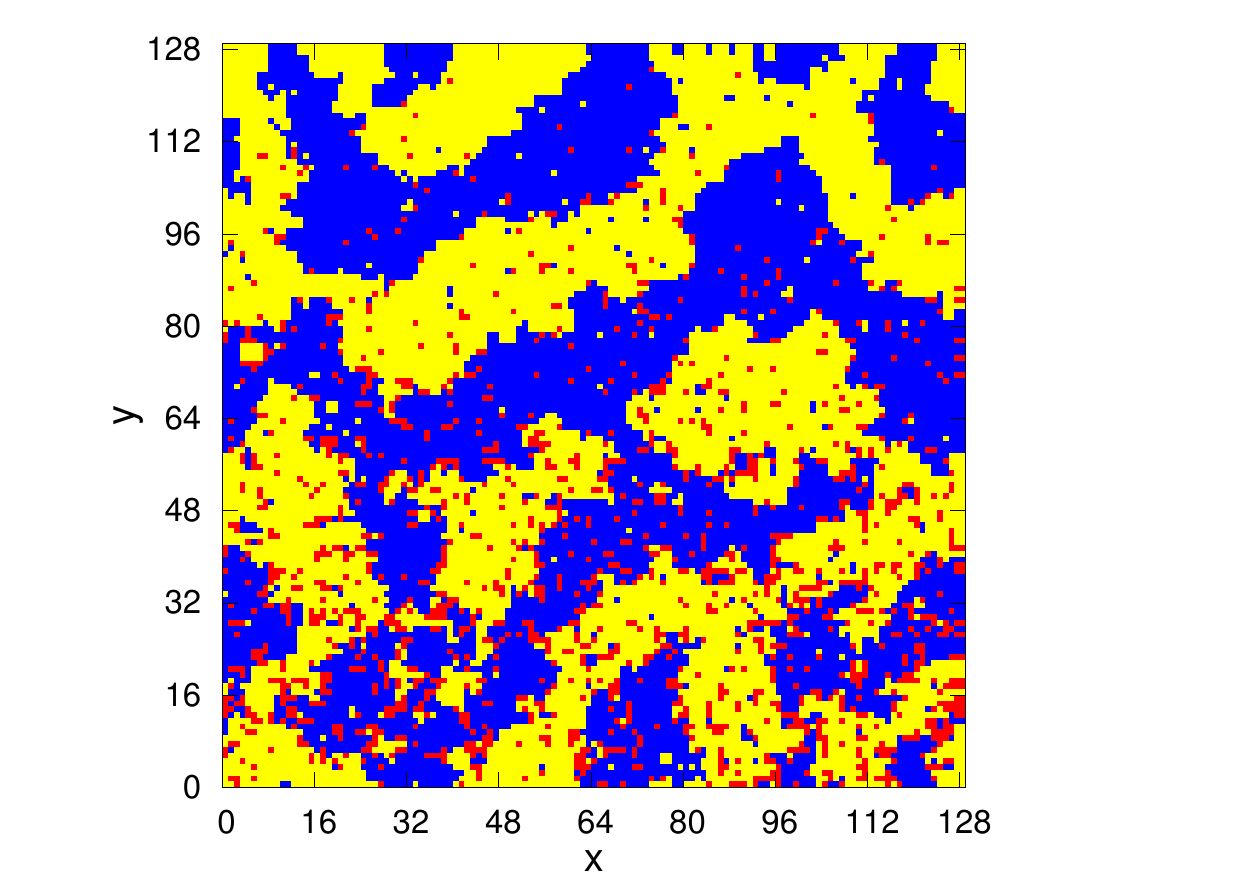}
\\[0.5cm]
\includegraphics[width = 0.33\textwidth]{con-128-128-0600-zz0-mz1-pz1-mm0-pp0-mp6-04-eps-converted-to.pdf} &
\includegraphics[width = 0.33\textwidth]{con-128-128-0600-zz0-mz1-pz1-mm0-pp0-mp6-03-eps-converted-to.pdf}  & 
\includegraphics[width = 0.33\textwidth]{con-128-128-0600-zz0-mz1-pz1-mm0-pp0-mp6-01-eps-converted-to.pdf}
\\[0.5cm]
\includegraphics[width = 0.33\textwidth]{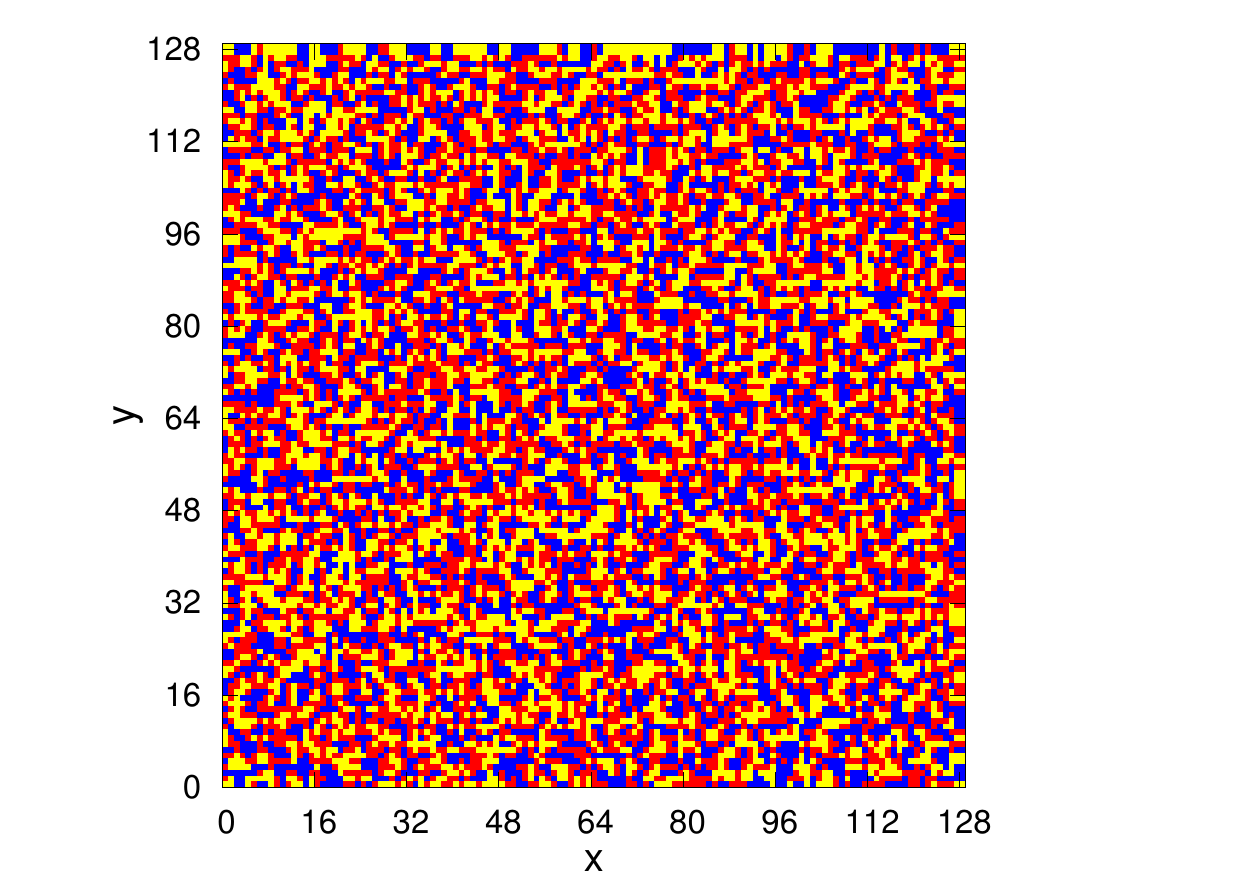} &
\includegraphics[width = 0.33\textwidth]{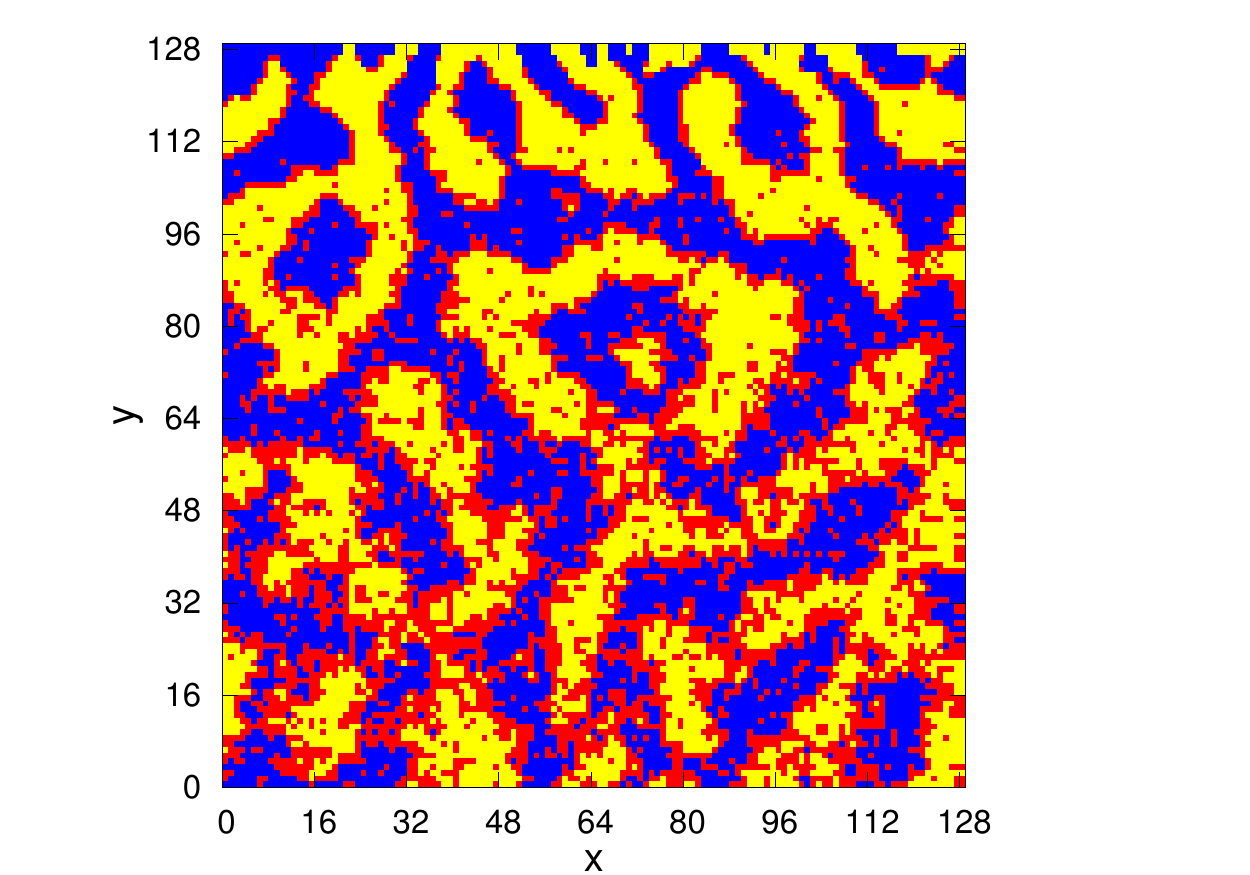}  & 
\includegraphics[width = 0.33\textwidth]{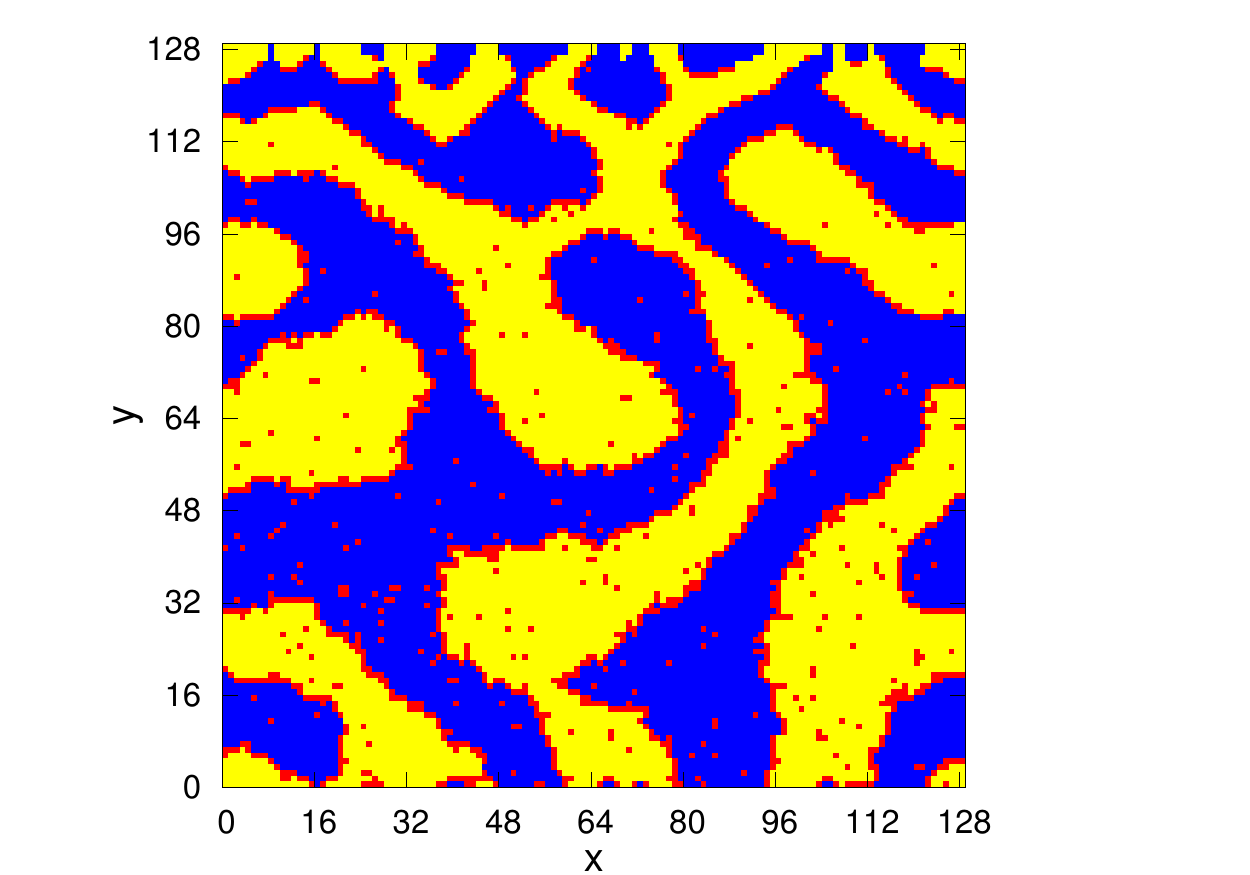}
\\[0.5cm]
\end{tabular}
\caption{Effect of the interaction strength between phases on microscopic configurations for $\beta=0.6$, with $J_{0,0}=J_{+1,+1}=J_{-1,-1}=0$, $J_{+1,0}=J_{-1,0}=1$ for different fractions of residual solvent (same as in Fig. \ref{fig:fig1}).  From top to bottom, we have, respectively,  $J_{+1,-1}=2$, $J_{+1,-1}=6$ and $J_{+1,-1}=10$. The effect of increasing the interaction strength between the spins $+1$ and $-1$ is especially visible in the last column. In all rows, the fraction of residual solvent is equal to $0.4$, $0.3$ and $0.1$, respectively (from left to right). The blue, yellow and red pixels represent the sites occupied by a ``$+1$'', ``$-1$'' or ``$0$'' spin, respectively.}
\label{fig:fig4}
\end{figure}

The shielding effect of the solvent is clearly highlighted in Fig. \ref{fig:fig4} (check the rightmost column), in which we tuned the interaction parameter $J_{+1,-1}$. This parameter is proportional to the energy cost for forming the interface. In the first row of Fig. \ref{fig:fig4} the parameter $J_{+1,-1}$=2 and the interface is rough, hence larger and not ``shielded'' by the solvent. In the third row we have $J_{+1,-1}$=10 and we observe a much smoother interface and the residual solvent positioned at the interface.

\begin{figure}[h!]
\centering
\begin{tabular}{ccc}
\includegraphics[width = 0.33\textwidth]{con-128-128-0600-zz0-mz1-pz1-mm0-pp0-mp6-04-eps-converted-to.pdf} &
\includegraphics[width = 0.33\textwidth]{con-128-128-0600-zz0-mz1-pz1-mm0-pp0-mp6-03-eps-converted-to.pdf}  & 
\includegraphics[width = 0.33\textwidth]{con-128-128-0600-zz0-mz1-pz1-mm0-pp0-mp6-01-eps-converted-to.pdf}
\\[0.5cm]
\includegraphics[width = 0.33\textwidth]{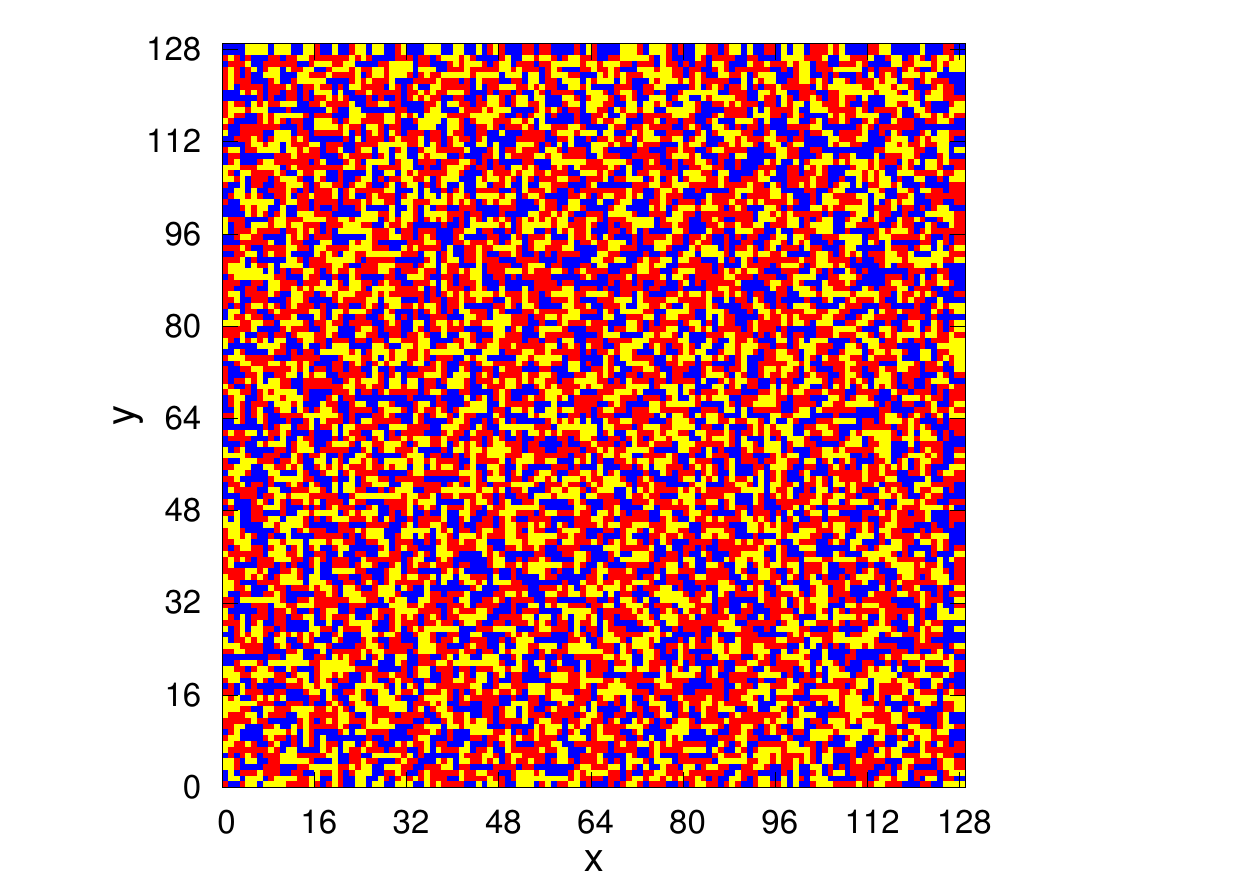} &
\includegraphics[width = 0.33\textwidth]{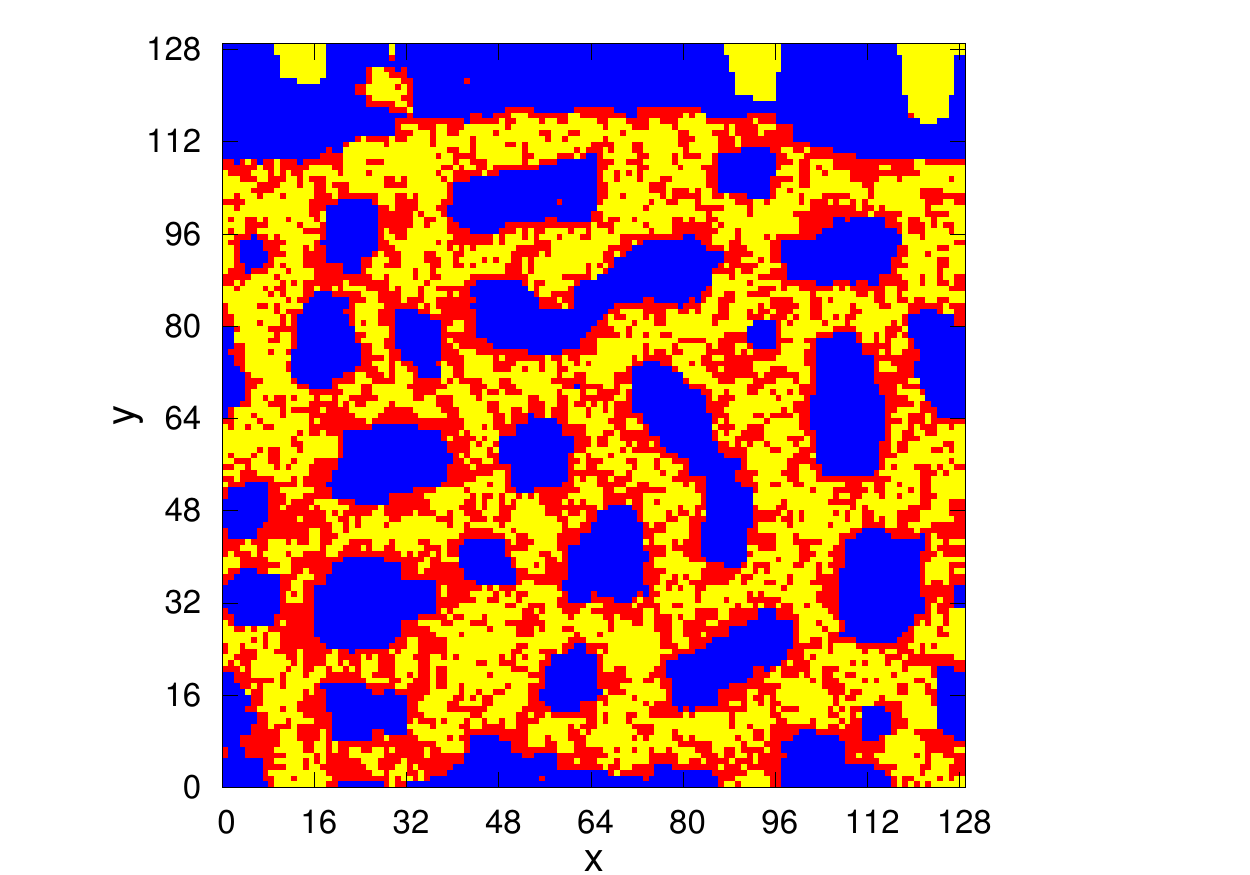}  & 
\includegraphics[width = 0.33\textwidth]{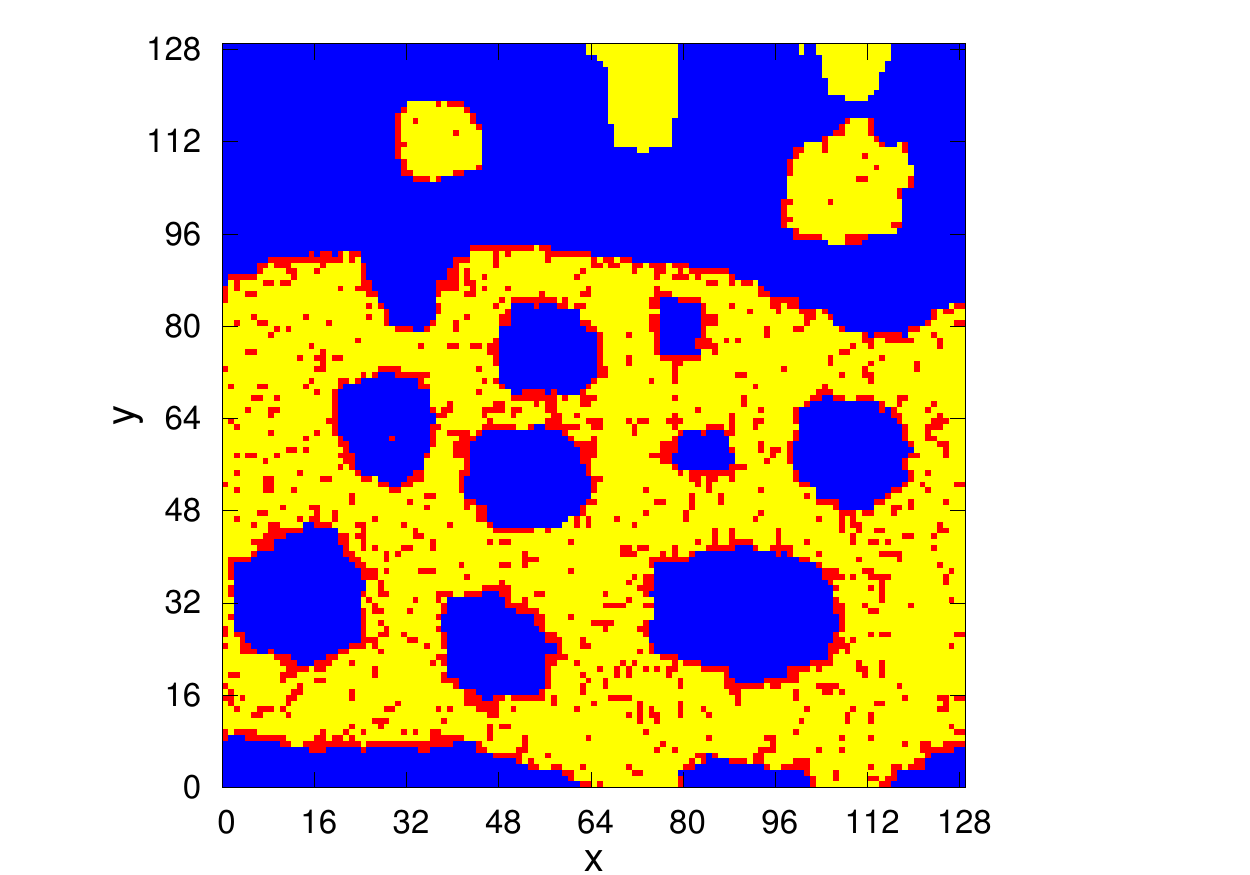}
\\[0.5cm]
\includegraphics[width = 0.33\textwidth]{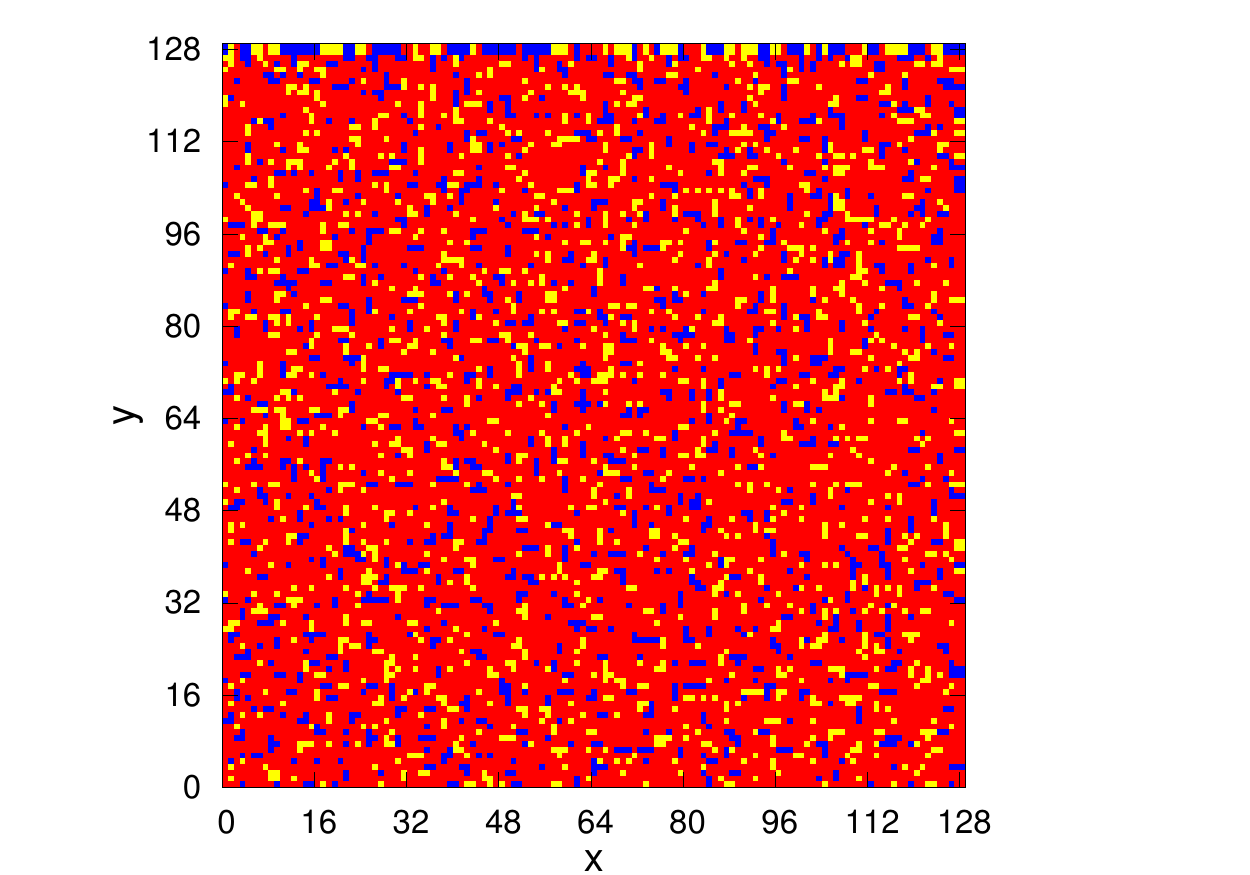} &
\includegraphics[width = 0.33\textwidth]{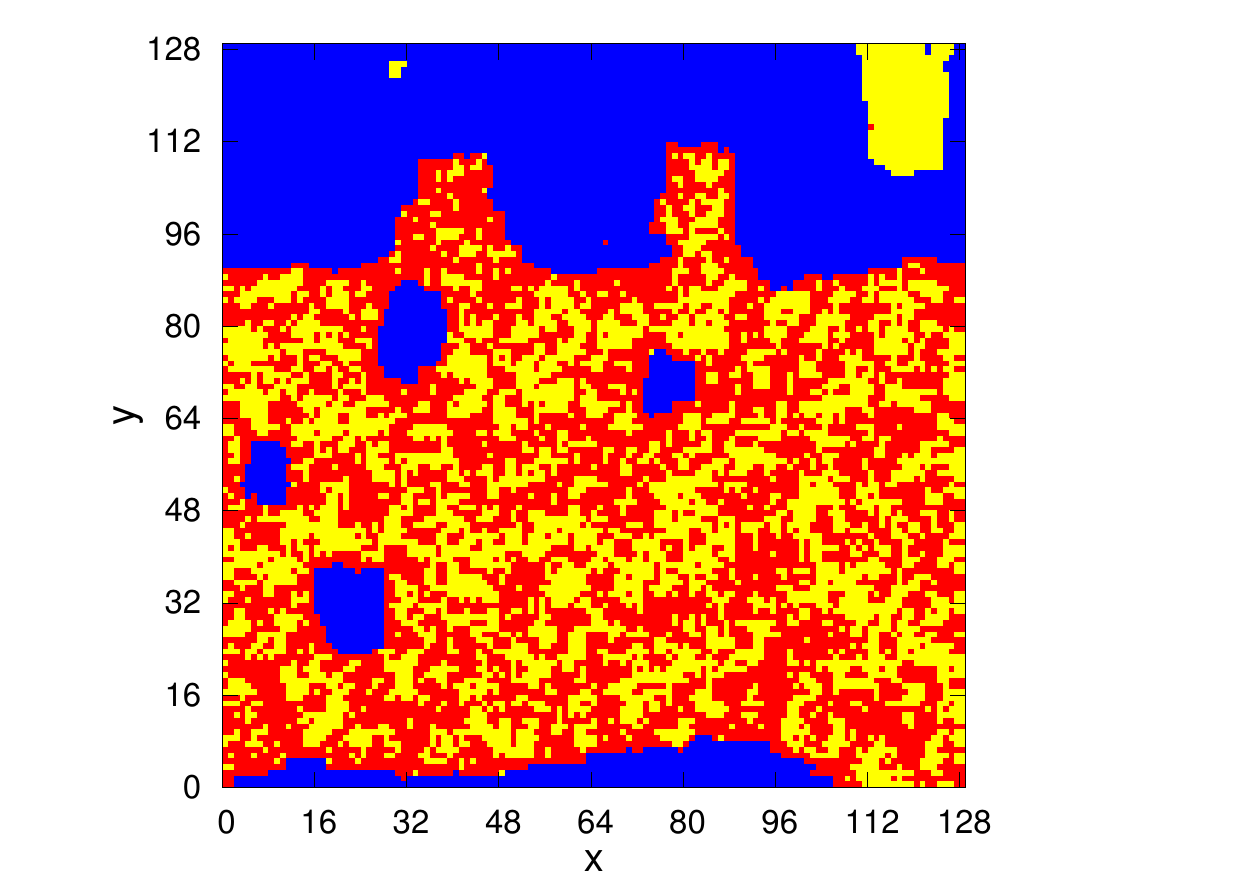}  & 
\includegraphics[width = 0.33\textwidth]{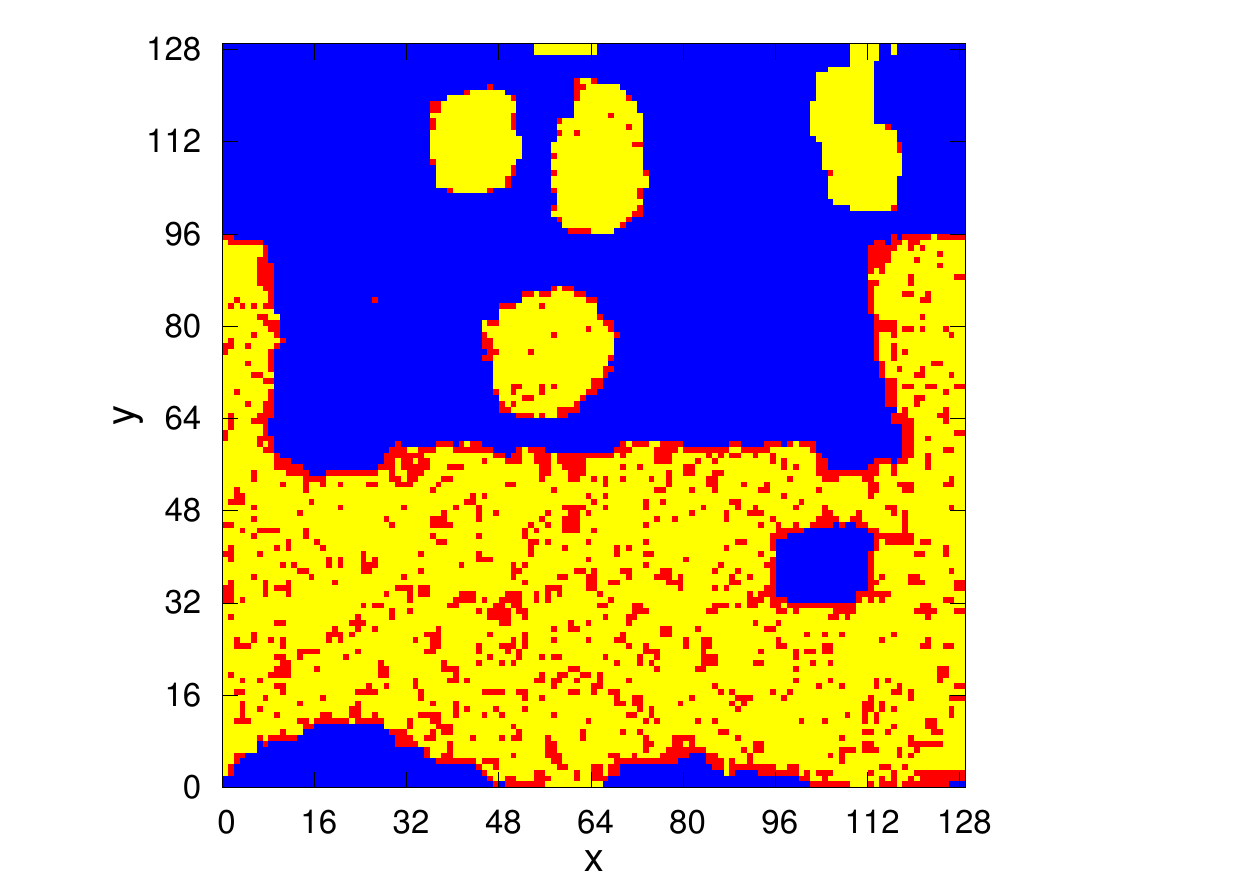}
\\[0.5cm]
\end{tabular}
\caption{Effect of solvent interaction mismatch on microscopic configurations for $\beta=0.6$ and $J_{0,0}=J_{+1,+1}=J_{-1,-1}=0$, $J_{-1,0}=1$ and $J_{+1,-1}=6$, and for different fractions of residual solvent (same as in Fig. \ref{fig:fig1}). First row:  $J_{+1,0}=1$. Second row: $J_{+1,0}=3$. Note in the 2nd and 3rd column that blue phase acts in the top layer as a barrier for the evaporation. In the two top rows, the fraction of residual solvent is equal to $0.4$, $0.3$ and $0.1$, respectively (from left to right). Bottom row: parameters as for the second row, but initial composition is $80:10:10$. The pictures represent the systems with residual solvent ratio $0.8$, $0.4$ and $0.1$, respectively (from left to right). The blue, yellow and red pixels represent the sites occupied by a ``$+1$'', ``$-1$'' or ``$0$'' spin, respectively.}
\label{fig:fig5}
\end{figure}

Figure \ref{fig:fig5} clarifies the effect of introducing a ``mismatch'' between the coefficients $J_{+1,0}$ and $J_{-1,0}$. As the interaction between $+1$ spins and the $0$ spins is now unfavored, the simulations reveal the onset of a pure phase constituted only by the spins ``$+1$''. Remarkably, due to the evaporation and the consequent replacement with spins $+1$ and $-1$, such phase is located on the top layer of the lattice and behaves as a sort of ``cap'', in that it hinders the evaporation of the solvent, thus ultimately slowing down the dynamics. This resembles the wetting layers discussed in \cite{Lyons2}. The difference between the system with initial composition $40:30:30$ and $80:10:10$ (second and third rows in Fig. \ref{fig:fig5}) is rather quantitative and is determined by the longer time necessary for the system, starting with composition $80:10:10$, to reach the state with 10\% residual solvent content and hence allowing the system to reduce the interface more. Due to the asymmetry in the interaction coefficients $J_{+1,0}$ and $J_{-1,0}$, we observe the formation of a large phase of spins $+1$ (blue), with the residual solvent positioned at the interface and in the phase of $-1$ spins (yellow). Phase purity was also discussed in \cite{Lyons1}.

\begin{figure}[h!]
\centering
\begin{tabular}{ccc}
\includegraphics[width = 0.33\textwidth]{con-128-128-0600-zz0-mz1-pz1-mm0-pp0-mp6-04-eps-converted-to.pdf} &
\includegraphics[width = 0.33\textwidth]{con-128-128-0600-zz0-mz1-pz1-mm0-pp0-mp6-03-eps-converted-to.pdf}  & 
\includegraphics[width = 0.33\textwidth]{con-128-128-0600-zz0-mz1-pz1-mm0-pp0-mp6-01-eps-converted-to.pdf}
\\[0.1cm]
\includegraphics[width = 0.33\textwidth]{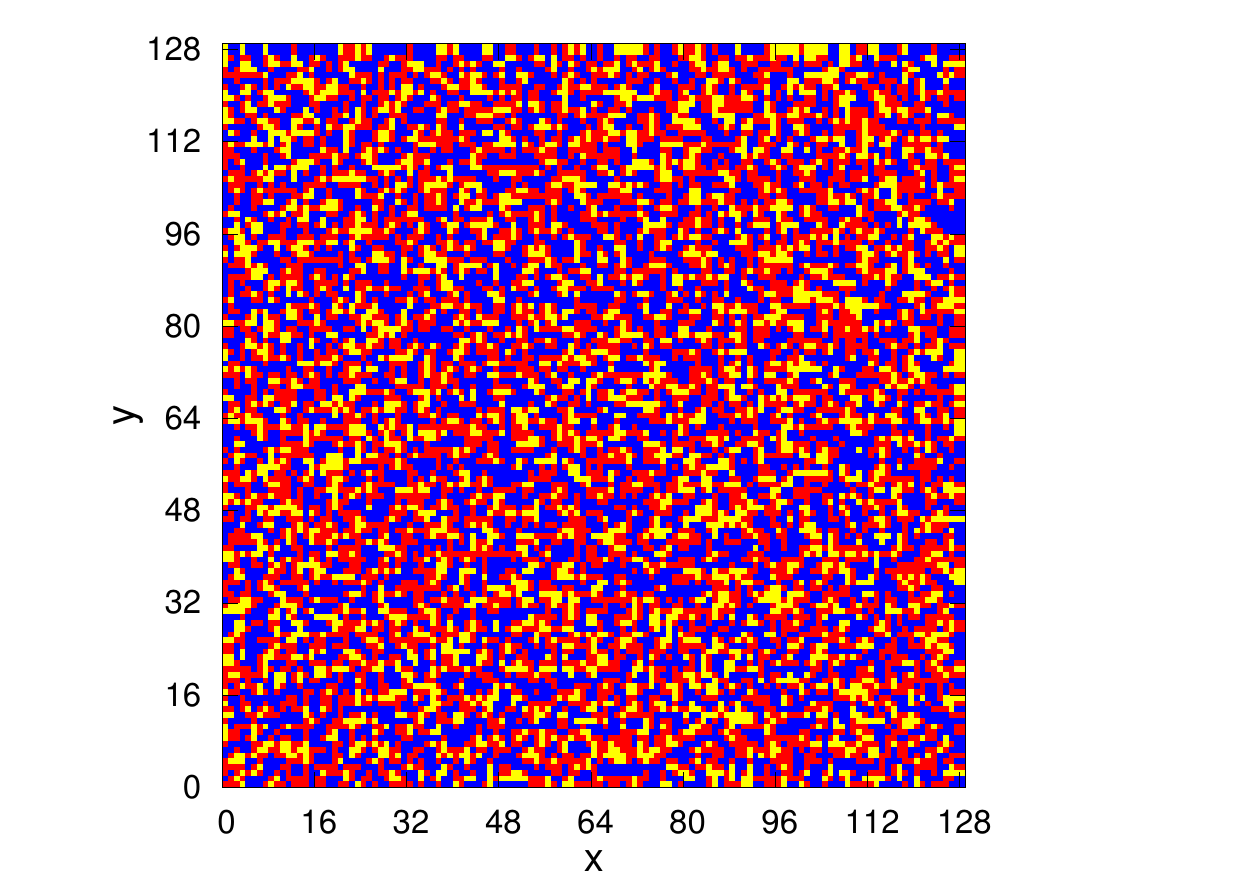} &
\includegraphics[width = 0.33\textwidth]{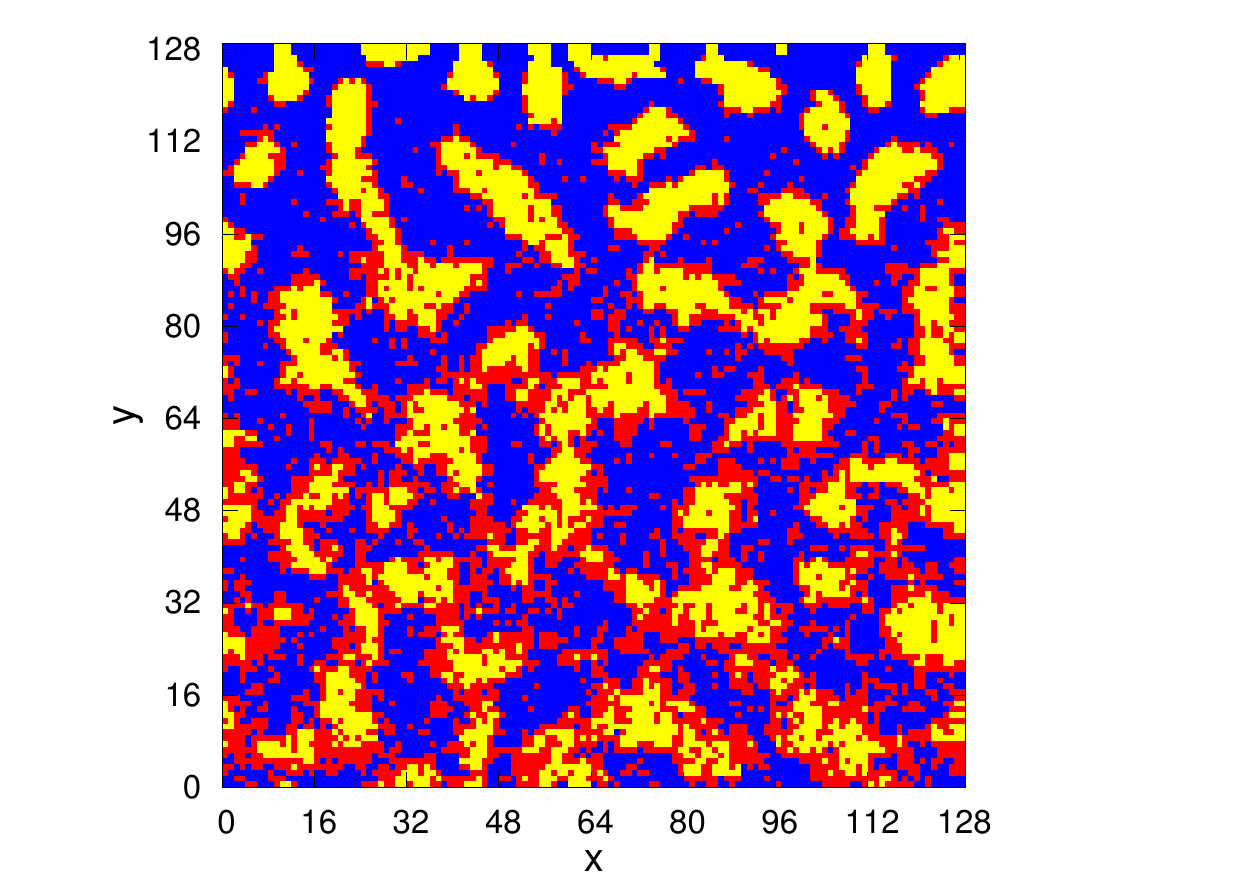}  & 
\includegraphics[width = 0.33\textwidth]{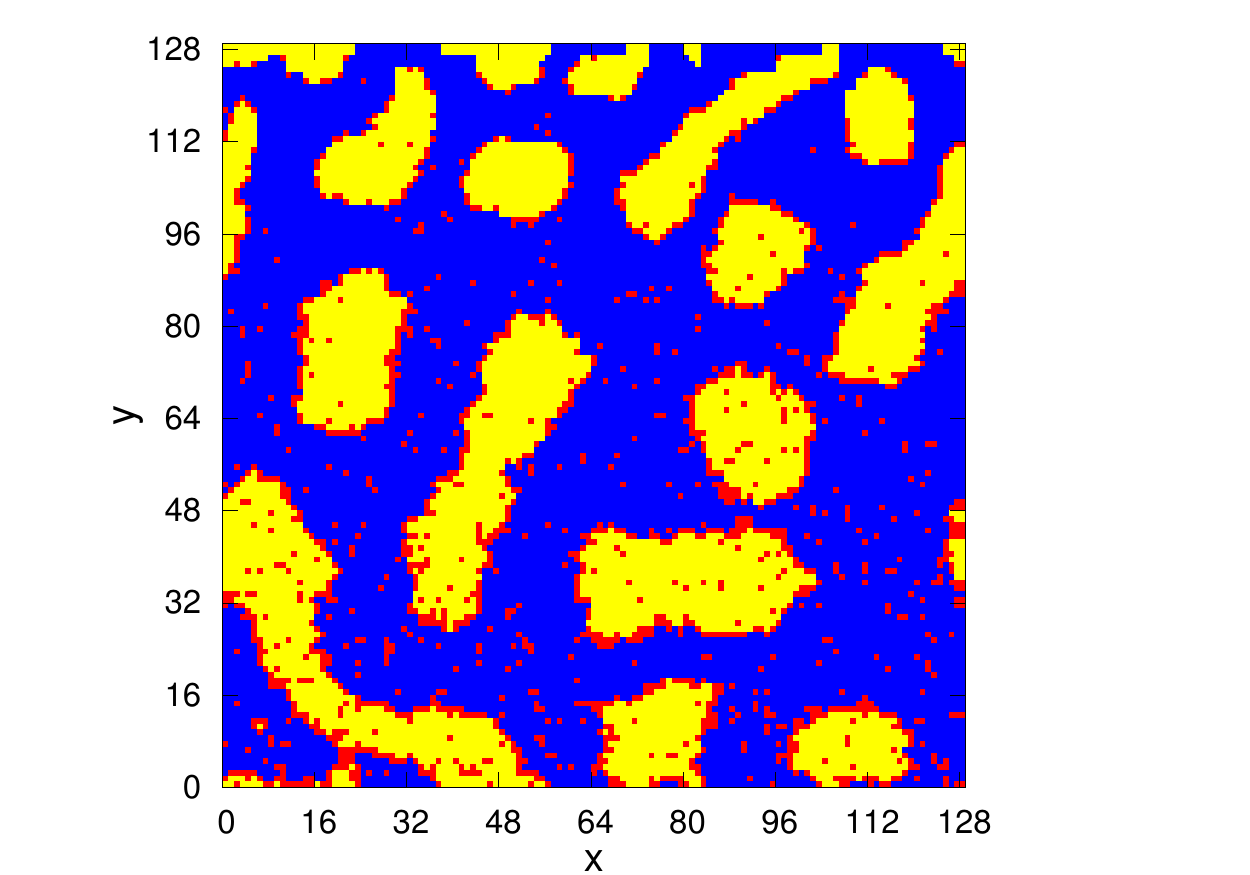}
\\[0.1cm]
\includegraphics[width = 0.33\textwidth]{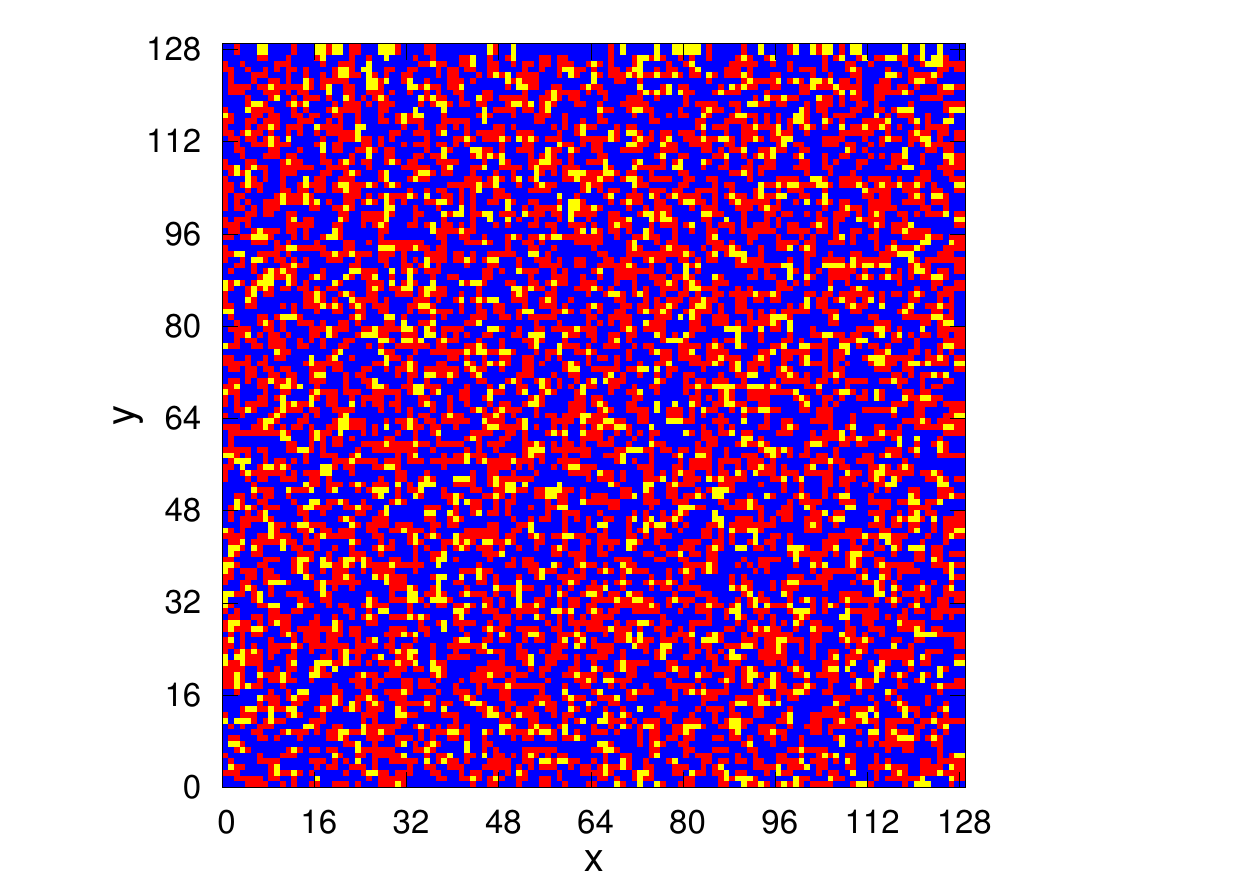} &
\includegraphics[width = 0.33\textwidth]{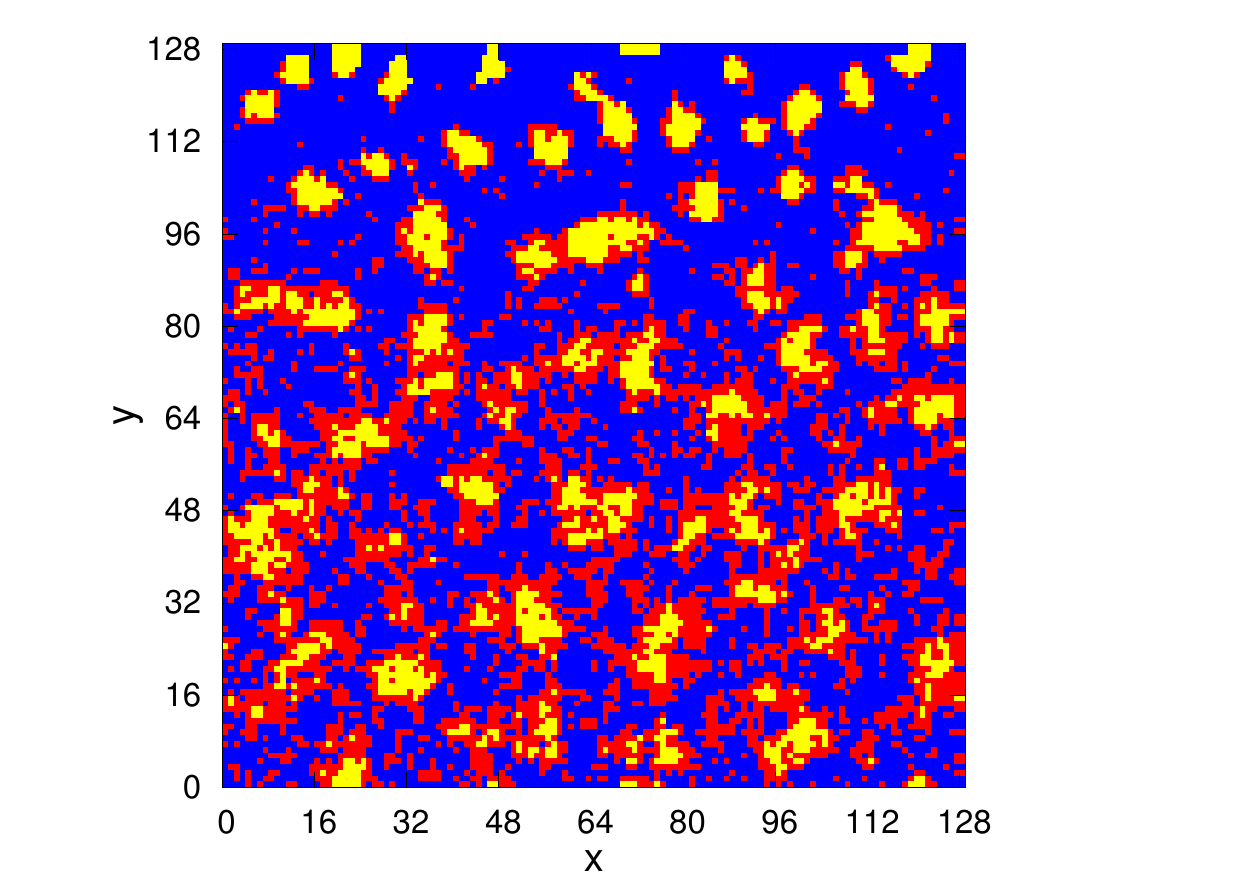}  & 
\includegraphics[width = 0.33\textwidth]{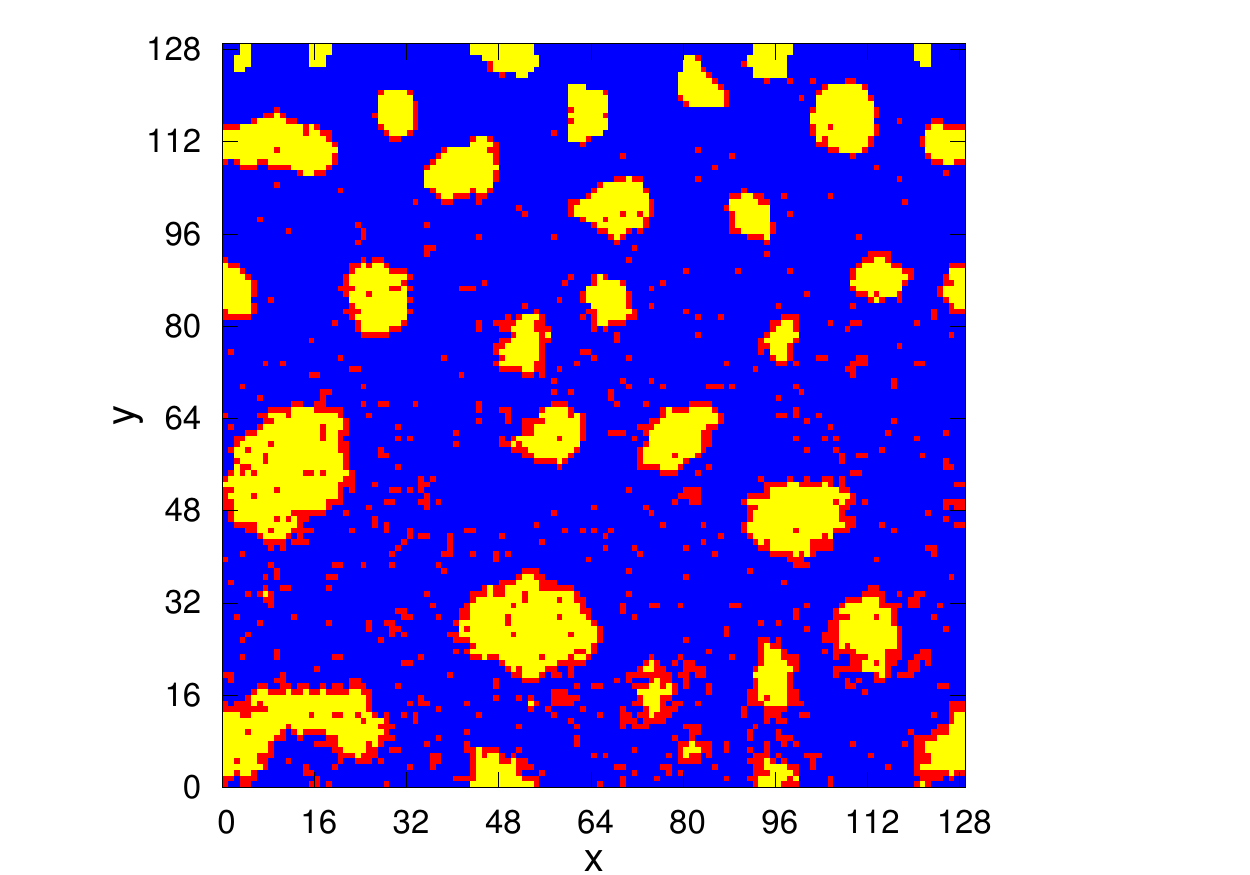}
\\[0.1cm]
\includegraphics[width = 0.33\textwidth]{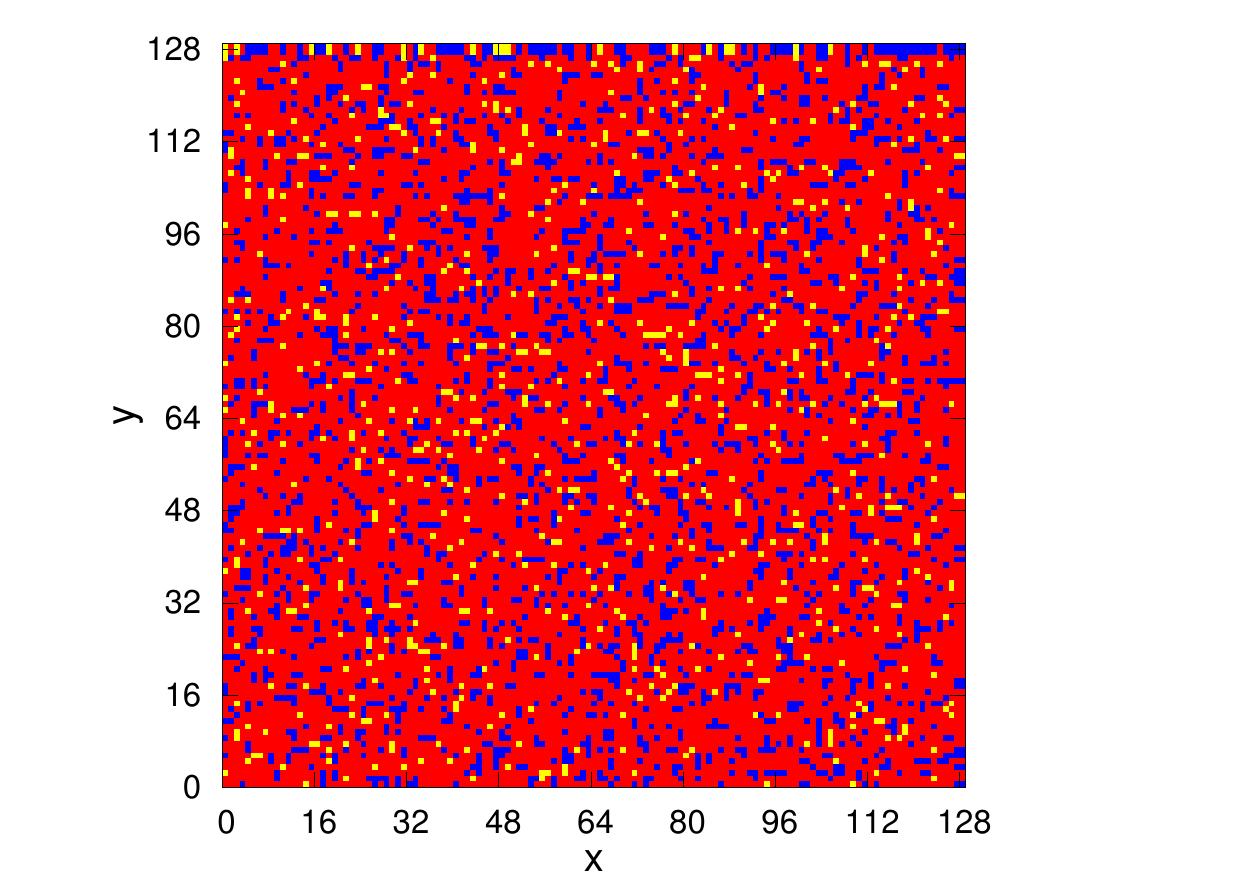} &
\includegraphics[width = 0.33\textwidth]{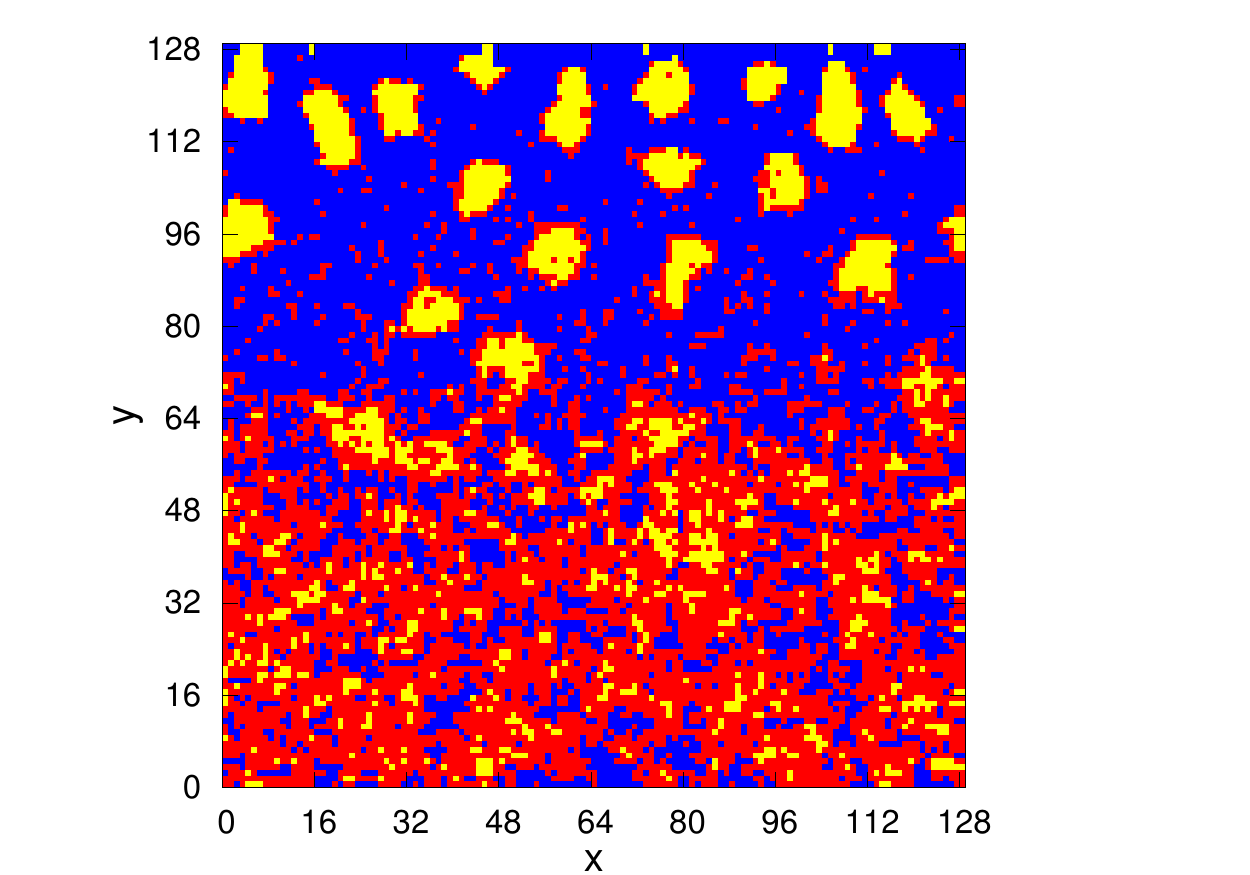}  & 
\includegraphics[width = 0.33\textwidth]{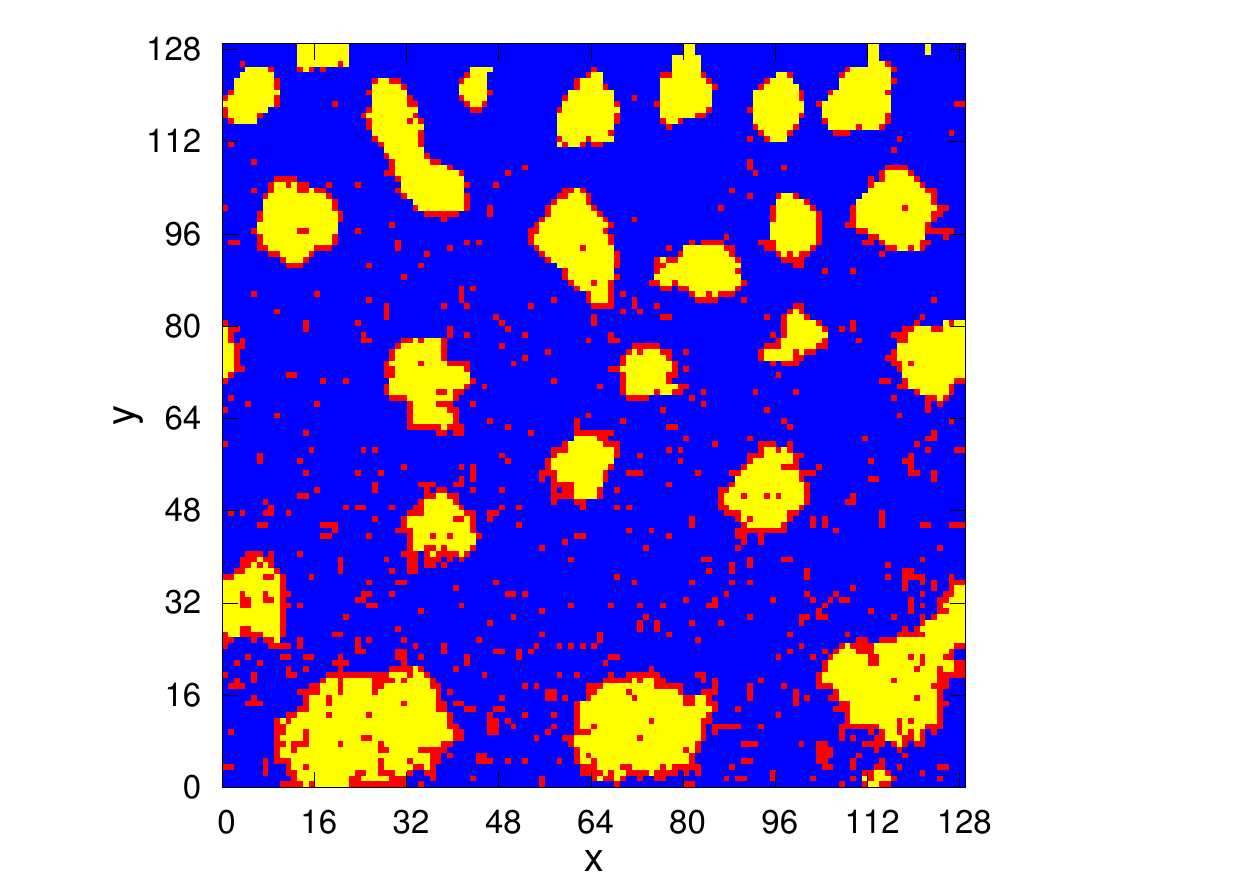}
\\[0.1cm]
\end{tabular}
\caption{Effect of initial proportions of mixture's components on configurations for different fractions of residual solvent (as in Fig.~\ref{fig:fig1}). 
We fix $J_{0,0}=J_{+1,+1}=J_{-1,-1}=0$, $J_{+1,0}=J_{-1,0}=1$ and $J_{+1,-1}=6$. The initial proportion of the three species $0,+1,-1$ is, respectively, $40:30:30$ (top row) , $40:40:20$ (second row) and $40:50:10$ (third row). The bottom row refers to the parameters as for the third row with initial composition $80:10:10$. In the three top rows, the fraction of residual solvent is equal to $0.4$, $0.3$ and $0.1$, respectively (from left to right). The bottom row pictures represent the systems with residual solvent ratio $0.8$, $0.4$ and $0.1$, respectively (from left to right). The blue, yellow and red pixels represent the sites occupied by a ``$+1$'', ``$-1$'' or ``$0$'' spin, respectively.}
\label{fig:fig6}
\end{figure}

The effect induced by varying the initial proportion between the three species of spins is portrayed in Fig. \ref{fig:fig6}, in which, in the first three rows, we fixed the initial amount of solvent and increased (respectively decreased) the amount of spins ``$+1$'' (respectively ``$-1$''). In the last row we show the results for the same parameters as the first three rows, but initial composition $80:16:04$, thus ratio of ``$+1$'' and ``$-1$'' spins comparable with the third row. We observe that the shape of the two phases changes with the initial composition, from interpenetrated fingers to islands of the less abundant phase (yellow) immersed in the abundant phase (blue). A similar situation can be found in experiments such as in \cite{Walheim,DalnokiVeress}. We also observe that the size of the islands is larger in the bottom of the pictures, in the regions where the solvent stays longer, allowing the other two species to diffuse laterally and, hence, forming larger phases. This effect is more pronounced in the last row, with initial composition $80:16:04$, where the evaporation of the solvent takes even longer. We also observe a difference in domain sizes between top and bottom of the picture whenever the evaporation of the solvent takes a long enough time.

\begin{figure}[h!]
\centering
\begin{tabular}{ccc}
\includegraphics[width = 0.33\textwidth]{con-128-128-0600-zz0-mz1-pz1-mm0-pp0-mp6-04-eps-converted-to.pdf} &
\includegraphics[width = 0.33\textwidth]{con-128-128-0600-zz0-mz1-pz1-mm0-pp0-mp6-03-eps-converted-to.pdf}  & 
\includegraphics[width = 0.33\textwidth]{con-128-128-0600-zz0-mz1-pz1-mm0-pp0-mp6-01-eps-converted-to.pdf}
\\[0.5cm]
\includegraphics[width = 0.33\textwidth]{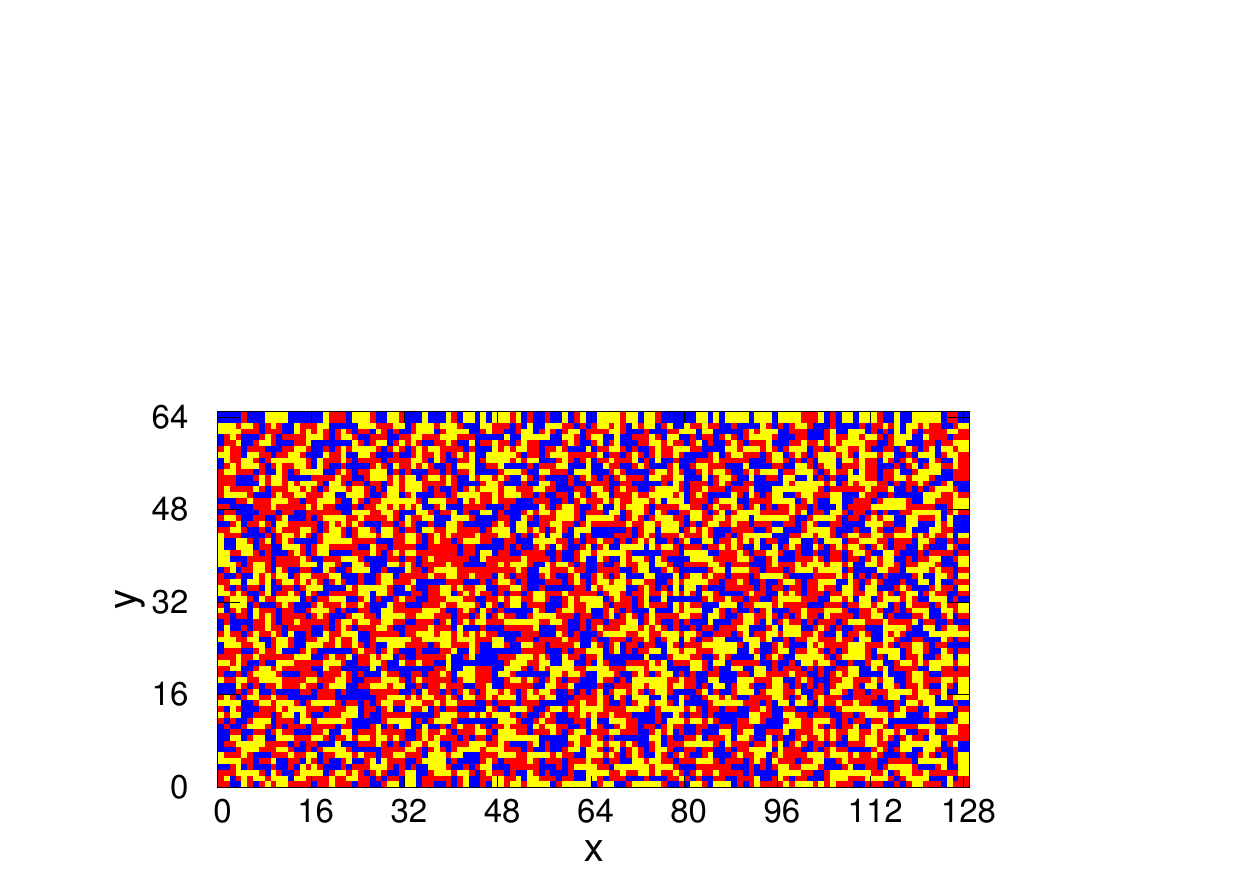} &
\includegraphics[width = 0.33\textwidth]{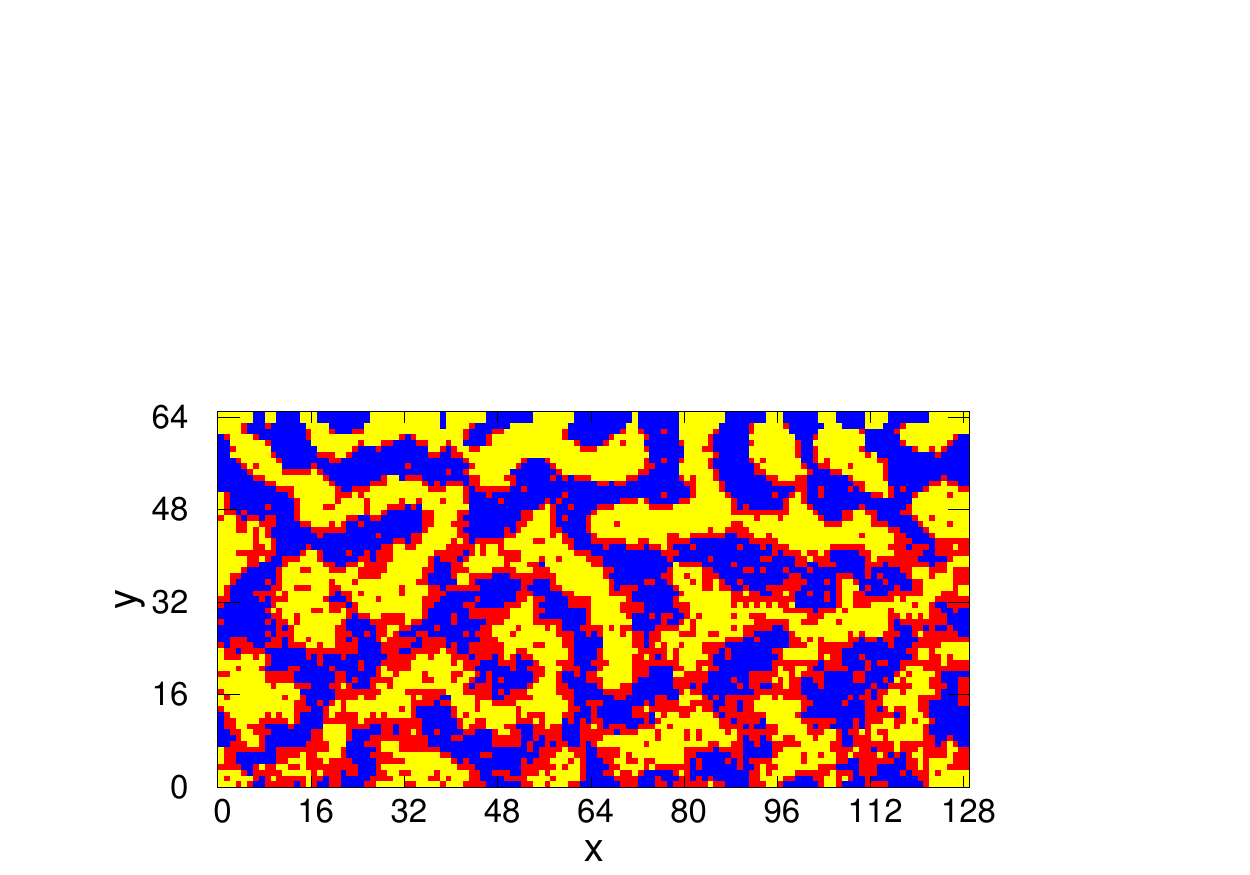}  & 
\includegraphics[width = 0.33\textwidth]{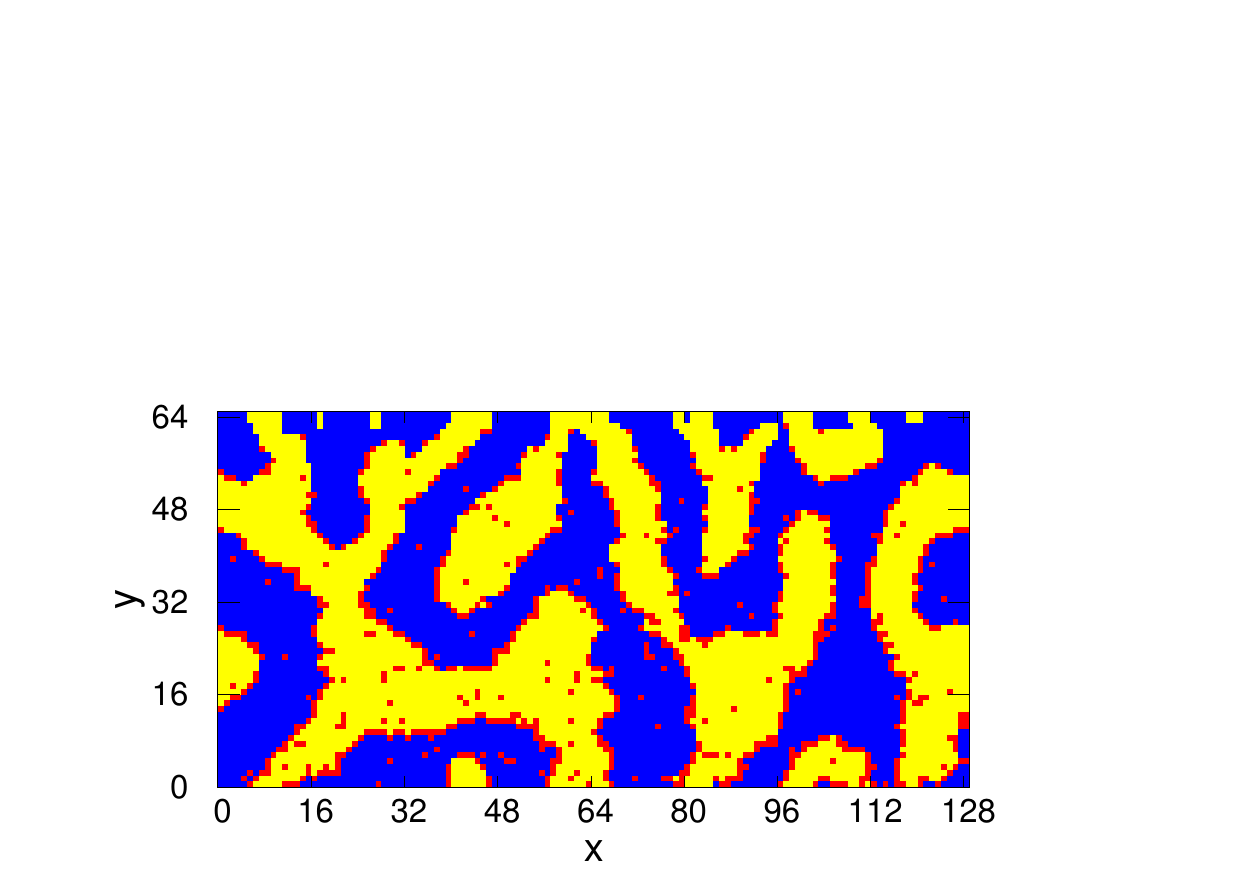}
\\[0.5cm]
\includegraphics[width = 0.33\textwidth]{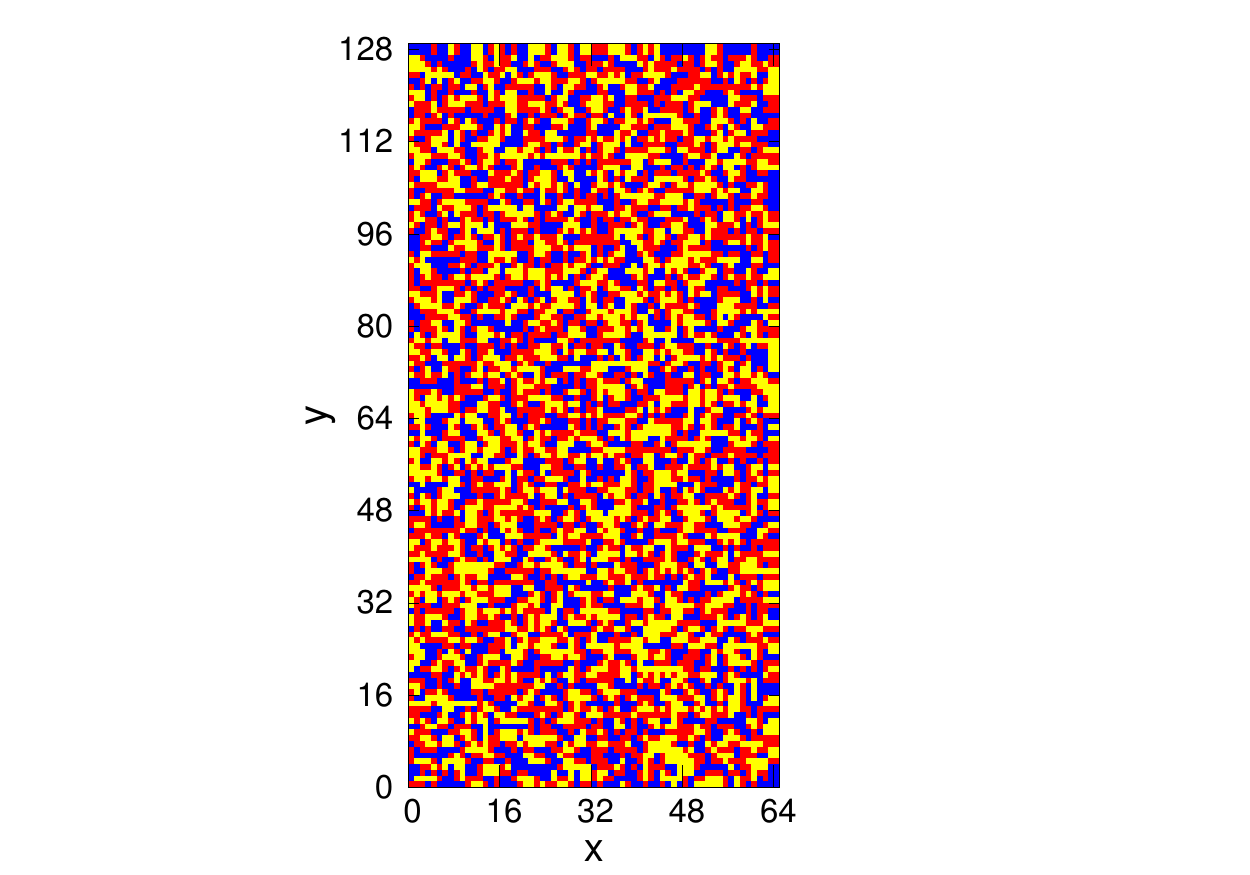} &
\includegraphics[width = 0.33\textwidth]{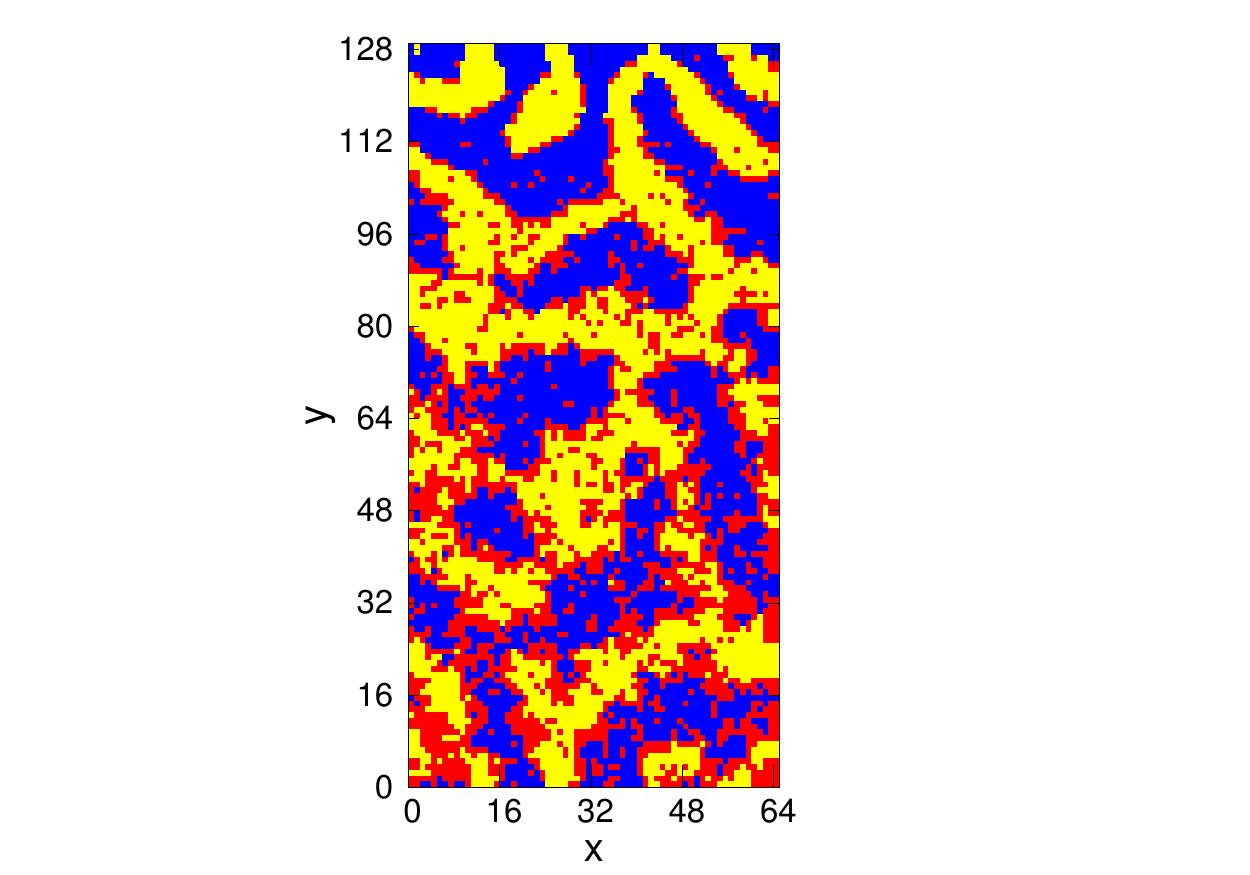}  & 
\includegraphics[width = 0.33\textwidth]{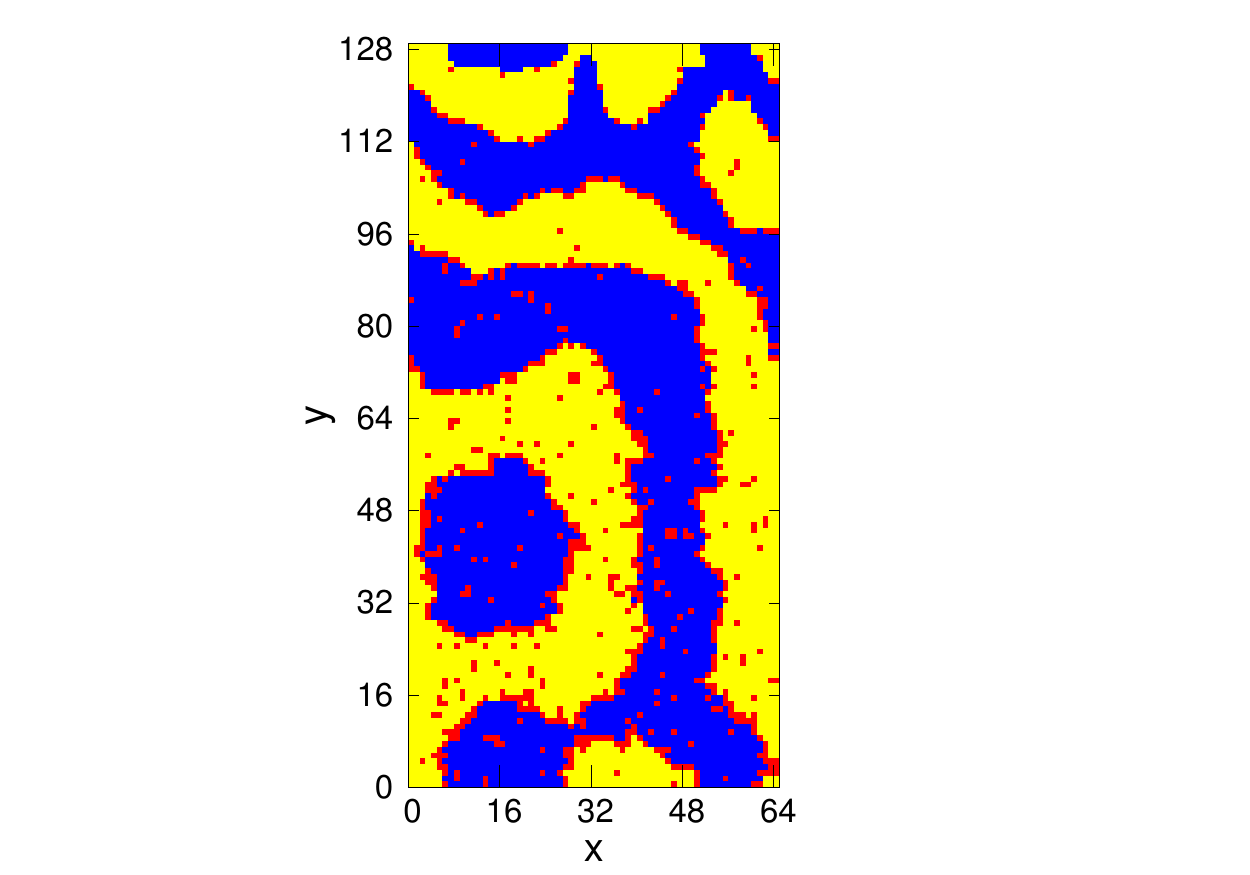}
\\[0.5cm]
\end{tabular}
\caption{Effect of box size on microscopic configurations, for different fractions of residual solvent  and with the same set of parameters used in Fig. \ref{fig:fig1}. The size is expressed in lattice points and is shown on the axes of coordinates in each picture. In all rows, the fraction of residual solvent is equal to $0.4$, $0.3$ and $0.1$, respectively (from left to right). The blue, yellow and red pixels represent the sites occupied by a ``$+1$'', ``$-1$'' or ``$0$'' spin, respectively.}
\label{fig:fig7}
\end{figure}

We also considered different lattice sizes for our model, the results obtained for a square or, rather, a rectangular box are portrayed in Fig. \ref{fig:fig7}.

\begin{figure}[h!]
\centering
\begin{tabular}{ccc}
\includegraphics[width = 0.33\textwidth]{con-128-128-0600-zz0-mz1-pz1-mm0-pp0-mp6-04-eps-converted-to.pdf} &
\includegraphics[width = 0.33\textwidth]{con-128-128-0600-zz0-mz1-pz1-mm0-pp0-mp6-03-eps-converted-to.pdf}  & 
\includegraphics[width = 0.33\textwidth]{con-128-128-0600-zz0-mz1-pz1-mm0-pp0-mp6-01-eps-converted-to.pdf}
\\[0.5cm]
\includegraphics[width = 0.33\textwidth]{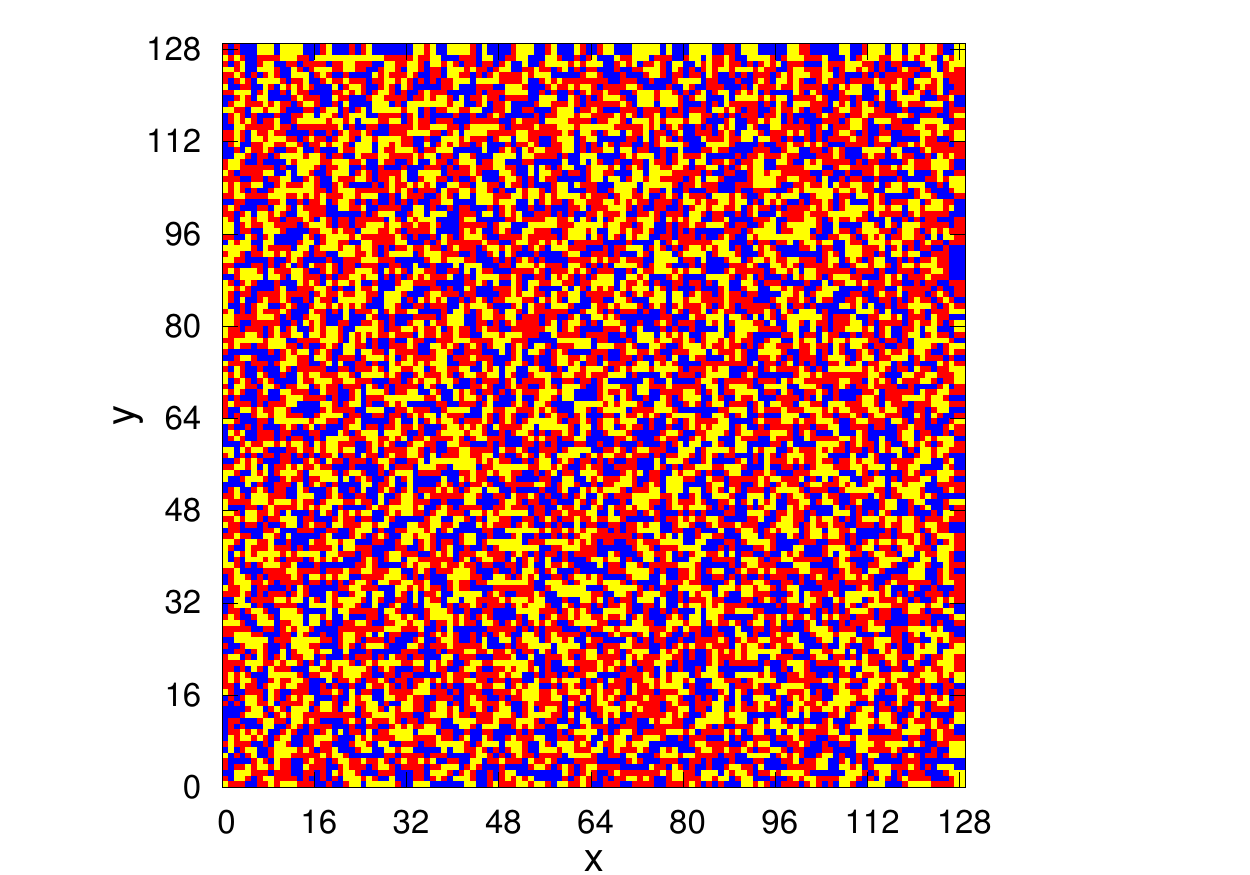} &
\includegraphics[width = 0.33\textwidth]{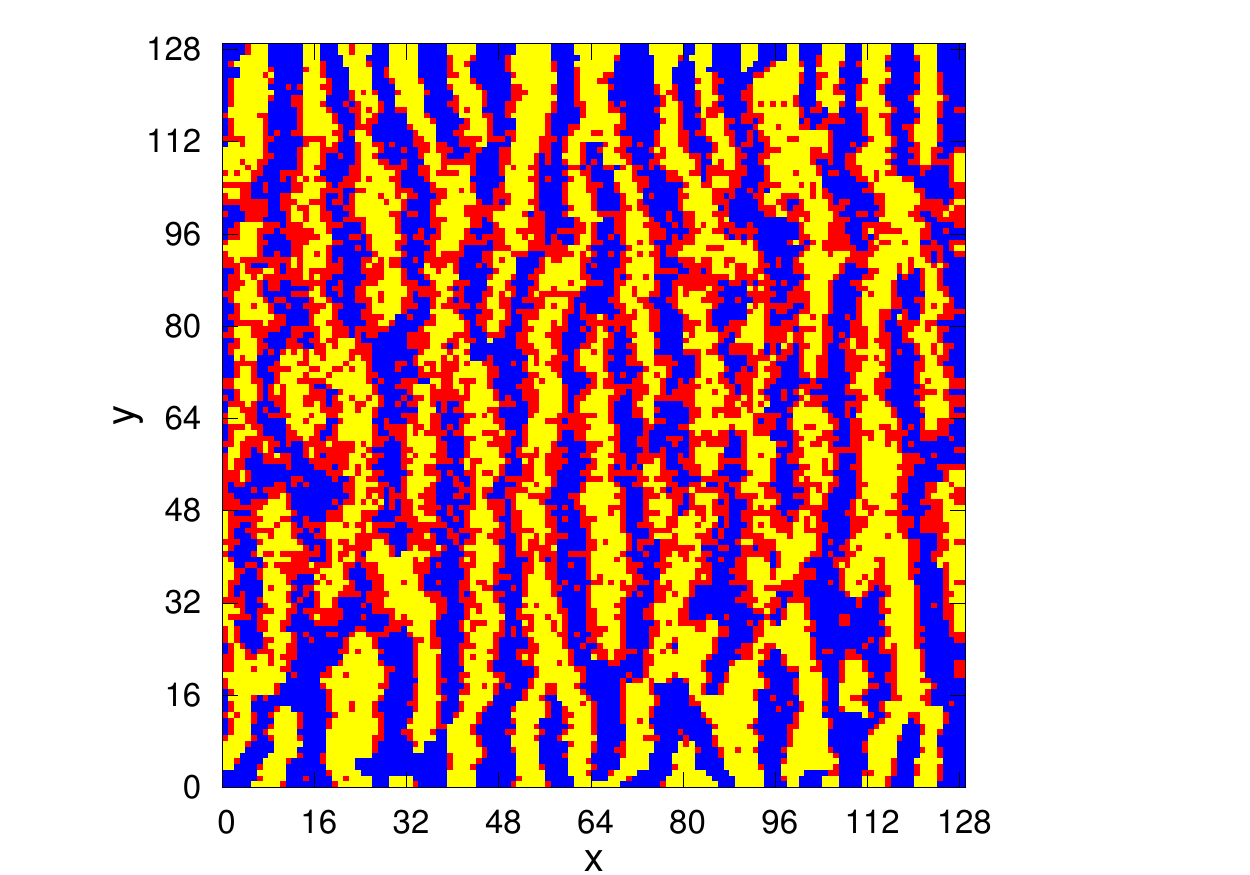}  & 
\includegraphics[width = 0.33\textwidth]{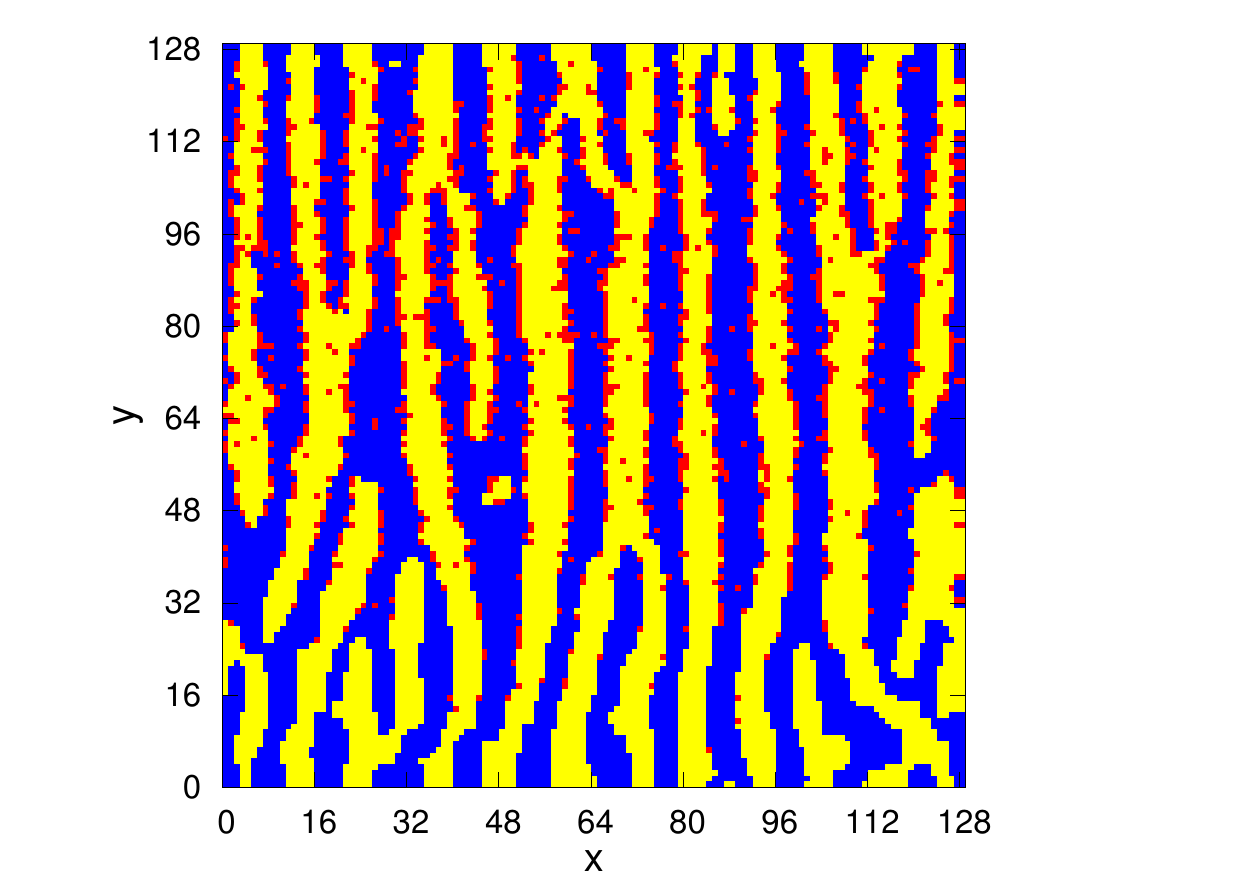}
\\[0.5cm]
\includegraphics[width = 0.33\textwidth]{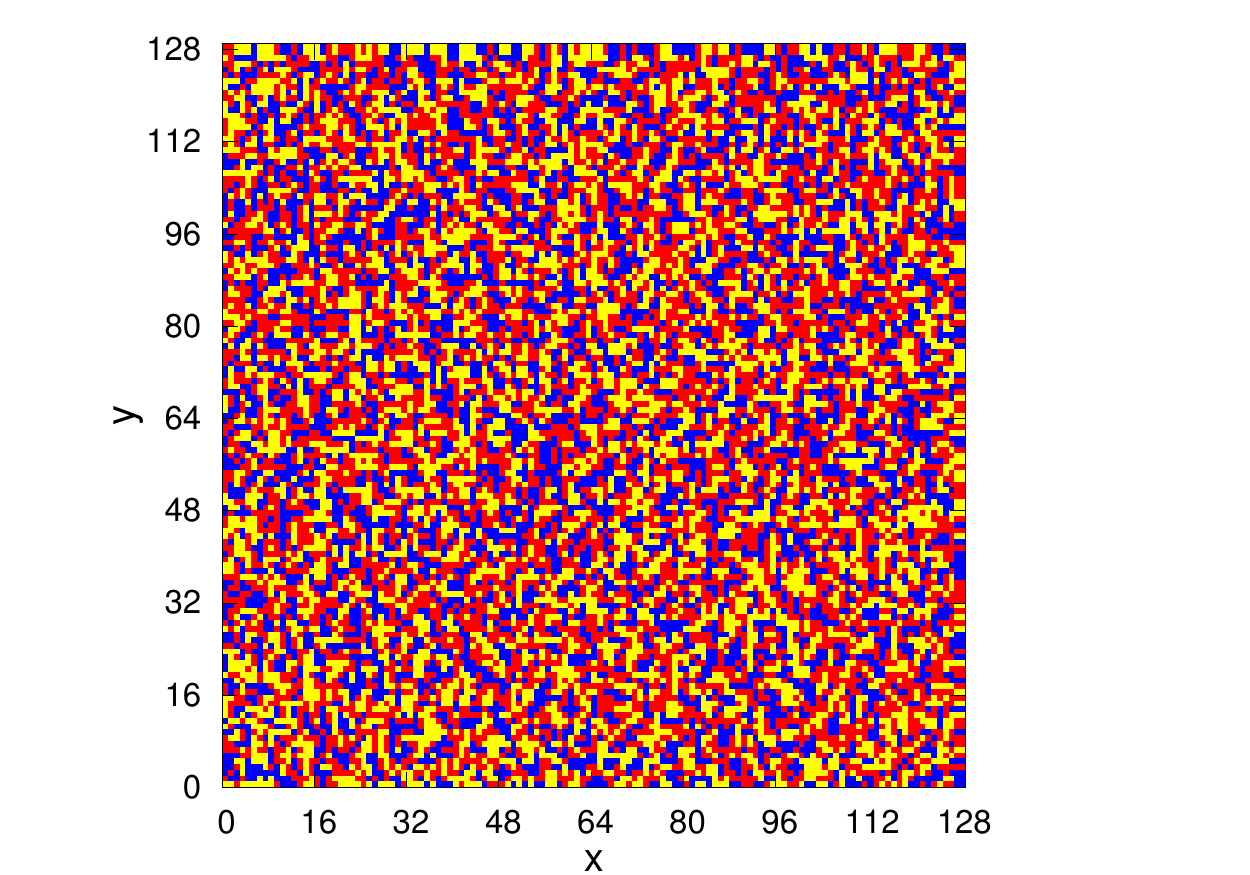} &
\includegraphics[width = 0.33\textwidth]{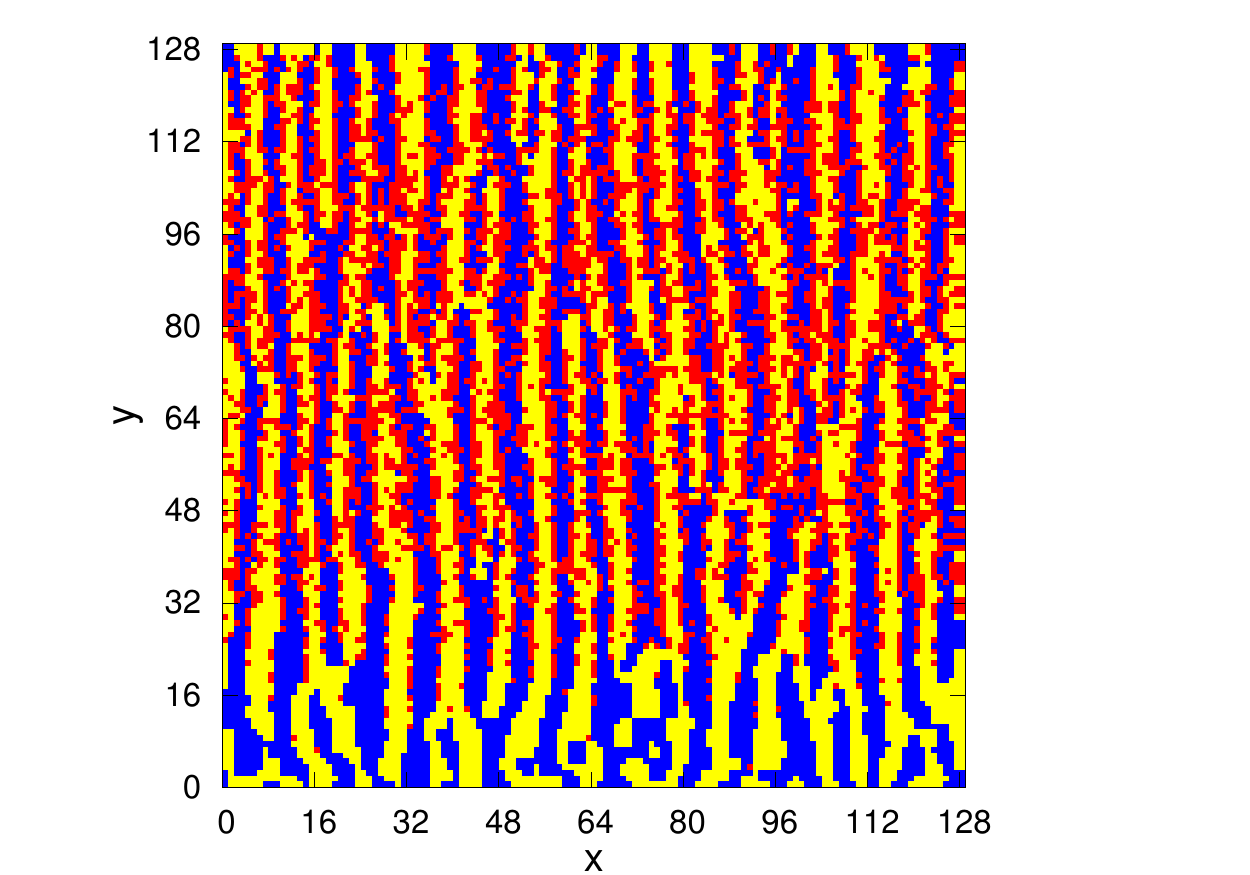}  & 
\includegraphics[width = 0.33\textwidth]{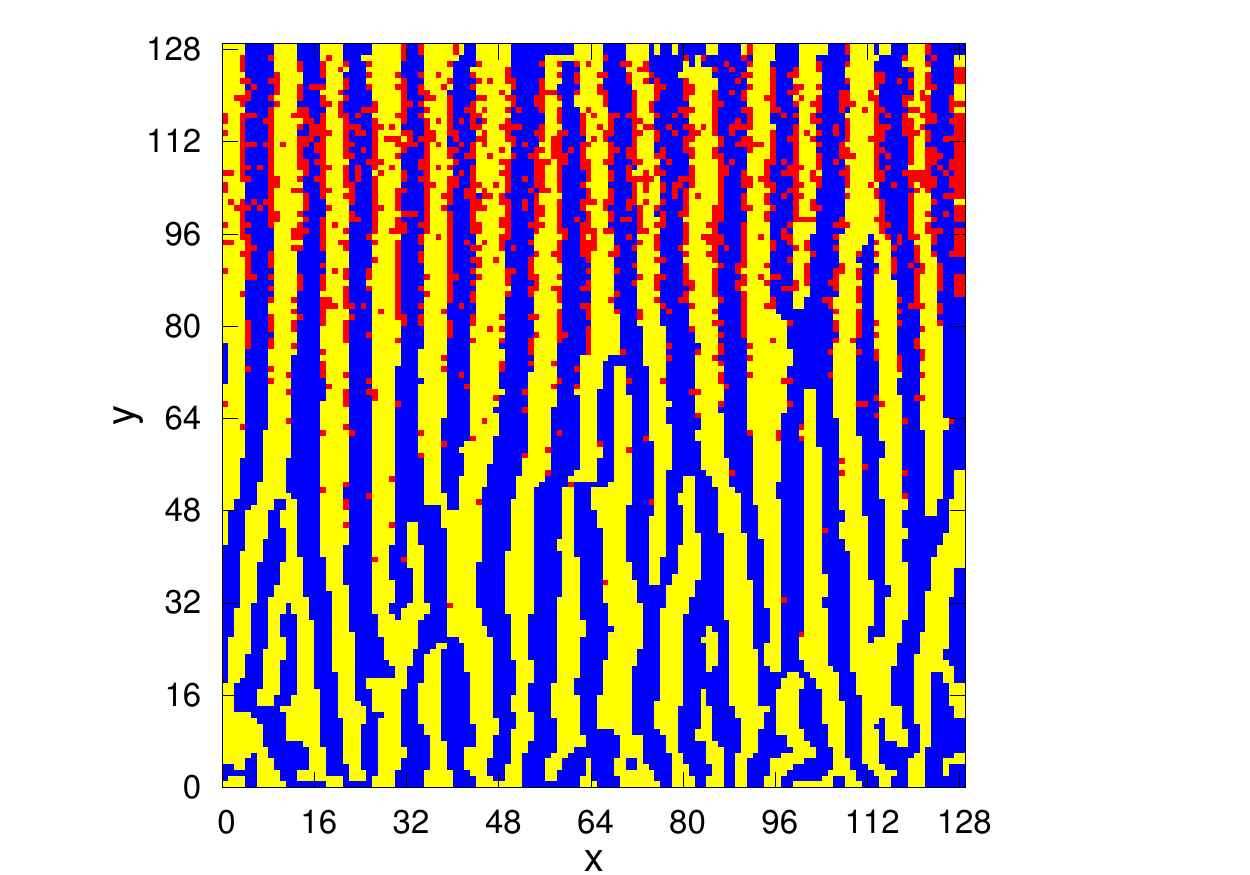}
\\[0.5cm]
\end{tabular}
\caption{Effect of the volatility of the solvent on the microscopic configurations, for different fractions of residual solvent and with the same set of parameters used in Fig. \ref{fig:fig1}. Shown are the configurations obtained for volatility parameters $\phi$=0 (first row), $\phi$=0.1 (second row) and $\phi$=1 (third row). In all rows, the fraction of residual solvent is equal to $0.4$, $0.3$ and $0.1$, respectively (from left to right). The blue, yellow and red pixels represent the sites occupied by a ``$+1$'', ``$-1$'' or ``$0$'' spin, respectively.}
\label{fig:fig8}
\end{figure}

The effect induced by the volatility parameter $\phi$ is highlighted in Fig. \ref{fig:fig8}. We see here the microscopic configurations obtained for different values of volatility, namely $\phi$=0, $\phi$=0.1 and $\phi$=1 in the first, second and third row, respectively. Remarkably, even small values of volatility give rise to vertical channels, the thickness of the channels decreasing with increasing volatility. For the chosen parameters, the solvent evaporates by diffusing upwards, in the bulk, mainly along the interfaces between the yellow and blue laminar structures. We conjecture that a suitable tuning of this parameter may help to account for the effect of the buoyancy, in presence of gravity.

\begin{figure}[h!]
\centering
\begin{tabular}{ccc}
\includegraphics[width = 0.33\textwidth]{con-128-128-0600-zz0-mz1-pz1-mm0-pp0-mp6-04-eps-converted-to.pdf} &
\includegraphics[width = 0.33\textwidth]{con-128-128-0600-zz0-mz1-pz1-mm0-pp0-mp6-03-eps-converted-to.pdf}  & 
\includegraphics[width = 0.33\textwidth]{con-128-128-0600-zz0-mz1-pz1-mm0-pp0-mp6-01-eps-converted-to.pdf}
\\[0.5cm]
\includegraphics[width = 0.33\textwidth]{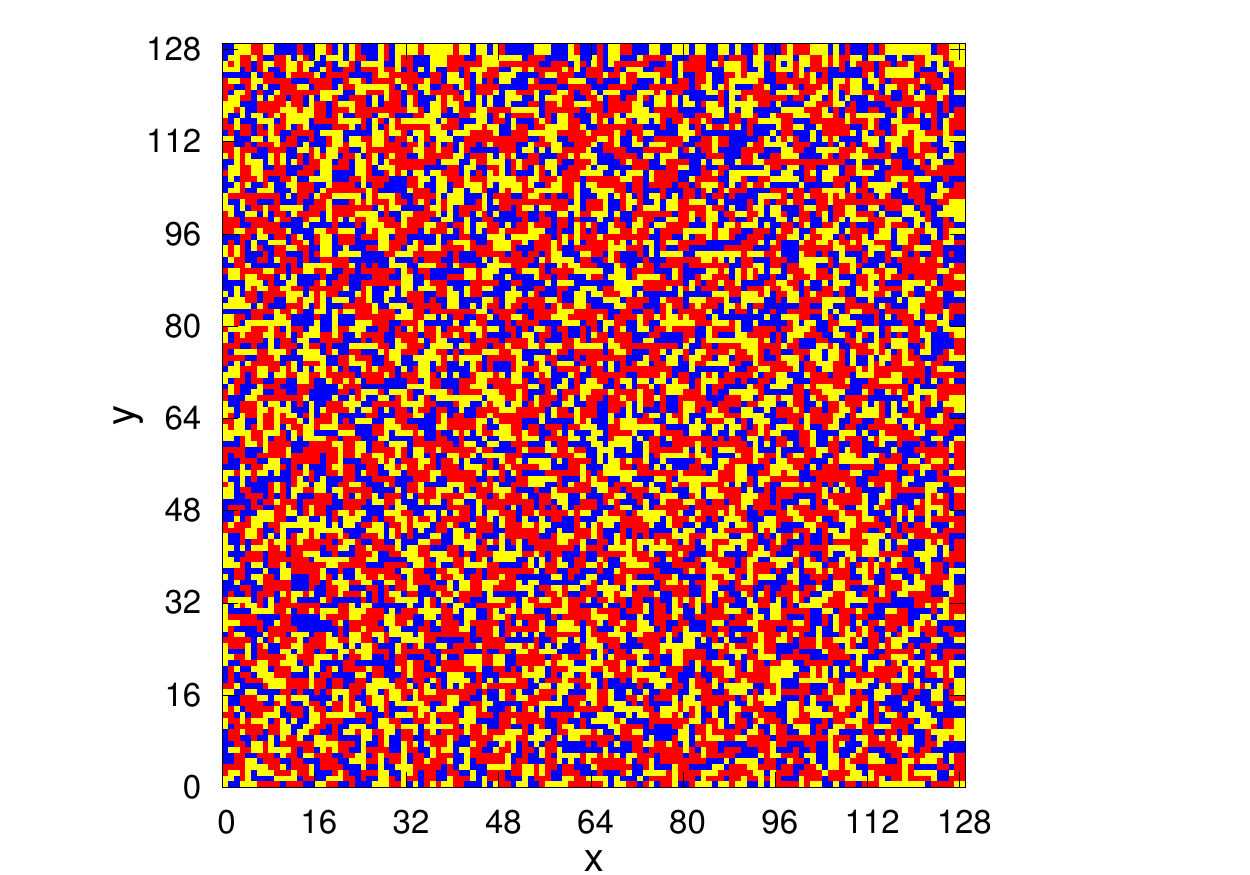} &
\includegraphics[width = 0.33\textwidth]{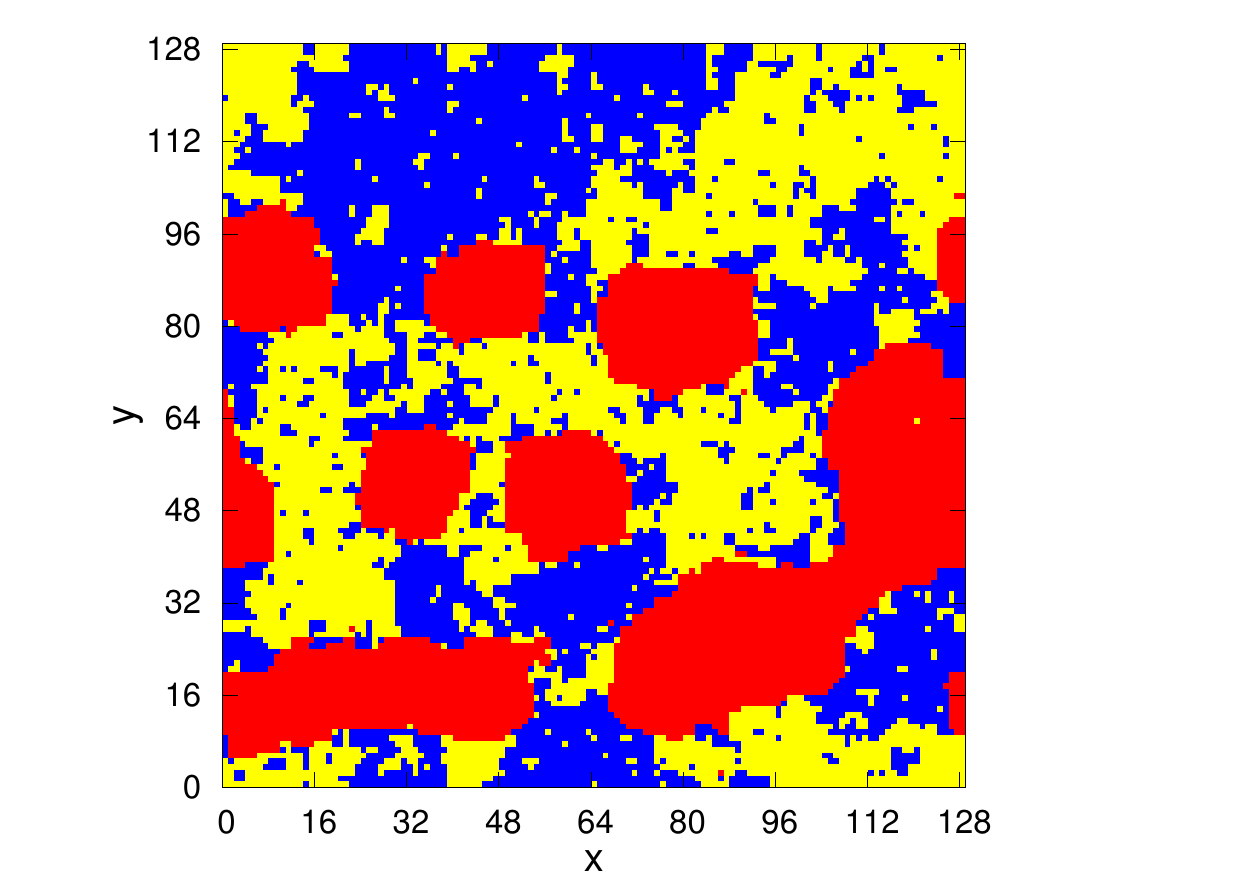}  & 
\includegraphics[width = 0.33\textwidth]{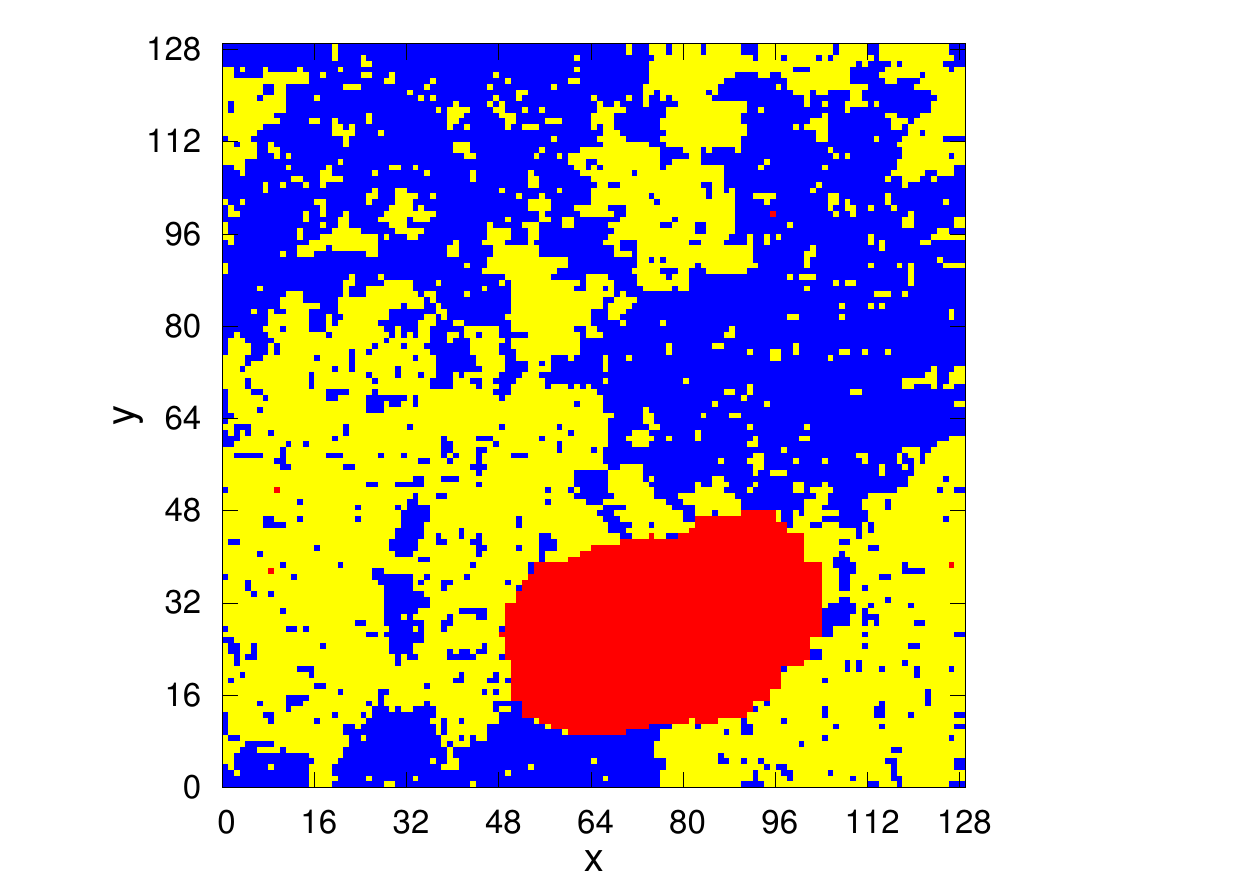}
\\[0.5cm]
\includegraphics[width = 0.33\textwidth]{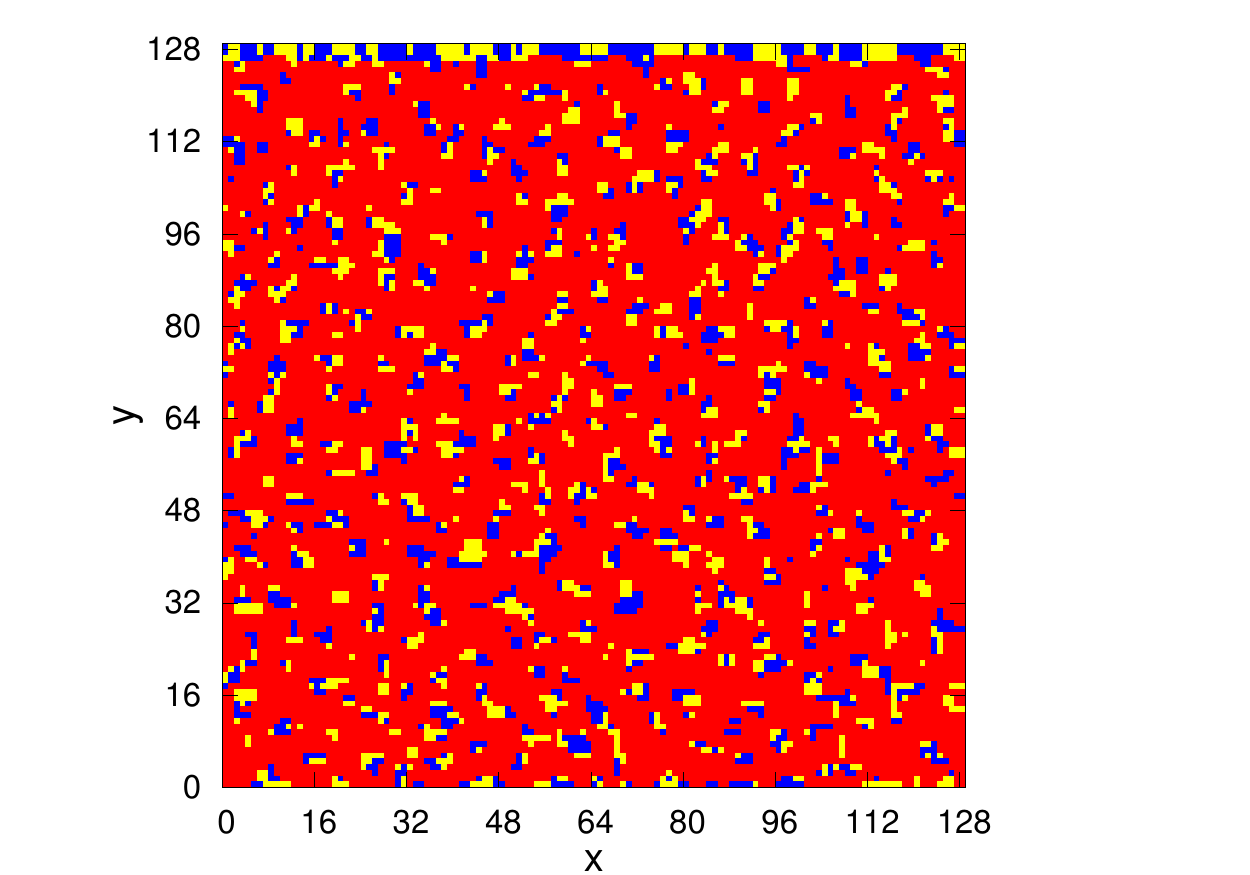} &
\includegraphics[width = 0.33\textwidth]{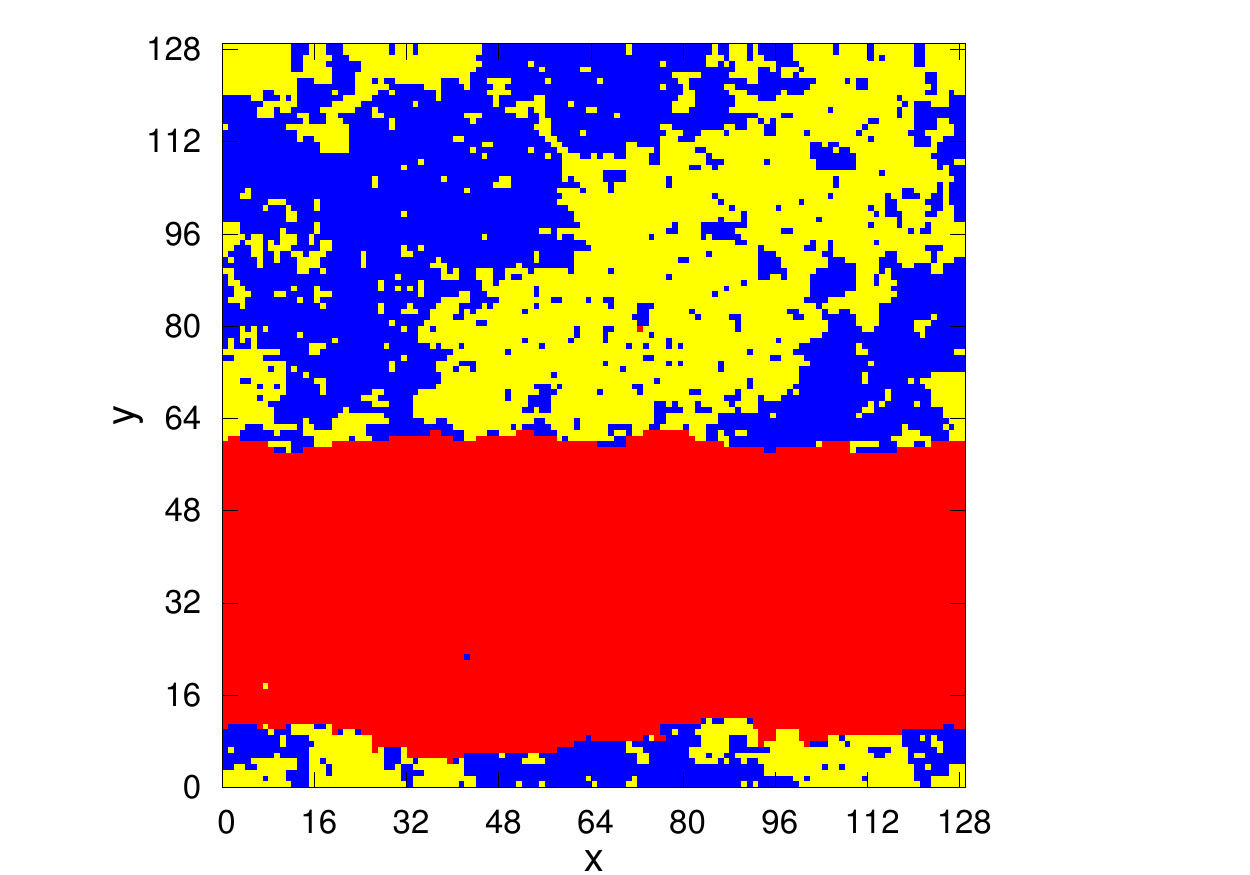}  & 
\includegraphics[width = 0.33\textwidth]{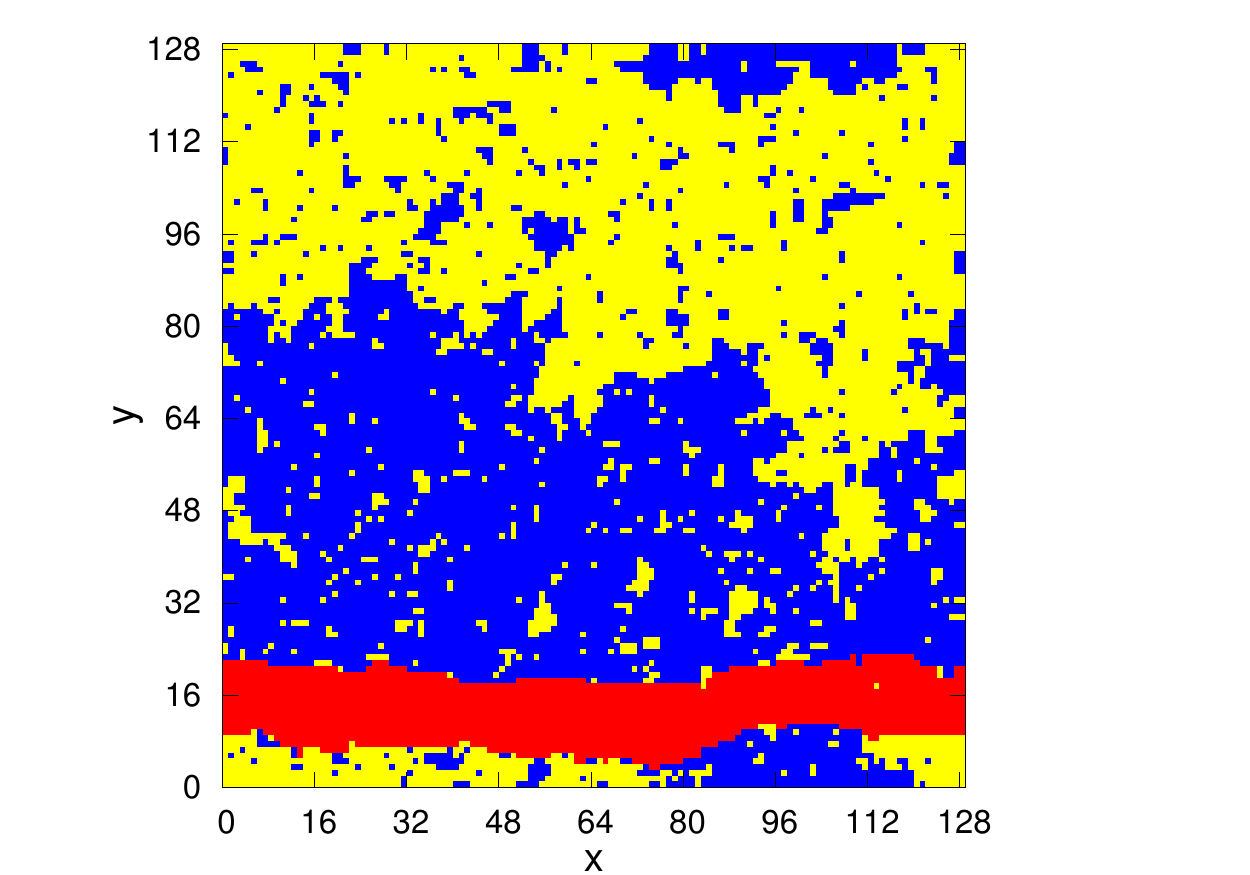}
\\[0.5cm]
\end{tabular}
\caption{Effect of strong repulsion between spins ``$0$'' and spins ``$-1$'' and between spins ``$0$'' and spins ``$+1$'', as well as weak repulsion  between spins ``$+1$'' and spins ``$-1$'' on the microscopic configurations, for different fractions of residual solvent (same as in Fig. \ref{fig:fig1}). (i) Top row: $\beta=0.6$, $J_{0,0}=J_{+1,+1}=J_{-1,-1}=0$, $J_{+1,0}=J_{-1,0}=1$, and $J_{+1,-1}=6$; (ii) Middle row: $\beta=0.6$, $J_{0,0}=J_{+1,+1}=J_{-1,-1}=0$, $J_{+1,0}=J_{-1,0}=35$, and $J_{+1,-1}=15$; (iii) Bottom row: parameters as for the middle row, but the initial composition is $80:10:10$. In the two top rows, the fraction of residual solvent is equal to $0.4$, $0.3$ and $0.1$, respectively (from left to right). The bottom row pictures represent the systems with residual solvent ratio $0.8$, $0.4$ and $0.1$, respectively (from left to right). The blue, yellow and red pixels represent the sites occupied by a ``$+1$'', ``$-1$'' or ``$0$'' spin, respectively.}
\label{fig:fig9}
\end{figure}

Finally, Fig. \ref{fig:fig9} illustrates the effect of an unfavorable interaction between the ``$0$'' spins and the spins ``$+1$'' or ``$-1$''. While in the first row, where $J_{+1,0}=J_{-1,0}=1$, $J_{+1,-1}=6$, the ``$0$'' spins acted as a sort of shield between the spins ``$+1$'' and ``$-1$'', in the second and third rows the situation is somehow reversed: the high energy cost of a spin ``$0$'' neighbor to the spins ``$+1$'' or ``$-1$'' induces an interesting phenomenon of coalescence of the ``$0$'' spins, that form a single, large drop immersed in a sea of ``$+1$'' or ``$-1$'' spins. The difference between the second and the third row is the initial composition, hence the longer time needed for the residual solvent to reach 10\%. This results in the formation of a compact layer of residual solvent, rather than individual drops.

\clearpage
\newpage

\section{Conclusions}
\label{sec:4}
\par\noindent
This work is a first step of a program that aims at describing the morphology observed in thin film evaporation experiments on the basis of lattice gas models that are amenable to an analytical or, typically, a numerical investigation. We showed that even this relatively simple model can be used to understand the influence of the multiple external and internal parameters on the morphology formation in ternary systems upon evaporation of one component. We observed that for low temperatures the phase separation appear for all range of interaction parameters considered. The ratio of the components, the strength of interaction, and the volatility of the solvent influence the shapes of the formed phases.

In future works we plan to consider also the effect of an intermediate, mesoscopic scale of interaction, that is expected to be vastly separated from the microscopic scale (conventionally taken as the unity) corresponding to the distance between two neighboring sites on the lattice, and also from the macroscopic scale describing the length of the box \cite{P2008}. A promising route points towards the use of Kac interaction potentials, following the analysis traced in \cite{CDMP2016,CDMP2017,CDMP17bis}. Also, it is worth investigating the same system in a multiscale perspective, aiming to obtain hierarchical structure formation as shown in Figure 3c in \cite{Sprenger}.
Finally, it would also be of interest to consider the effect of local heterogeneities in the lattice gas dynamics \cite{CC17,CCM16,CCM16bis,CCM17} on the characteristic time scales of morphology formation.

\begin{acknowledgments}
ENMC thanks FFABR 2017 for financial support. MC acknowledges financial support from FFABR 2017 and from the Lerici Foundation. SAM and JvS acknowledge the funding from the Swedish National Space Agency under contracts 148/16 and 185/17.
\end{acknowledgments}

\end{document}